\documentclass[final,11pt, 3p]{elsarticle}
\usepackage{graphics, color, hyperref, soul}
\soulregister\cite7
\soulregister\ref7
\usepackage{subfig, ulem}
\usepackage{lineno}
\usepackage{multirow}
\usepackage{amssymb}

\renewcommand{\emph}[1]{\textit{#1}}
 
\def\bbn{$\beta\beta$-0$\nu$}

\def\Fe{$^{55}$Fe}
\def\Re{$^{187}$Re}
\def\Ho{$^{163}$Ho}
\def\Dy{$^{163}$Dy}

\def\agre{AgReO$_4$}

\def\mne{$m_{\nu_e}$}

\def\mnesq{$m_{\nu_e}^2$}
\def\mug{$\mu$g}
\def\mum{$\mu$m}

\def\mus{$\mu$s}

\def\de{$\Delta E$}
\def\fwhm{$_{\mathrm{FWHM}}$}
\def\Aec{$A_\mathrm{EC}$}

\def\fpp{$f_{pp}$}
\def\Nev{$N_{ev}$}
\def\ero{Er$_2$O$_3$}

\def\Ho{$^{163}$Ho}
\def\Dy{$^{163}$Dy}
\def\Er{$^{162}$Er}

\begin{document}
\journal{Advances in High Energy Physics}

\title{The use of low temperature detectors for direct measurements of the mass of the electron neutrino}

\address[Milano]{Dipartimento di Fisica, Universit\`a di Milano-Bicocca, Milano I-20126 - Italy}
\address[INFNMiB]{INFN - Sezione di Milano Bicocca, Milano I-20126 - Italy}
\author[Milano,INFNMiB]{A.\,Nucciotti}

\date{\today}

\begin{abstract}
Recent years have witnessed many exciting breakthroughs in neutrino physics. 
The detection of neutrino oscillations has proved that neutrinos are massive particles but the assessment of their absolute
mass scale is still an outstanding challenge in today particle physics and cosmology.
Since low temperature detectors were first proposed for neutrino physics experiments in 1984,
there have been tremendous technical progresses: today this technique offers the high energy resolution and scalability required
to perform competitive experiments challenging the lowest electron neutrino masses.
This paper reviews the thirty-year effort aimed at realizing a calorimetric measurements with sub-eV neutrino mass sensitivity using low temperature detectors.
\end{abstract}

%
%
%

\maketitle

\tableofcontents

\section{Introduction}
Almost two decades ago, the discovery of neutrino flavor oscillations firmly demonstrated that neutrinos are massive particles \cite{PDG14}. This was a crucial breach in the Standard Model of fundamental interactions which assumed massless neutrinos. 
Flavor oscillations show that the three active neutrino flavor states ($\nu_e$, $\nu_\mu$, and $\nu_\tau$) are a superposition of three  mass states ($\nu_1$, $\nu_2$, and $\nu_3$), and allow to measure the difference between the squared mass of the neutrino mass states; but they are not at all sensitive  
to the absolute masses of the neutrinos. 

Today, assessing the neutrino mass scale is still an outstanding task for particle physics, 
as the absolute value of the neutrino mass would provide an important parameter to extend the Standard Model of particle physics and understand the origin of fermion masses beyond the Higgs mechanism. 
Furthermore, due to their abundance as big-bang relics, neutrinos 
strongly affect the large-scale structure and dynamics of the universe by means of their gravitational interactions, which hinder 
the structure clustering with an effect that is dependent on their mass \cite{Boy08,Han10}. In the framework of $\Lambda$CDM cosmology (the model with Cold Dark Matter and a cosmological constant $\Lambda$), the scale dependence of clustering observed in the Universe can indeed be used to set an upper limit on the neutrino mass sum $m_\Sigma=\sum_i m_i$, where $m_i$ is the mass of the $\nu_i$ state.
Depending on the model complexity and the input data used, this limit spans in the range between about 0.3 and 1.3\,eV \cite{Aba11}; more recently, by combining cosmic microwave background data with galaxy surveys and data on baryon acoustic oscillations a significantly lower bound on the neutrino mass sum of 0.23\,eV has been published \cite{Planck14},
although this value is strongly model-dependent.

The oscillation discovery and the accurate cosmological observations revived and boosted the interest in neutrino physics\footnote{This is also confirmed by the Nobel Prizes in Physics awarded in the years 2002, 2008 and, very recently, 2015.}, with the start of many ambitious experiments for different high precision measurements and the rate of publishing papers increased by almost an order of magnitude; but in spite of the enhanced experimental efforts very little is known about neutrinos and their properties.
Several crucial pieces are still missing, in particular: the absolute neutrino mass scale, the neutrino mass ordering (the so-called \textit{mass hierarchy}), the neutrino nature (Dirac or Majorana fermion), the magnitude of the CP (charge and parity) violation phases, and the possible existence of sterile neutrinos.

This paper is devoted to the assessment of the absolute neutrino mass scale, and in particular to the direct measurement of the electron neutrino mass via calorimetric experiments.
After a brief overview of 
our present picture for massive neutrinos, I will introduce both the theoretical and the experimental issues involved in the direct determination of the neutrino mass, and discuss past and current calorimetric experiments, with a focus on experiments with low temperature detectors.

\section{The neutrino mass pattern and mixing matrix}
Most of the existing experimental data on neutrino oscillations can be explained by assuming a three-neutrino framework, where
 any  flavor state $\nu_l$ ($l=e, \mu, \tau$) is described as a superposition of mass states $\nu_i$ ($i=1, 2, 3$), or
\begin{equation}
|\nu_l \rangle = \sum_i U_{li} |\nu_i \rangle
\end{equation}
where $U_{li}$ is the $3\times3$ Pontecorvo-Maki-Nakagawa-Sakata unitary 
mixing matrix  (see e.g. \cite{Giunti-Kim}). As a consequence, the neutrino flavor is no longer a conserved quantity and for neutrinos propagating in vacuum the amplitude of the process $\nu _l \to \nu_{l'}$ is not vanishing.

The $U_{li}$ mixing matrix
is parametrized by three angles, conventionally denoted as $\Theta _{12}$, $\Theta _{13}$ and $\Theta _{23}$, one CP violation phase $\delta$, and two Majorana phases $\alpha _1$, $\alpha _2$ -- these two have physical consequences only if neutrinos are Majorana particles, i.e. identical to their antiparticles, but do not affect neutrino oscillations.
To these six parameters -- three angles and three phases -- the three mass values $m_i$ must be added also, for a total of nine unknowns altogether. 
In the years, oscillation experiments measuring the flux of solar, atmospheric, reactor, and accelerator neutrinos 
have contributed to precisely determine many of these unknowns.

\begin{figure}[hbt]
\begin{center}
 \includegraphics[width=0.4\linewidth]{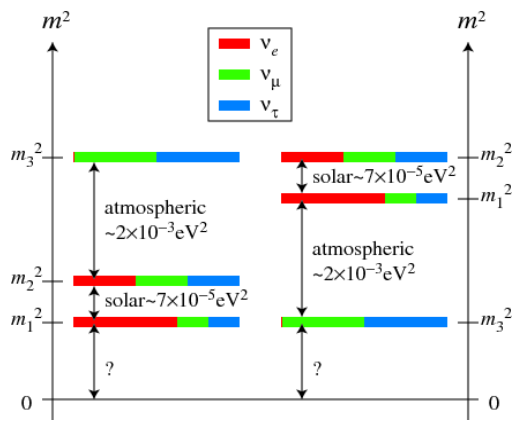}
      \caption{\label{fig:mass} In a three-neutrino scheme, the oscillation parameters measured by the various experiments paint two possible scenarios regarding the neutrino mass ordering: 
Normal Hierarchy (left)  and Inverted Hierarchy (right). The absolute scale is not accessible by presently available data.}
\end{center}
\end{figure}

The oscillation probabilities depend, in general, on the neutrino energy, on the source-detector distance, on the elements of the mixing matrix, and on the neutrino mass squared differences $\Delta m^2_{ij} \equiv m^2_{j} - m^2_{i}$.
At present,  the three mixing angles and the two mass splittings, conventionally $\Delta m^2_{21}$ (from solar neutrino oscillations) and  $\Delta m^2_{31}$ (from atmospheric neutrino oscillations),  have been determined with reasonable accuracy \cite{PDG14}.
However, the available data are not yet able to discriminate the neutrino mass ordering. It proves convenient to assume  $m_1<m_2$, so $\Delta m^2_{21}>0$; with these choices, $\Delta m^2_{21} \ll |\Delta m^2_{31}| \cong |\Delta m^2_{32}|$ and we are left with two possibilities: either $m_1<m_2<m_3$ (normal ordering, i.e. $\Delta m^2_{31}>0$) or $m_3<m_1<m_2$ (inverted ordering, i.e. $\Delta m^2_{31}<0$)\footnote{Compare also with Fig.\,\ref{fig:mass-ordering}.}. In both schemes, there is a Quasi-Degeneracy (QD) of the three neutrino masses when $m_1 \simeq m_2 \simeq m_3 $, with $m_{i} \gg \sqrt{\Delta m^2_{31}} \simeq 5\times 10^{-2}$\,eV.
Depending on the value of the lightest of the mass values, the neutrino mass ordering can also follow a Normal Hierarchy (NH), with $m_1 \ll m_2 \ll m_3$ (in which $m_2 \simeq  \sqrt{\Delta m^2_{21}}$  and $m_3 \simeq  \sqrt{\Delta m^2_{31}}$), or an Inverse Hierarchy (IH), with $m_3 \ll m_1 < m_2$ (in which $m_1$ and $m_2$ are quasi-degenerate) -- see Fig.\,\ref{fig:mass}. As a final remark, as shown in Fig.\,\ref{fig:mass-ordering}, independently of the mass scheme, oscillation results state that at least two neutrinos are massive, with masses larger than $ \sqrt{\Delta m^2_{21}} \simeq 8.7 \times 10^{-3}$\,eV.

Most of the oscillation data are well described by the three-neutrino schemes. However, there are a few anomalous indications (the so-called \textit{reactor neutrino anomaly}) \cite{Men11} that cannot be accommodated within this picture. If confirmed, they would indicate the existence of  additional neutrino families, the sterile neutrinos. These neutrinos do not directly participate in the standard weak interactions and would manifest themselves only when mixing with the familiar active neutrinos. Future reactor experiments will test this fascinating possibility.

Assessing the 
neutrino mass ordering, i.e. the sign of $|\Delta m^2_{31}|$, is of fundamental importance not only because it would address the correct theoretical extension of the Standard Model, but also because it can impact on many important processes in particle physics (like neutrinoless double beta decay).
In addition, the phase $\delta$ governing CP violation in the flavor oscillation experiments remains unknown, and a topic of considerable interest \cite{Qia15}. 
A worldwide research program is underway to address these important open issues in the near future by precise study of the various oscillation patterns.
 
The oscillation experiments, however, 
are not able to access the remaining unknown quantities, i.e. the absolute mass scale and the two Majorana phases. 
 Their determination is the ultimate goal of nuclear beta decay end-point experiments and neutrinoless double beta decay searches.
 
\section{Weak nuclear decays and neutrino mass scale}
Fundamental neutrino properties, in particular its absolute mass and its nature, can be investigated by means of suitable weak decays, where flavor state neutrinos are emitted along with charged leptons and/or pions. There are two complementary approaches for the measurement of the neutrino mass in laboratory experiments: the precise spectroscopy of beta decay at its kinematical \textit{end-point}, and the search for neutrinoless double beta decay. Though the expected effective mass sensitivity for neutrinoless double beta decay search is higher, this process implies a strong model-dependence since it requires the neutrino to be a Majorana particle.

Direct neutrino mass measurement, by analyzing the kinematics of electrons emitted in a beta decay, is the most sensitive model independent method to assess the neutrino mass absolute value\footnote{Analogue measurements involving pion or tau decays give much weaker limits on  $m_{\nu_{\mu}}$ or $m_{\nu_{\tau}}$.}. 
The beta decay is a nuclear transition involving two nuclides $(A,Z-1)$ and $(A,Z)$
\begin{equation}
(A,Z-1) \rightarrow (A,Z) + e^- + \overline\nu_e
\end{equation}
where $A$ and $Z$ are, respectively, the mass and atomic numbers of the involved nuclei. Neglecting the nuclear recoil, the kinetic energy $E_0$ available to the electron and anti-neutrino in the final state is given by
\begin{equation}
E_0 = E_\beta + E_{\overline\nu} = M(A,Z-1) - M(A,Z) = Q
\end{equation}
where $M$ indicates the mass of the {\it atoms} in the initial and final state.

In practice, this method exploits only momentum and energy conservation: it measures the minimum energy carried away by the neutrino -- i.e. its rest mass -- by observing the highest energy electrons emitted in this three body decay.
To balance the energy required to create the emitted neutrinos, the highest possible kinetic energy $E_\beta$ of the electrons is slightly reduced. This energy deficit may be noticeable when measuring with high precision the higher energy end (the so-called \textit{end-point}) of the emitted electron kinetic energy distribution $N_\beta(E_\beta,m_{\nu_e})$. If one neglects the nucleus recoil energy, $N_\beta(E_\beta,m_{\nu_e})$ is described in the most general form by

\begin{equation}
N_\beta(E_\beta,m_{\nu_e}) = p_\beta E_\beta (E_0 - E_\beta)\sqrt{(E_0 - E_\beta)^2 - m_{\nu_e}^2} F(Z,E_\beta) S(E_\beta) [1+\delta_R(Z,E_\beta)]\theta(E_0 - E_\beta -m_{\nu_e})
\label{eq:SPEBETA}
\end{equation}
where $F(Z,E_\beta)$ is the Coulomb correction (or Fermi function) which accounts for the effect of the nuclear charge on the wave function of the emitted electron, $S(E_\beta)$ is the {\it form factor}  which contains the nuclear matrix element $\cal{M}(E_\beta)$ of the electroweak interaction  and can be calculated using the V-A theory, and $\delta_R(Z,E_\beta)$ is the radiative electromagnetic correction, usually neglected due to its exiguity.
$\theta$ is the Heaviside step function, which confines the spectrum in the physical region $(E_0 - E_\beta -m_{\nu_e})>0$.
The term $p_\beta E_\beta (E_0 - E_\beta)\sqrt{(E_0 - E_\beta)^2 - m_{\nu_e}^2}\ $
is the phase space term in a three-body decay, for which the nuclear recoil has been neglected; $p_\beta$ is the electron momentum.
For the sake of completeness, it is worth noting that the particle emitted in the experiments considered here is the electron anti-neutrino $\overline{\nu}_e$. Since the CPT theorem assures that particle and antiparticle have the same rest mass, from now on I will speak simply of ``neutrino mass'' both for $\nu_e$ and $\overline{\nu}_e$.
Moreover, it must be stressed that since the effect of the neutrino mass in nuclear beta decay is due purely to kinematics, this measurement does not give any information on the Dirac or Majorana origin of the neutrino mass.

\begin{figure}[!t]
\begin{center}
 \includegraphics[width=0.7\textwidth]{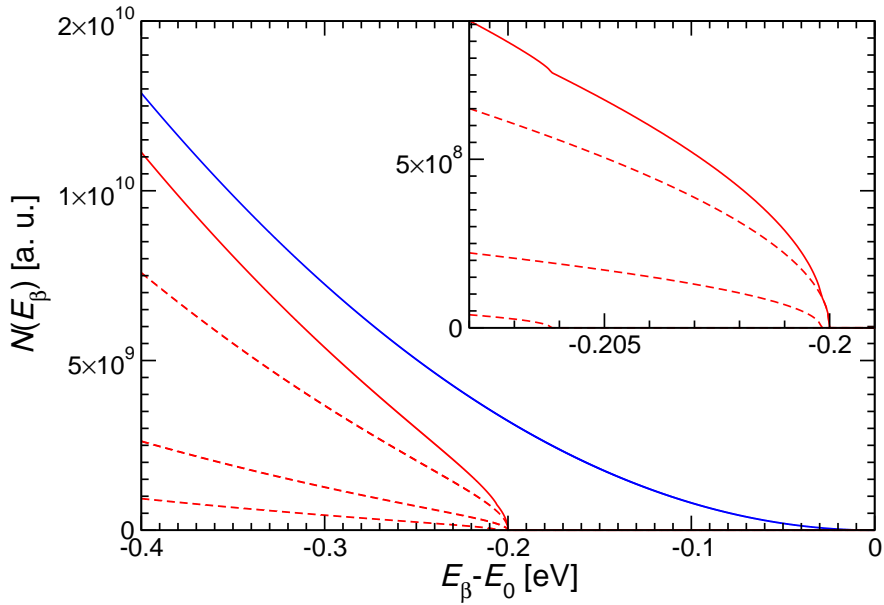}
      \caption{\label{fig:spectrum-mixing} Expected electron spectrum following $\beta$ decay. Blue curve: expected spectrum in the case $m _{\nu _e}=0$. Red curve: expected spectrum in the case $m _{\nu _e}=200$\,meV. The red dashed curves show the contributions to the total spectrum from the three mass states, with the mass differences driven by present values of $ \sqrt{\Delta m^2_{21}}$ and $ \sqrt{\Delta m^2_{31}}$ \cite{PDG14}.}
\end{center}
\end{figure}

From oscillation experiments, we know that any neutrino flavor state is a superposition of mass states. Therefore, (\ref{eq:SPEBETA}) can be generalized as

\begin{equation}
N_\beta(E_\beta,m_{\nu_e}) = R(E_\beta) \sum_{i=1}^3 |U_{ei}|^2 \sqrt{(E_0 - E_\beta)^2 - m_i^2}\theta(E_0 - E_\beta -m_i)
\label{eq:SPEBETA-mixing}
\end{equation}
where 
$R(E_\beta)$ is a term which groups all terms in (\ref{eq:SPEBETA}) which do not depend on the neutrino mass, $|U_{ei}|$ is the electron row of the neutrino mixing matrix, and $m_i$ are the masses of the neutrino mass states.  The square root term is the part of the phase space factor 
sensitive to the neutrino masses. An example of the resulting spectrum is shown in Fig.\,\ref{fig:spectrum-mixing}.

\begin{figure*}[!t]
\begin{centering}
  \includegraphics[width=.7\textwidth]{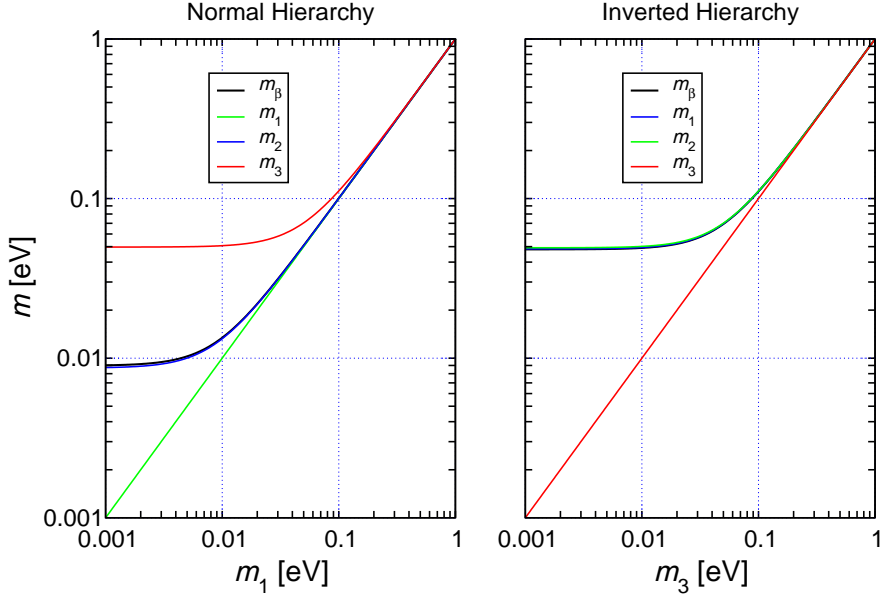}
  \caption{\label{fig:mass-ordering} Effective electron neutrino mass $m_{\beta}$ as a function of the lightest mass in both NH and IH mass schemes. The values of all three mass states are also plotted for comparison.}
\end{centering}
\end{figure*}

Since the individual neutrino masses are too close to each other to be resolved experimentally, the measured spectra can still be analyzed with (\ref{eq:SPEBETA}), but the quantity
\begin{equation}
m _{\nu _e} \equiv m_\beta = \sqrt {\sum _{i=1} ^3  | U _{ei} | ^2 m_i^2}.
\label{eq:effective-mnue}
\end{equation}
should now be interpreted as an effective electron neutrino mass,  where the sum is over all mass values $m _i$. Therefore, a limit on $m_{\nu_e}$ implies trivially an upper limit on the minimum value $m_{min}$ of all $m_i$, independent of the mixing parameters $U_{ei}$: $m_{min} \le m_{\nu_e}$, i.e. the lightest neutrino cannot be heavier than $m_{\nu_e}$. 
By using currently available information from oscillation data \cite{PDG14}, it is possible to formulate the values of the neutrino masses (and of \mne\ as well) as a function of the lightest mass, i.e. $m_1$ in the Normal Hierarchy (NH) and $m_3$ in the Inverted one (IH). This is done in Fig.\,\ref{fig:mass-ordering}, which shows that in the case of NH the main contribution to \mne\ is mainly due to $m_2$.
In the case of IH, \mne\ has practically the same value of $m_1$ and $m_2$. Finally, in the case of QD spectrum $m_{\nu_e} \simeq m_1 \simeq m_2 \simeq m_3 $ in both schemes. From the figure it is also clear that the allowed values for \mne\ in the two mass schemes are quite different: in the case of IH there is a lower limit for \mne\ of about 0.04\,eV, while in the NH this limit is of about 0.01\,eV. Therefore, if a future experiment will determine an upper bound for \mne\ smaller than 0.04\,eV, this would be a clear indication in favor of the NH mass pattern. Finally, Fig.\,\ref{fig:mass-ordering} shows that the ultimate sensitivity needed for a direct neutrino mass measurement is set at about 0.01\,eV, the lower bound in case of NH.
However, if experiments on neutrino oscillations provide us with the values of all neutrino mass-squared differences $\Delta m_{ij}^2$ (including their signs) and the mixing parameters $|U_{ei}|^2$, and the value of $m_{\nu_e}^2$ has been determined in a future search, then the individual neutrino mass squares can be determined
\begin{equation}
m_j  ^2  =  m_{\nu_e} ^2  - \sum _{i=1} ^3 | U_{ei} | ^2 \Delta m_{ij}^2  \qquad (\Delta m_{ij} ^2 = m_i ^2 - m_j^2) 
\end{equation}
On the other hand, if only the absolute values $| \Delta m_{ij} ^2 |$ are known (but all of them), a limit on $m_{\nu_e}^2$ from beta decay may be used to define an upper limit on the maximum value $m_{max}$ of $m _i$
\begin{equation}
m_{max}^2 \le  m_{\nu_e} ^2  + \sum _{i < j} | \Delta m_{ij}^2 |
\end{equation}
In other words, knowing $ | \Delta m _{ij} ^2  | $ one can use a limit on $m_{\nu _e}$ to constrain the heaviest active neutrino.

At present, the most stringent experimental constraint  on $m _{\nu _e}$ is the one obtained by the Troitzk \cite{Ase11} and the Mainz \cite{Kra05} neutrino mass experiments, $m _{\nu _e} < 2.05$\,eV at 95\%\,C.L.: this falls in the QD region for both mass schemes.  

Another type of weak process sensitive to the neutrino mass scale is the neutrinoless double beta decay  (\bbn), a second order weak decay that violates the total lepton number conservation by two units, and whose existence is predicted for many even-even nuclei

\begin{equation}
(A,Z) \to (A,Z + 2) + e^-_1 + e^-_2 
\end{equation}
The search for \bbn\ is the only available experimental tool to demonstrate the Majorana character of the neutrino (i.e. $\nu\equiv\bar{\nu}$). In fact, the observation of \bbn\ always requires and implies that neutrinos are massive Majorana particles \cite{Sch82}. However, there are many proposed mechanisms
which could contribute to the \bbn\ transition amplitude, and only when the \bbn\ is mediated by a light mass Majorana neutrino the observed decay is useful for determining the neutrino mass. In this case the measured decay rate is 
given by 

\begin{equation}
\frac{1}{\tau_{1/2}^{0\nu}}= \frac{m_{\beta\beta}^2}{m_e^2} F_N 
\end{equation}
where $\tau_{1/2}^{0\nu}$ is the \bbn\ decay half-life, $m_e$ is electron mass, and $m_{\beta\beta}$ is the effective Majorana mass, defined below.
The nuclear structure factor $F_N$ is given by
\begin{equation}
F_N = G^{0\nu}(Q_{\beta\beta},Z) |M^{0\nu}|^2
\end{equation}
where $G^{0\nu}$ is the accurately calculable phase space integral, and $M^{0\nu}$  is the nuclear matrix element which is subject to uncertainty \cite{Ell02}.
At present, the discrepancies among different nuclear model calculations of  $M^{0\nu}$ amount to a factor of about 2 to 3. 
These  reflect on $F_N$ and are an unavoidable source of systematic uncertainties
in the determination of $m_{\beta\beta}$ from the experimental data. 
Measuring the lifetime of different isotopes would allow to disentangle the model dependency linked to the exact mechanism causing the
\bbn\  and to reduce the systematic uncertainties on  $m_{\beta\beta}$.

If the \bbn\ decay is observed, and the nuclear matrix elements are known, one can deduce the corresponding $m_{\beta\beta}$ value, which in turn is related to the oscillation parameters through
\begin{equation}
m_{\beta\beta} =  \left | \sum _{i=1} ^3 |U_{ei}|^2 m_i e^{i \alpha _i} \right |
\label{eq:effective-majorana-mass}
\end{equation}
Due to the presence of the unknown Majorana phases $\alpha _i$, cancellation
of terms in (\ref{eq:effective-majorana-mass}) is possible, and $m_{\beta\beta}$ could be smaller than any of the $m_i$. 
Therefore, unlike the direct neutrino mass measurement, a limit on $m_{\beta\beta}$ does not allow to constrain the individual mass values $m_i$ even when the mass differences $\Delta m_{ij}^2$ are known. 
On the other hand, the observation of the \bbn\ decay and the accurate determination of the $m_{\beta\beta}$ value, would not only establish that neutrinos are massive Majorana particles, but would contribute considerably to the determination of the absolute
neutrino mass scale. Moreover, if the neutrino mass scale would be known from independent measurements, one could possibly obtain from the measured $m_{\beta\beta}$ also some information about the CP violating Majorana phases \cite{pascoli2002cp}.

\begin{figure*}[!htb]
\begin{centering}
  \includegraphics[width=.8\textwidth]{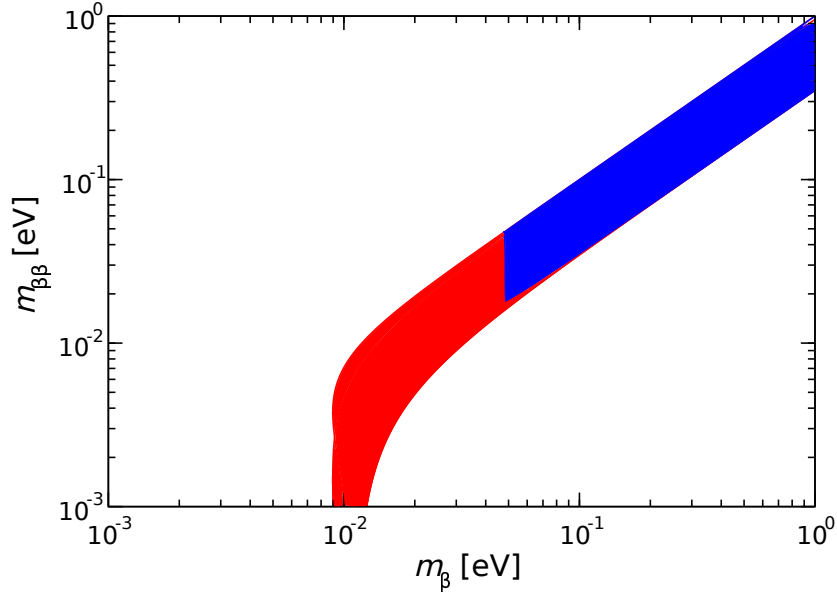}
  \caption{\label{fig:mbb-mb} Relationship between the effective Majorana mass $m_{\beta\beta}$ and the effective electron neutrino mass $m_{\beta}$ for both IH (blue) and NH (red) mass schemes. The width of the bands is caused by the unknown Majorana phases.}
\end{centering}
\end{figure*}

Given the present knowledge of the neutrino oscillation parameters, it is possible to derive the relation between the effective Majorana mass and the lightest neutrino mass in the different neutrino mass schemes. This is done in a number of papers (see e.g. \cite{Del14}). Fig.\,\ref{fig:mbb-mb} shows the effective Majorana mass as a function of the effective electron neutrino mass in both the NH and IH mass schemes, demonstrating the complementarity of the two methods.

As a final remark, \bbn\ and $\beta$ decays both depend on different combinations of the neutrino mass values and oscillation parameters. The \bbn\ decay rate is proportional to the square of a coherent sum of the Majorana neutrino
masses because the process originates from exchange of a {\it virtual} neutrino. On the other hand, in beta decay one can determine an incoherent sum because a {\it real} neutrino is emitted. That shows clearly that a complete neutrino physics program cannot renounce either of these two experimental approaches. The various methods that constrain the neutrino absolute mass scale are not redundant but rather complementary. If, ideally, a positive measurement is reached in all of them (\bbn\ decay, $\beta$ decay, cosmology) one can test the results for consistency and with a bit of luck determine the Majorana phases.

\section{The direct neutrino mass measurement via single nuclear beta decay}
\label{sec:betadecay}
As already pointed out, the most useful tool to constrain kinematically the neutrino mass is the study of the ``visible'' energy in single beta decay. 
The experimental beta spectra are normally analyzed by means of a transformation which produces a quantity generally linear with the kinetic energy $E_\beta$ of the emitted electron
\begin{equation}
K(E_\beta)\equiv \sqrt{\frac{N_\beta(E_\beta,m_{\nu_e})}{p_\beta E_\beta F(Z,E_\beta) S(E_\beta) [1+\delta_R(Z,E_\beta)]}}
=(E_0-E_\beta)\left( 1 - \frac{m_{\nu_e}^2}{(E_0-E_\beta)^2}\right)^{1/4}
\end{equation}
The graph of this quantity as a function of $E_\beta$ is named Kurie plot. In a Kurie plot, each bin has the same error bar and therefore the same statistical weight.

\begin{figure}[!t]
\begin{center}
 \includegraphics[width=0.7\linewidth]{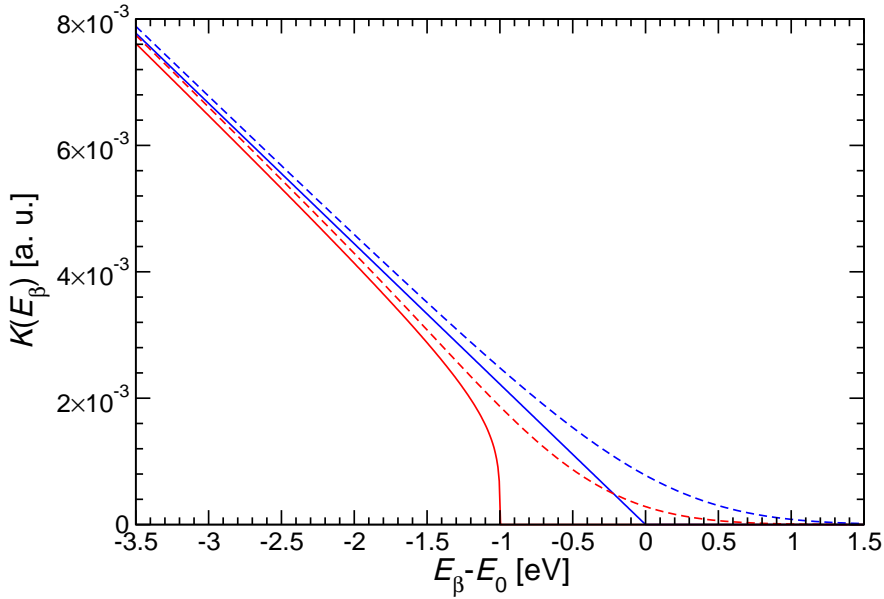}
      \caption{\label{fig:KURIE} $^3$H Kurie plot close to the end-point, computed for neutrino masses equal to 0 (blue) and 1\,eV (red), and an energy resolution $\sigma_E$ of 0 (full line) and 0.5\,eV (dashed line).}
\end{center}
\end{figure}

Assuming massless neutrinos and infinite energy resolution, the Kurie plot is a straight line intersecting the X-axis at the transition energy $E_0$. In case of massive neutrino, the Kurie plot is distorted close to the end-point and intersects with vertical tangent the X-axis at the energy $E_0-m_{\nu_e}$. The two situations are depicted in Fig.\,\ref{fig:KURIE}.

Most of the information on the neutrino mass is therefore contained in the final part of the Kurie plot, which is the region where the counting rate is lower. In particular, the relevant energy interval is \mbox{$\delta \! E \approx m_{\nu_e}$} and the fraction of events occurring here is
 \begin{equation}
F(\delta \! E) =\int^{E_0}_{E_0-\delta \! E} N_\beta(E,m_{\nu_e} \! = \! 0) d \! E \approx
C\left( \frac {\delta \! E} {E_0} \right)^3
\end{equation}
where $C$ is a constant of order unity which depends on the details of the beta transition.
From this it is apparent that kinematical mass measurements require beta decaying isotopes with the lowest end-point energy.
Tritium is one of the best and most used isotopes thanks to its very low transition energy, $E_0 = 18.6$\,keV; nonetheless, the fraction of events falling in the last 5\,eV of the tritium spectrum is only $4\times 10^{-11}$.

Every instrumental effect such as energy resolution or background will tend to hinder or even wash out this tiny signal. In Fig.\,\ref{fig:KURIE} the effect on the spectral end-point of an energy resolution of 0.5\,eV is shown.
This distorts the Kurie plot in the opposite way with respect to the neutrino mass effect. It is therefore mandatory to evaluate and/or measure the detector response function, which includes the energy resolution but is not entirely determined by it. Finally, the analysis of the final part of the Kurie plot is complicated by the background due to cosmic rays and environmental radioactivity. Because of the low beta counting rate in the interesting region, spurious background counts may affect the neutrino mass determination. 

The possibility to use beta decay to directly measure the neutrino mass was first suggested by Fermi \cite{fermi1934attempt} in 1934, but the first sensitive experiments where performed only in the '70s. The first experiments were the one of Bergkvist \cite{bergkvist1972high1,bergkvist1972high2}
and the one of the ITEP group \cite{lyubimov1980estimate}, both of which used magnetic spectrometers to analyze the electrons emitted by tritium sources. This experimental approach has clear advantages such as 1) the high specific activity of tritium, 2) the high energy resolution and luminosity of spectrometers, and 3) the possibility to select and analyze only the electrons with kinetic energies close to $E_0$. 

\begin{figure}[!t]
\begin{center}
 \includegraphics[width=0.7\linewidth]{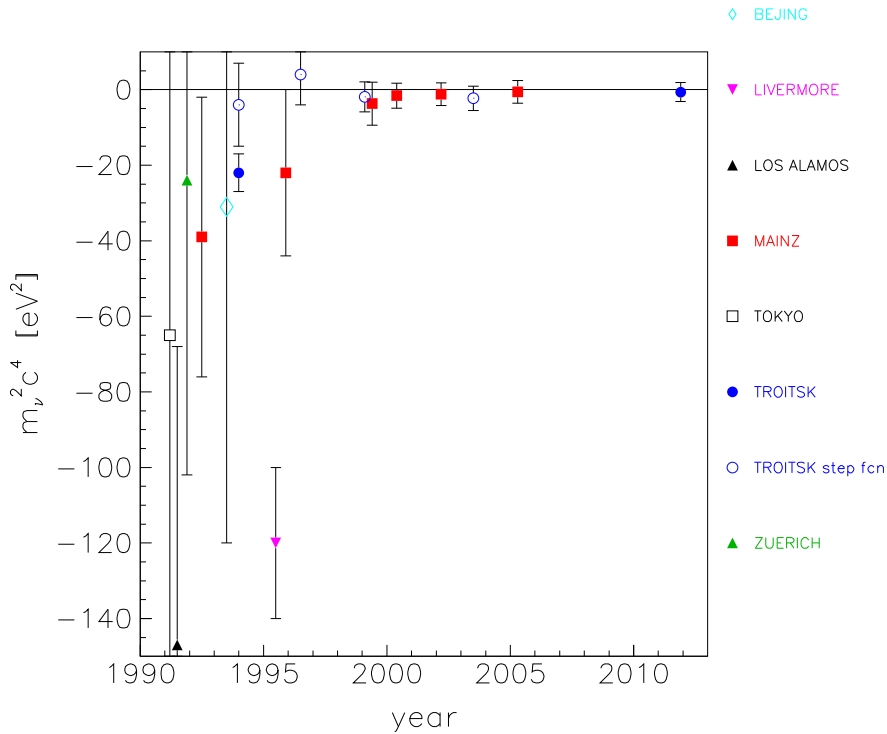}
      \caption{\label{fig:negmass2} Historical trend of the \mnesq\ measured with spectrometer using tritium (taken from \cite{drexlin2013current}).}
\end{center}
\end{figure}

In the '80s and through the '90s, experiments with spectrometers using tritium were reporting largely negative \mnesq\ \cite{robertson1988direct} (see Fig.\,\ref{fig:negmass2}) or even an unlikely finite value of about 35\,eV  \cite{lyubimov1980estimate}. 
These were all signs of under- or over- corrected instrumental effects which were causing systematic shifts \cite{bergkvist1985questioning,bergkvist1997effects}.
In fact, despite the relative conceptual simplicity of the kinematic direct determination of the neutrino mass, it has been soon recognized that there are many subtle effects which threaten the accuracy of these measurements. 
Some are related to beta decay itself, since the atom or the molecule containing the decaying nucleus can be left in an excited state, leading even in this case to dangerous distortions of the Kurie plot (see \S\,\ref{sec:calo}).
Other are due to the scattering and absorption of the electrons in the source itself. And last but not least, systematic effects are also caused
by the imperfect characterization of the detector response.
In the past 30 years many experiments using tritium were performed. Starting from the '90s, magnetic spectrometers were gradually abandoned for electrostatic retarding spectrometers with adiabatic magnetic collimation \cite{weinheimer1993improved,belesev1995results}. Many improvements in the detectors, in the tritium source, and in the data analysis and processing allowed to constantly improve the statistical sensitivity and to minimize the systematic uncertainties, as it is shown in Fig\,\ref{fig:negmass2}.
Today, owing to a continuous and strenuous investigation of all experimental effects and systematic uncertainties, 
the \mnesq\ measurements reported by the two most sensitive experiments \cite{Ase11,Kra05} are compatible with a zero mass, with the systematic errors reduced to the same level of statistical ones.

Nevertheless, today direct neutrino mass measurements remain affected by an intrinsic potential bias. As it  already happened in the past, in a sensitive experiment small miscorrections of instrumental effects may again either mimic or cancel the traces of a small positive neutrino mass. A weak unexpected effect not included in the data analysis may compensate and hide the signal of a small mass within the statistical sensitivity of an experiment, which would therefore report a \mnesq\ nicely compatible with the null hypothesis and thus quote just an upper limit. On the other hand, in a future experiment with a statistical sensitivity approaching the range predicted by oscillation parameters a slightly excessive correction for an expected effect could mimic the signal for a tiny mass which would not contradict the community expectations. For these reasons, direct neutrino mass measurements call for a continuous crosscheck from different independent experiments to confirm both positive and negative findings.   

Already in the '80s when the negative squared masses and the positive claim from ITEP were puzzling the neutrino community, A. De Rujula proposed the use of other beta decaying isotopes with low decay energy. In \cite{derujula1981new} it was noticed the \Re\ has an endpoint around 2\,keV, much more favorable than the one of tritium. This isotope was at that time discarded because of its long half life around $10^9$\,years. 
The focus of \cite{derujula1981new} was therefore on the isotope \Ho, which decays by Electron Capture (EC) with a very low transition energy.
In the EC process \cite{Bambynek}
\begin{equation}
(A,Z) + e^- \rightarrow (A,Z-1) + \nu_e
\end{equation}
the available decay energy 
is 
\begin{equation}
Q = M(A,Z) - M(A,Z-1) 
\end{equation}
where $M$ indicates the mass of the {\it atoms} in the initial and final state.
Neglecting the nuclear recoil, the energy $Q$ is shared between the neutrino and the radiation emitted in the de-excitation of the daughter atom 
\begin{equation}
\label{eq:q_ec}
Q= E_\nu + E_X 
\end{equation}
Here $E_X$ includes the energy of X-rays, Inner Bremsstrahlung photons, and Auger and Coster-Kronig electrons emitted in the atomic de-excitation
of the daughter atom and adds up the binding energy of the captured electron, allowing for a small indetermination due to the natural width of the atomic energy levels.
Because of energy conservation, the end-points of the spectra of these electrons or photons -- where the massive neutrino emitted in the EC is at rest -- is sensitive to the neutrino mass. 
It is worth noting here that the kinematics of EC decay probes the mass the neutrino $\nu_e$ whereas the regular beta decays which probes the mass of the anti-neutrino $\overline{\nu}_e$ but, as already recalled above, the CPT theorem warrants the equality of these quantities.

In particular, two measurements were discussed in 1981 for the \Ho\ isotope: the end-point of the IBEC (Inner Bremsstrahlung in EC) spectrum \cite{derujula1981new}
\begin{equation}
\label{eq:ibec}
^{163}\mathrm{Ho} \rightarrow (^{163}\mathrm{Dy^{H^*}} + \nu_e) \rightarrow ^{163}\mathrm{Dy^{H^\prime}} + \gamma(k) + \nu_e
\end{equation}
and the end-point of SEEEC (Single Electron Ejection in EC) spectrum \cite{derujula1983seeec}
\begin{equation}
\label{eq:seeec}
^{163}\mathrm{Ho} \rightarrow (^{163}\mathrm{Dy^{H^*}} + \nu_e) \rightarrow ^{163}\mathrm{Dy^{H_1H_2}} + e^- + \nu_e
\end{equation}
Even if at that time the $Q$ of \Ho\ EC was largely unknown, this decay was already considered very promising for a sensitive neutrino mass measurement, since it was clear that  the $Q$ value is one of the lowest available.
Both processes (\ref{eq:ibec}) and (\ref{eq:seeec}) start with a first intermediate atomic vacancy H$^*$ caused by the EC, where the $^*$ reminds that the state is not necessarily on-shell. The energy of the vacant state has its own natural width.  
Because of the low $Q$ value, this first vacancy H can be created only in one of the M1, M2, N1, N2, O1, O2, or P1 shells of the Dy daughter atom.

In the IBEC process (\ref{eq:ibec}) a photon is emitted during the virtual transition of an electron from H$^\prime$ to the intermediate state H$^*$, from which the electron was captured.
For each possible final vacancy H$^\prime$, and for $m_{\nu_e}=0$, the spectrum of the emitted photons is not made of mono-energetic lines at $k=E(\mathrm{H^\prime})-E(\mathrm{H^*})$, where $E(\mathrm{H})$ is the ionization energy of the H shell in Dy, but is a continuum with a kinematic limit $k\le k^{\mathrm{H^\prime}}_{max} =  Q - E(\mathrm{H^\prime})$, and the total photon spectrum is therefore a superposition of several spectra with different end-points.
The spectral end-points follow the three-body statistical shape
\begin{equation}
N^{\mathrm{H^\prime}}(k) \propto k (k^{\mathrm{H^\prime}}_{max}-k)\sqrt{(k^{\mathrm{H^\prime}}_{max}-k) - m_{\nu_e}^2}
\end{equation}
In general, since the IBEC is a second-order effect, its intensity is very low. However, the photon emission may experience large resonant enhancements for photons with energies $k_{res}$ equal to the ones of the characteristic X-ray transitions of the daughter atom.   
In particular, A. De Rujula has shown  for \Ho\ that when H$^\prime$ is one of the N1, N2, O1, O2 shells then the dominant resonance close to the end-point is with the X-ray transitions H$^\prime\rightarrow$M1, that is when the intermediate vacancy of the virtual transition H corresponds to the M1 shell. In this case the distance between the resonance and the end-point is $k^{\mathrm{H^\prime}}_{max}-k_{res}=Q-E(\mathrm{M1})$, which for \Ho\ is equal to a few hundreds electronvolts.
Unfortunately, calculations \cite{riisager1985internal,springer1987measurement} showed that with a $Q$ at around 2.8\,keV, an IBEC measurement with \Ho\ is not going to be statistically competitive with the tritium experiments, also because of complex destructive interference patterns. 

The SEEEC process (\ref{eq:seeec}) is the analogous of the IBEC with the role of the IB photon  played by an Auger (or Coster-Kronig) electron.
The spectrum  of the ejected electron is a continuum with an end-point at $E^{\mathrm{H1H2}}_{max} =  Q - E(\mathrm{H1})-E(\mathrm{H2})$, for $m_{\nu_e}=0$.
Also in this case, the kinematics of a 3-body decay process applies, and a phase space term $p_\nu E_\nu$ appears in the spectral shape of ejected electrons.
The continuous spectra show many resonances for different combinations of H,H1 and H2, but close to the end-point the dominant ones result from the M1 capture and are at $E_{res}=E(\mathrm{M1})-E(\mathrm{H1})-E(\mathrm{H2})$. These resonances provide an enhancement of the spectrum close to end-point, thereby increasing the statistical sensitivity to $m_{\nu_e}$. The inclusive spectrum of all the ejected electrons is quite complicate because of the many possible end-points $E^{\mathrm{H1H2}}_{max}$ and resonance peaks: nevertheless, the authors in \cite{derujula1983seeec} argue that the end-point region of this spectrum is unaffected by all the atomic details, since it is dominated by the upper tails of few resonances and maintains its usable sensitivity to $m_{\nu_e}$, although the estimated $F(\Delta E)$, depending on the $Q$ value, may be substantially lower than for tritium. 

One stressed advantage of IBEC and SEEEC measurements is that, unlike what happens in tritium beta decay, the probability of atomic excitations in the final state -- such as shake-up or shake-off processes -- is strongly suppressed and estimated to be $<1/Z^2$ (see also \S\,\ref{sec:ho163spe}).  

More than 30 years later, none of the above suggestions has been successfully exploited to perform an experiment with a competitive sensitivity on $m_{\nu_e}$.
Of the various attempts to perform an IB end-point measurements \cite{jonson1983determination,springer1987measurement,yasumi1983mass,yasumi1986mass,yasumi1994mass}, only the one of P.T.\,Springer \cite{springer1987measurement} reported a limit on \mne\ of about 225\,eV obtained by fitting the end-point of the X-ray spectrum.  

Most of the measurements performed on \Ho\ to directly measure the neutrino mass followed instead another proposal from C.L.\,Bennett et al. \cite{bennett1981x} in 1981.
In \cite{bennett1981x} it is suggested that \mne\ and the transition energy $Q$ can be determined or constrained by measuring the ratios of absolute capture rates\footnote{A better treatment includes $C_i$ factors for the nuclear shape factor.}   
\begin{equation}
\label{eq:ratios}
\lambda_i/\lambda_j=(n_i p_{\nu,i} E_{\nu,i} \beta^2_i B_i)/(n_j p_{\nu,j} E_{\nu,j} \beta^2_j B_j)
\end{equation}
where neutrino momentum $p_{\nu,i}$ is given by
\begin{equation}
p^2_{\nu,i} = (Q - E_i)^2 - m_{\nu_e}^2=  E^2_{\nu,i} - m^2_{\nu_e}
\end{equation}
$n_i$ is the fraction of occupancy of the $i$-th atomic shell, 
$\beta_i$ is the Coulomb amplitude of the electron radial wave function 
(essentially, the modulus of the wave function at the origin), and $B_i$ 
is an atomic  correction for electron exchange and overlap. 
Following this idea, practically all the experimental researches on EC of \Ho\ so far focused on the atomic 
emissions - photons and electrons contributing to $E_X$ in (\ref{eq:q_ec}) - following the EC and used the capture ratios to determine the $Q$ value \cite{jonson1983determination,Laegsgaard154030,hartmann1985observation,springer1987measurement,yasumi1983mass,yasumi1986mass,yasumi1994mass, hartmann1992high}.
Unfortunately the accuracy achieved for \mne\ and $Q$ with this method is adversely affected by the limited knowledge of the atomic parameters
in (\ref{eq:ratios}).

As repeatedly underlined by A. De Rujula and M. Lusignoli \cite{de1982calorimetric}, there is one experimental approach to the measurement of the neutrino mass from the \Ho\ EC which overcomes all the difficulties above: the calorimetric measurement of all the energy released in the EC ($E_X$ in (\ref{eq:q_ec})) except for the energy of the neutrino. This will be discussed in \S\,\ref{sec:holmium}.

Today all expectations for a new direct measurement of the neutrino mass with a substantially improved statistical sensitivity 
are directed to the KATRIN experiment \cite{KATRIN}. KATRIN uses a large electrostatic spectrometer which will analyze the tritium beta decay end-point with an energy resolution of about 1\,eV and  with an expected  statistical sensitivity of about 0.2\,eV. KATRIN reaches the maximum size and complexity practically achievable for an experiment of its type and no further improved project can be presently envisaged. 
As an alternative for the study of tritium end-point, Project8 proposes a new experimental approach based on the 
detection of the relativistic cyclotron radiation detection emitted by the beta electrons \cite{monreal2009relativistic}, which	
is presently under development \cite{asner2015single}.

\section{Calorimetric measurements}
\subsection{General considerations}
\label{sec:calo}
In the global effort to cure the weaknesses of direct neutrino mass measurements with spectrometers yielding negative \mnesq\ which started to show up since the '80, J.J.\,Simpson first proposed the calorimetric approach \cite{simpson1981limits}.
In an ideal calorimetric experiment, the source is embedded in the detector and therefore only the neutrino energy escapes to detection. The part of the energy spent for the excitation of atomic or molecular levels is measured through the de-excitation of these states, provided that their lifetime be negligible with respect to the detector time response. In other terms, the kinematical parameter which is effectively measured is the neutrino energy $E_\nu$ (or $E_0-E_\nu$), in the form of a missing energy, a common situation in experimental particle physics. 
The advantages of a calorimetric measurement are 1) the measurement of all the energy temporarily stored in excited states, 2) the absence 
of source effects -- such as self-absorption, and 3) the lack of backscattering from the detector.
The effect of final states on the tritium beta spectrum was discussed throughly in many works \cite{bergkvist1997effects,ching1984determination,robertson1988direct}. In the following for simplicity we consider the so-called sudden approximation or first order perturbation of an atomic tritium beta decay, neglecting the sum over the mass eigenstates $m_i$.
Due to the excited final states, the measured beta spectrum is a combination of different spectra characterized by different transition energies $E_0 - V_i$, where $V_i$ is the energy of the $i$-th excited state
\begin{equation}
\label{eq:speb2}
N_\beta(E_\beta,m_{\nu_e}) \approx
\sum_i{w_i p_\beta E_\beta 
(E_0 - E_\beta - V_i)^2 \left( 1 - \frac{m_{\nu_e}^2 }{(E_0 - E_\beta -
V_i)^2}\right)^{1/2}F(Z,E_\beta) S(E_\beta) }
\end{equation}
with $w_i$ describing the transition probability to the final $i$-th excited state. The spectral shape induced by the excited states is misleading when one tries to extract the value of the neutrino mass. Assuming  that the neutrino mass is null and summing up over all the final states, from  equation (\ref{eq:speb2}) one obtains
\begin{equation}
\label{eq:speb2m0}
N_\beta(E_\beta,0) \approx 
p_\beta E_\beta (E_0 - E_\beta - \langle V_i \rangle)^2
\left( 1 + \frac {\langle V_i^2 \rangle - \langle V_i \rangle ^2} {(E_0 - E_\beta - \langle V_i \rangle)^2}\right)
F(Z,E_\beta) S(E_\beta)
\end{equation}\noindent
which approximates the single beta spectrum (\ref{eq:SPEBETA}) with a negative squared neutrino mass equal to \mbox{$-2 \sigma^2<0$}, where $\sigma$ is the variance of the final state spectrum  given by \mbox{$\sigma^2 = \langle V_i^2 \rangle - \langle V_i \rangle ^2$} (Fig.\,\ref{fig:kp_spettr}), and
with an end-point shifted by $\langle V_i \rangle$. 
\begin{figure*}[!t]
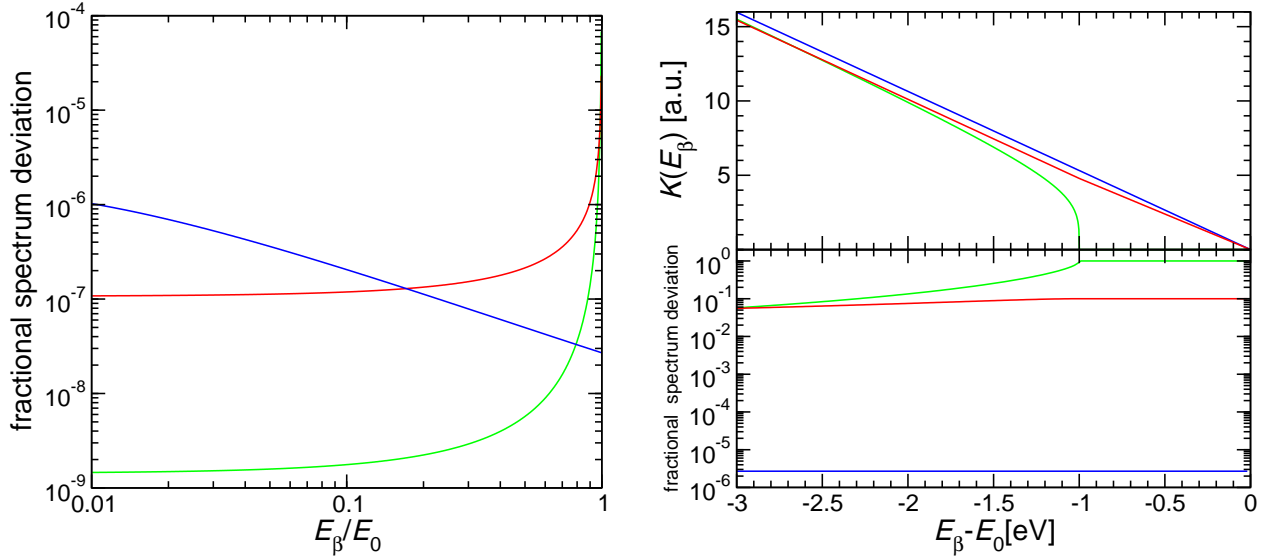

\begin{centering}
  \includegraphics[width=.48\textwidth]{deviation_spec_calo.eps}
  \hfill
  \includegraphics[width=.48\textwidth]{cfr_K_spec_calo.eps}
  \caption{\label{fig:spectrum}
      \label{fig:kp_calo} 
These plots compare the effect of a final excited state ($V_1=1$\,eV and $\omega_1=0.1$\,eV) on the beta spectrum as measured with a calorimeter (blue) and with a spectrometer (red), and the effect of \mnesq$=1$\,eV (green). The fractional spectrum deviation is the quantity $1-N^*(E_\beta)/N(E_\beta)$, where $N^*(E_\beta)$ and $N(E_\beta)$ are respectively the observed beta spectra with and without excited final state.}
\end{centering}
\end{figure*}

In case of tritium atom, the Shr\"{o}dinger equation can be solved analytically and one gets $\sigma^2=$740.5\,eV$^2$. In all the other cases, the final state distributions are estimated numerically.

The situation changes completely in the calorimetric approach. Even in this case the observed spectrum is a combination of different spectra. It can be obtained by operating the following replacements
\begin{equation}
\label{eq:eshift}
\begin{array}{c}
E_\beta \rightarrow  E_\beta^\prime = E_\beta - V_i \\ [5.0mm]
p_\beta = (E_\beta^2 - m_e^2)^{1/2} \rightarrow  p_\beta^\prime = \left((E_\beta - V_i)^2 - m_e^2 \right)^{1/2}
\end{array}
\end{equation}
motivated by the distinguishing feature of the calorimeters to measure simultaneously the beta electron energy {\it and} the de-excitation energy $V_i$ of the final state.

By combining (\ref{eq:speb2}) and (\ref{eq:eshift}) one gets
\begin{eqnarray*}
\label{eq:spebetacalo0}
N_\beta(E_\beta,m_{\nu_e}) 
& \approx & (E_0 - E_\beta)^2 \left(\displaystyle 1 - \frac{m_{\nu_e}^2}{(E_0 -E_\beta)^2}\right)^{1/2} \\
&         & \sum_i{w_i (E_\beta - V_i) \left((E_\beta - V_i)^2 - m_e^2\right)^{1/2} \,\, 
            F(Z,E_\beta -V_i) S(E_\beta-V_i) } 
\end{eqnarray*}
Observing that $F(Z,E_\beta -V_i) S(E_\beta-V_i)\approx F(Z,E_\beta) S(E_\beta)$ and expanding in a series of powers of $V_i/E_\beta$, one obtains
\begin{eqnarray}
\label{eq:spebetacalo1}
N_\beta(E_\beta,m_{\nu_e})
& \approx & p_\beta E_\beta (E_0 - E_\beta)^2 \left(\displaystyle 1 - \frac{m_{\nu_e}^2}{(E_0 -E_\beta)^2}\right)^{1/2}
            F(Z,E_\beta) S(E_\beta) \nonumber\\
&         & \sum_i{w_i \left( 1 - \frac{V_i}{E_\beta} - \frac{V_i E_\beta}{E_\beta^2 - m_e^2} +
            \frac{V_i^2}{2(E_\beta^2 - m_e^2)}\right) } 
\end{eqnarray}
Apart from the sum term, for a null neutrino mass the equation  (\ref{eq:spebetacalo1}) 
describes a beta spectrum with a linear Kurie plot in the final region ($E_\beta \gg V_i$);
Fig.\,\ref{fig:kp_calo} shows as the influence of the excited final states on the calorimetric beta spectrum is confined at low energy. 
Therefore, a calorimeter provides a faithful reconstruction of the beta spectral shape over large energy range below the end-point . This is not true for spectrometers for which the measured spectrum at the end-point presents a deviation of the same size of that caused by a finite neutrino mass. Furthermore it is apparent from Fig.\,\ref{fig:kp_calo} as the presence of an excited state causes the spectrum of a spectrometer to mimic a lower $E_0$ along with a negative \mnesq.   
The possibility to observe a substantial undistorted fraction of the spectrum is very useful to check systematic effects and to prove the general reliability of a calorimetric experiment.

As a general drawback, calorimeters present a major inconvenience which may be  a serious limitation for the approach. 
In a calorimeter, the whole beta spectrum is acquired and the detector technology poses important restrains to the source strength.
This in turns limits the statistics that can be accumulated. The consequences on the achievable statistical sensitivity are discussed in the next section.
First of all the counting rate must be controlled to avoid distortions of the spectral shape due to pile-up pulses. 
Then the concentration of the decaying isotope maybe not freely adjustable.
For example at the time of J.J.\,Simpson experiments, the only way to make a sensitive calorimetric measurement was to ion implant tritium  
in semiconductor ionization detectors such as Si(Li) or High Purity Ge. There is however a trade off between the required tritium implantation dose 
-- i.e. the tritium concentration -- and the acceptable radiation damage. The tritium activity is then limited by the detector size in relation to 
its energy resolution.

This first generation of calorimetric experiments exploited Si(Li) or Ge detectors with implanted tritium, but suffered for their intrinsic energy resolution which is limited to about 200\,eV at 20\,keV. With these experiments a limit on \mne\ of about 65\,eV was set \cite{simpson1981limits}. 
At the same time, these experiments showed that the calorimetric approach does not cancel all the systematic uncertainties. As it was already recognized by J.J.\,Simpson in \cite{simpson1981limits}, one source of systematic uncertainty relates to the precise evaluation of the resolution function of these solid states detectors. The resolution function is obtained through X-ray irradiation from an external source. The response of the detector may be different for X-rays entering the detector from one direction and the betas emitted isotropically within the detector volume. Moreover, the beta emission is localized in the deep region of the detector where an incompletely recovered irradiation damage may lead to incomplete charge collection, while X-ray interactions are distributed in the whole detector volume. 

Soon it became clear that calorimeters may also be affected  by solid states effect. The ``17\,keV neutrino saga''  \cite{franklin1995appearance,wietfeldt199617} started off from an unexpected feature observed first by J.J.\,Simpson in the low energy part of the tritium spectrum measured with the implanted Si(Li) detectors \cite{simpson1985evidence}.
While a neutrino with a mass of 17\,keV was finally deemed inexistent and the observed kink ascribed to a combination of various overlooked instrumental effects in spectrometric experiment \cite{abele1993origin}, the evidence in calorimetric measurements remained unexplained.
The invoked explanations include environmental effects in silicon and germanium  and remain of interest for future calorimetric experiments.
One of these solid state effects was first described by S.E.\,Koonin in 1991 \cite{koonin1991environmental}: it is a solid state effect known as Beta Environmental Fine Structure (BEFS), which introduces oscillatory patterns in the energy distribution of the electrons emitted by a beta isotope in a lattice.
It is an effect analogous to the Extended X-ray Absorption Fine Structure (EXAFS) and it will be addressed in more detail in \S\,\ref{sec:stat-sys}.

So far only tritium beta decay was considered, but all the arguments above apply to other isotopes undergoing nuclear beta decay. 
In particular, as it will be shown quantitatively in next section, isotopes with a transition energy lower than that of tritium are better suited
for a calorimetric experiment. The rest of the present work will focus on two such isotopes: \Re\ and \Ho, which have a $Q$ around 2.5\,keV.
In fact, already in the '80s many authors realized that low temperature detectors could offer a solution for making calorimetric measurements with high energy resolution and could be used either for tritium or, better, the lower $Q$ beta emitters \Re\ and \Ho\ (\S\,\ref{sec:ltd}).

A final remark from the discussion above is that the spectrometer and the calorimeter methods have both complicated but totally different systematic effects. Therefore, once that it is demonstrated that the achievable sensitivities are of the same order of magnitude in the two cases, it is scientifically very sound  to develop complementary experiments exploiting these two techniques. 

\subsection{Sensitivity of calorimeters: analytical evaluation}
\label{sec:calosens}
\begin{figure*}[!t]
\begin{centering}
  \includegraphics[width=.8\textwidth]{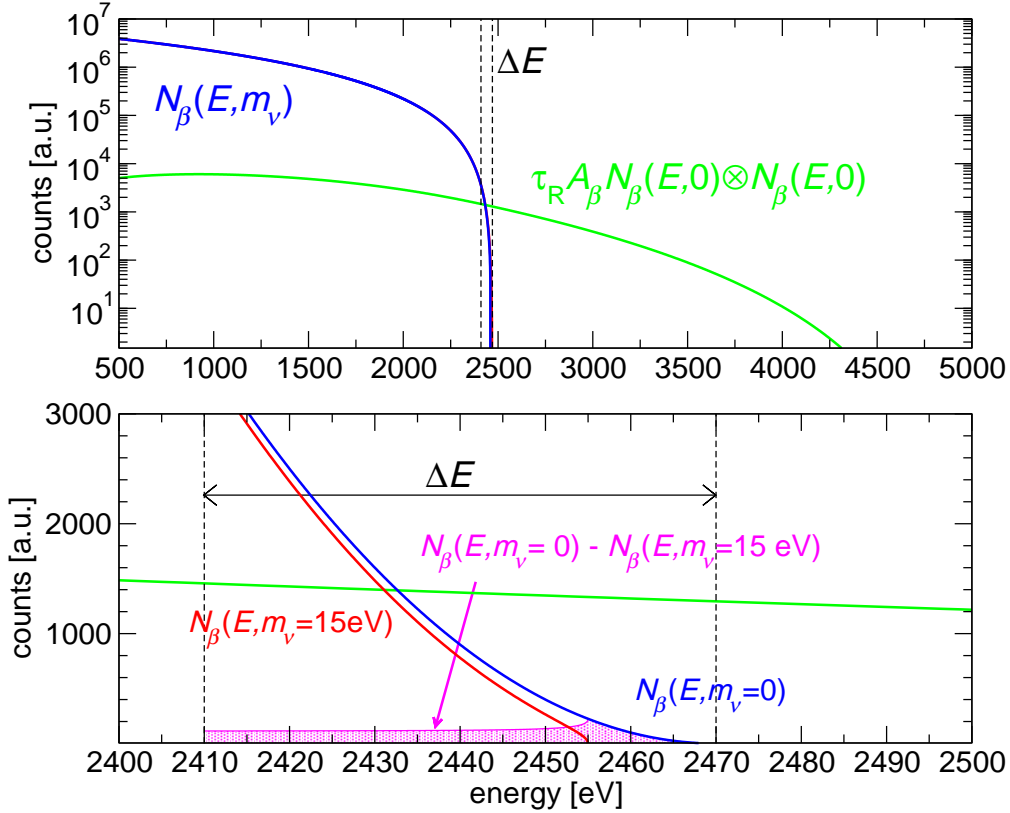}
\caption{\label{fig:sensi} Effects of pile-up on the experimental energy spectrum of \Re\ beta decay. Higher panel: beta spectrum compared with pile-up spectrum. Lower panel: zoom around the end-point, with a comparison between 0 and finite neutrino mass beta spectra}
\end{centering}
\end{figure*}

It is useful to derive an approximate analytic expression for the statistical sensitivity of a calorimetric neutrino mass experiment (see for example \cite{nucciotti2010expectations}).
The primary effect of a finite mass $m_{\nu_e}$ on the beta spectrum is to cause the
spectrum to turn more sharply down to zero at a distance $m_{\nu_e}$ below the endpoint
$E_0$ (higher panel of Fig.\,\ref{fig:sensi}). 
To rule out such a mass, an experiment must be sensitive to the number of counts expected in this interval. 
The fraction of the total spectrum within $\Delta E$ of the endpoint $E_0$ is given by
\begin{equation}\label{}
F_{\Delta \! E}(m_{\nu_e}) =\int^{E_0}_{E_0-\Delta \! E} N_\beta(E,m_{\nu}) d \! E 
\end{equation}
For $m_{\nu_e}=0$ this is approximately 
\begin{equation}
F_{\Delta \! E}(0) \approx \left( \frac {\Delta \! E} {E_0} \right)^3
\end{equation}\noindent 
For a finite mass it is found also
\begin{equation}
F_{\Delta \! E}(m_{\nu_e}) \approx F_{\Delta \! E}(0) \left(1 - \frac {3 m_{\nu}^2} {2 \Delta \! E^2} \right)
\end{equation}\noindent 
In addition to the counting statistics, the effect must be detected in the presence of an external background, 
and of the background due to undetected pile-up of two events.
Decays which occur within a definite time interval cannot be resolved by a calorimetric detector, giving rise to the phenomenon of pile-up. 
This implies that a certain fraction of the detected events is the sum of two or more single events. In particular, two low energy events can sum up and contribute with a count in the region close to the transition energy, contaminating the spectral shape in the most critical interval.
In a first approximation the external background can be neglected.
The pile-up spectrum can then be approximated by
assuming a constant pulse-pair resolving time, $\tau_R$, such that events with greater separation are
always detected as being doubles, while those at smaller separations are always interpreted as
singles with an apparent energy equal to the sum of the two events. In reality, the resolving time
will depend on the amplitude of both events, and the sum amplitude will depend on the
separation time and the filter used, so a proper calculation would have to be done through a Monte
Carlo applying the actual filters and pulse-pair detection algorithm being used in the experiment. However, this
approximation is good enough to get the correct scaling and an approximate answer.
In practice, $\tau_R$ depends on the high frequency signal-to-noise ratio but it is of the order of the detector rise time.

With these assumptions, for a pulse-pair resolving time of the detector $\tau_R$, the fraction of events which suffer with not-identified pile-up of two events for a Poisson time distribution is
\begin{equation}
\label{eq:PPU}
P(\Delta t < \tau_R) = 1 - e^{-A_\beta \tau_R} \approx A_\beta \tau_R
\end{equation}
where $A_\beta$ is the source activity in the detector and $\Delta t$ is the time separation between two events.
The beta spectrum of the unresolved pile-up events is given by the convolution product
\begin{equation}
\label{eq:spebpu}
N_{pp}(E) = (1 - e^{-A_\beta \tau_R}) \int_0^{E_0} N_\beta(E^\prime,0)N_\beta(E-E^\prime,0)d \! E^\prime =  (1 - e^{-A_\beta \tau_R})  N_\beta(E,0) \otimes N_\beta(E,0)
\end{equation} 
The coincidence probability, in a first approximation, is given by $\tau_R A_\beta$.
As shown in the lower part of Fig.\,\ref{fig:sensi}, a fraction $F^{pp}_{\Delta E}$ these events will fall in the region
within  $\Delta E$ of the endpoint $E_0$, and can be approximated by
\begin{equation}
F^{pp}_{\Delta E} = \int^{E_0}_{E_0-\Delta} N_{pp}(E) d \! E \approx \tau_R A_\beta \int^{E_0}_{E_0-\Delta \! E} N_\beta(E,0) \otimes N_\beta(E,0)d \! E 
\end{equation}
\noindent 
Measuring for a length of time $t_M$, the signal-to-background ratio in the region
within  $\Delta E$ of the endpoint $E_0$ can be expressed as
\begin{equation}\label{eq:S2B}
\frac{signal}{background} = \frac{A_\beta N_{det} t_M |F_{\Delta \! E}(m_{\nu_e})-F_{\Delta \! E}(0)|}{\sqrt{A_\beta N_{det} t_M (F_{\Delta \! E}(0)+F^{pp}_{\Delta E})}} = \sqrt{A_\beta T}\frac{|F_{\Delta \! E}(m_{\nu_e})-F_{\Delta \! E}(0)|}{\sqrt{(F_{\Delta \! E}(0)+F^{pp}_{\Delta E})}}
\end{equation}\noindent 
where $N_{det}$ is the number of detectors and $T=N_{det} t_M$ is the exposure.
This ratio must be about 1.7 for a 90\% confidence limit.  Therefore, in absence
of background, an approximated expression for the 90\% C.L. limit on \mne\  -- $\Sigma(m_{\nu_e})_{90}$ -- can be written  as \cite{nucciotti2010expectations}
\begin{equation}
\label{eq:sensitivity}
\Sigma_{90}(m_\nu) = 1.13  \frac{E_0}{\sqrt[4]{t_{M}A_\beta N_{det}}} \left[ \frac{\Delta E}{E_0} + \frac{3}{10}\frac{E_0}{\Delta E} \tau_R A_\beta \right]^{\frac{1}{4}}
\end{equation}
The two terms in (\ref{eq:sensitivity}) arise from the statistical fluctuations of, respectively, the beta and pile-up
spectrum in (\ref{eq:S2B}). Equation (\ref{eq:sensitivity}) shows the importance of improving the detector energy resolution and of minimizing the pile-up by reducing the detector rise time. On the other hand it shows also that the largest reduction on the \mne\ limit can only come by
substantially increasing the total statistics $N_{ev} = t_{M}A_\beta N_{det}$. 

\begin{figure*}[!t]
\begin{centering}
  \includegraphics[width=.6\textwidth]{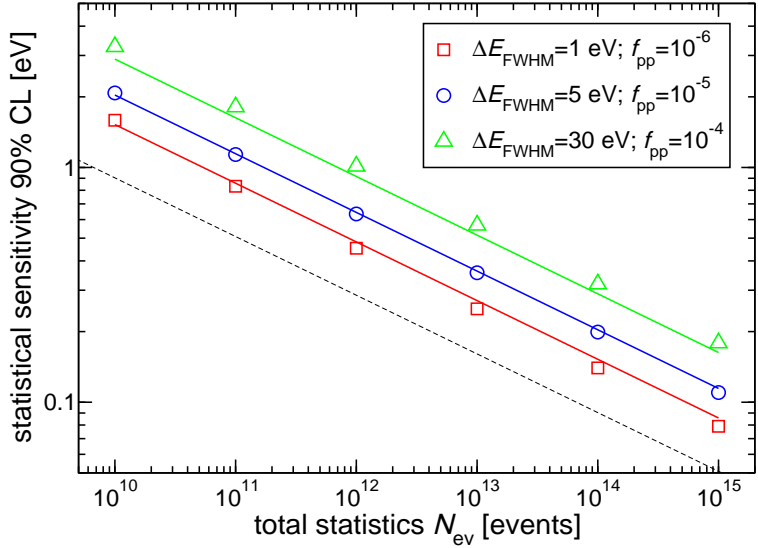}
\caption{\label{fig:sensre}Statistical neutrino mass sensitivity for calorimetric measurements of the \Re\ beta decay. Lines are obtained from (\ref{eq:sensitivity}) according to \cite{nucciotti2010expectations}. Symbols are obtained with Monte Carlo simulations (\S\,\ref{sec:stat-sys}).}
\end{centering}
\end{figure*}
If the pile-up is negligible, i.e. when the following condition is met
\begin{equation}\label{eq:condizione}
\tau_R A_\beta \ll \frac{10}{3} \frac{\Delta E^2}{E_0^2}
\end{equation}\noindent 
from (\ref{eq:sensitivity}) one can write the 90\% confidence limit sensitivity as 
\begin{equation}\label{eq:S90}
\Sigma_{90}(m_{\nu_e}) \approx1.13 \sqrt[4]{\frac{E_0^3 \Delta E}{ N_{ev} }}
\end{equation}\noindent
where energy interval $\Delta E$ in (\ref{eq:S90}) cannot be taken smaller than about 2 times the detector energy resolution
$\Delta E_{\rm FWHM}$. 

It is then apparent that to increase the sensitivity one has both to improve the energy resolution and to augment the statistics; however, there is a technological limit to the resolution improvements, thus the statistics $N_{ev} = t_{M}A_\beta N_{det}$ is in fact the most important factor in (\ref{eq:S90}). 
For a more complete treatment, also in presence of a not negligible pile-up, refer to
\cite{nucciotti2010expectations}.

A similar approach for assessing the statistical sensitivity of \Ho\ EC decay cannot be pursued with the same simplicity because of the more complex spectrum (see \S\,\ref{sec:holmium}).
Nevertheless it is worth anticipating that with some approximations -- discussed in  \S\,\ref{sec:holmium} -- one can at least
easily show that 
\begin{equation}\label{eq:sensho}
\Sigma_\mathrm{EC}(m_{\nu_e}) \propto E_0 - E_{i_{max}}
\end{equation}\noindent
where $E_{i_{max}}$ is the energy of the Lorentzian peak whose high energy tail dominates the end-point region, i.e. the M1 peak in (\ref{eq:E_c-distr}). 
Equation \ref{eq:sensho} is to be compared to equation \ref{eq:S90}, which gives 
\begin{equation}\label{eq:sensre}
\Sigma_\mathrm{\beta}(m_{\nu_e}) \propto E_0^{4/3} 
\end{equation}\noindent
From (\ref{eq:sensho}) it is apparent that for EC experiments in general -- and for \Ho\ in particular -- it is not only winning to have the lowest possible $Q$, but the end-point energy must be as close as possible to the binding energy of the deeper shell accessible to the EC.

\subsection{LTD for calorimetric neutrino mass measurements}
\label{sec:ltd}
In 1981 A.\,De Rujula was already discussing with E.\,Fiorini about the possibility of performing a calorimetric measurement of the electron capture process in \Ho, apparently without any useful conclusion. It was only 3 years later -- in 1984 -- that
two independent seminal papers proposed for the first time the use of 
phonon-mediated detectors operated at low temperatures (simply called here low temperature detectors, LTDs) 
for single particle detection with high energy resolution. E.\,Fiorini and T.\,Niinikoski \cite{fiorini1984low} proposed to apply these new detectors to various rare events searches in a calorimetric configuration, while D.\,McCammon et al. \cite{moseley1984thermal,mccammon1984experimental} initiated the application to X-ray detection. It was immediately clear to D.\,McCammon et al. that this could be extended to the spectroscopy of an internal beta source by realizing high energy resolution calorimeters with implanted tritium  \cite{barger1985neutrino}. 

In 1985, few years after A.\,De\,Rujula suggested the use of \Re\ and \Ho\ for a sensitive neutrino mass measurement in \cite{derujula1981new},
S.\,Vitale et al. came up with the first operative proposal for an experiment using LTDs to measure the \Re\ spectrum  calorimetrically \cite{vitale1985renio}. 
The same year, also N.\,Coron et al. started a research program aiming at exploiting LTDs to perform the calorimetry of \Ho\ EC decay \cite{coron1985composite}, which was soon discontinued for what concerns \Ho. 
In the following years, the Genova group pioneered the development of LTDs aimed at a direct neutrino mass measurement using the \Re\ beta decay.
The experiment was later called MANU and produced its first result in 1992.
Some years later, in 1993, the Milano group, mostly focused on carrying out a \bbn\ search with LTDs, also opened a research line to develop high energy resolution LTDs for a calorimetric measurement of \Re\ beta decay. This project was named MIBETA and came to the first measurement in 1999.
In 2005, the MANU and MIBETA experiments merged in the international project MARE.
In parallel to the work on \Re, starting from 1995 the Genova group was also carrying on a research for a calorimetric measurement of the \Ho\ EC decay.
This activity, later on, was first absorbed in MARE and then transferred into the HOLMES project.
In 2012, the Heidelberg group, former member of the MARE collaboration, presented its own R\&D program for a \Ho\ calorimetric experiment, ECHo.
Recently, also the Los Alamos group started a preliminary work for a \Ho\ experiment, with a project named NuMECS.
All these experiments and projects will be discussed in the next two sections.

It is worth mentioning two other groups that  in these three decades participated to the efforts to develop LTDs for neutrino mass measurements.
The Oxford group developed arrays of indium based Superconducting Tunnel Junctions (STJ) to search for the 17\,keV neutrino in the $^{63}$Ni beta decay \cite{gaitskell1996measurement} and to measure precisely the exchange effect in the low energy part of the spectrum of the same decay \cite{angrave1998measurement}.
The Duke University group developed Transition Edge Sensors (TESs) based detectors for measuring calorimetrically the tritium decay \cite{Deptuck200080,Erhardt200092},
but this project was abandoned before obtaining a statistically meaningful sample. 

All these activities were triggered in the early '80s by the lucky coincidence to have the need for a tool to perform calorimetric measurements of new low $Q$ beta isotopes just at the time when a new promising particle detection technology was appearing on the scene.
It took more than 20 years to the LTD technology to actually be mature enough to sustain the ambitions of calorimetric neutrino mass experiments 
Nowadays, LTDs can indeed deliver to this science case what they have been developed for. In particular, LTDs provide better
energy resolution and wider material choice than conventional detectors.
The energy resolution of few electronvolts is comparable to that of spectrometers and the restrictions caused by the full spectrum 
detection are lifted by the parallelization of the measurement with large arrays of detectors.
Still, the detectors time constants of the order of microseconds and, correspondingly, the read-out bandwidth remain the most serious 
technical constraint to the full exploitation of LTDs in this field. 

\subsubsection{LTD basic principles}
A complete overview of LTDs can be found in \cite{enss2005cryogenic}, while
the status-of-the-art is well summarized in the proceedings of the bi-yearly international workshop on Low Temperature Detectors \cite{ltd15proceedings}.   

LTDs were initially proposed  as perfect calorimeters, i.e. as devices able to thermalize thoroughly the energy released by the impinging particle.
In this approach, the energy deposited by a single quantum of radiation into an energy absorber (weakly connected to a heat sink) determines an increase of its temperature $T$. This temperature variation corresponds simply to the ratio between the energy $E$ released by the impinging particle and the heat capacity $C$ of the absorber, i.e. is given by $\Delta T = E/C$. The only requirements are therefore to operate the device at low temperatures (usually $< 0.1$\,K) in order to make the heat capacity of the device low enough, and to have a sensitive enough thermometer coupled to the energy absorber.
Often LTDs with a total mass not exceeding 1\,mg and few hundreds micron linear dimensions are called low temperature (LT) microcalorimeters.

In the above linear approximation, using simple statistical mechanics arguments, it can be shown  that the internal energy of an LTD weakly linked to
a heat sink fluctuates according to
\begin{equation}
\label{eq:thermodinamicallimit}
\Delta E_{rms}^2 = k_B T^2 C
\end{equation}
where $T$ is the equilibrium operating temperature, $k_B$ is the Boltzmann constant, and independent of the weak link thermal conductance $G$.
Eq.\,\ref{eq:thermodinamicallimit} is often referred to as the {\it thermodynamical limit} to the LTD sensitivity and the internal energy fluctuations as Thermodynamic Fluctuation Noise (TFN). Although, strictly speaking, (\ref{eq:thermodinamicallimit}) is not the best energy resolution achievable by an LTD,
it turns out that when a sensitive enough thermometer is considered and all sources of broadband noise are included in the calculation, the    
real {\it thermodynamical limit} of the energy resolution of an LTD can be expressed as \cite{moseley1984thermal}
\begin{equation}
\label{eq:TLriso}
\Delta E_{rms}^2 = \xi^2 k_B T_b^2 C_0
\end{equation}
where now $T_b$ is the heat sink temperature, $C_0$ is the heat capacity at $T_b$, and $\xi$ is a numerical parameter of order one which is derived 
from the LTD thermal details and for the optimal operating temperature. A detailed analysis of the optimal energy resolution for various thermometers
can be found in \cite{enss2005cryogenic}.

From the above and (\ref{eq:TLriso}) it is evident that the LTD absorber with its $C$ together with the thermometer with its sensitivity are the crucial
ingredients for obtaining high energy resolution detectors.
A sensitive thermometer is the one which allows to transduce the temperature fluctuations of the TFN to a signal larger than the other noise sources
intrinsic to the thermometer itself and to the signal read-out chain.
Today, this condition has been met -- and (\ref{eq:TLriso}) is achieved -- for LT microcalorimeters using at least three types of optimized thermometers: semiconductor thermistors, transition edge sensors (TES), and Au:Er metallic magnetic sensors. 
The thermal sensor of a LTD does not only affect the achievable energy resolution, but also determines the {\it speed} of the detector, i.e. it determines the time scale of the signal formation with the details of the thermal mechanisms entering in the temperature transduction.
Although the detector speed is a crucial parameter in calorimetric neutrino mass experiments, a complete technical treatment for the three sensor technologies is out of the scope of present work. Here it is enough to say that the three technologies above are sorted from the slowest to the fastest: the numerical values for the achievable speeds (from hundreds of nanoseconds to hundreds of microseconds) will be given in the following sections.
Each sensor technology has its pros and cons which have driven the choice for the various neutrino mass experiments. The traded off parameters, which include the achievable performances, the ease of fabrication, and the read-out technology, will be discussed in \S\,\ref{sec:sensors}.

Next section is dedicated to the other critical component, i.e. the absorber.

\subsubsection{Energy absorber and thermalization process}
Under many respects, the absorber of LTDs plays the most crucial role in calorimetric experiments.
First of all (\ref{eq:TLriso}) shows that the absorber heat capacity $C$ sets the achievable energy resolution.
When designing LTDs, usually the absorber is chosen to be made out of a dielectric and diamagnetic material, so that $C$ is described 
only by the Debye term, which is proportional to $(T / \Theta_D)^3$ at low temperatures, and can be extremely small for a good material with large
Debye temperature $\Theta_D$. Insulators and semiconductors are often good examples of suitable dielectric and diamagnetic materials.
Metals are instead discarded because of the electron heat capacity, which is proportional to $T$ and remains large also at very low temperatures, thereby dominating the total $C$ of the absorber.
Superconductors are in principle also suitable, since the electronic contribution to the specific heat vanishes exponentially below the critical temperature $T_c$, and only the Debye term remains.
For microcalorimeters, the situation is different because their reduced size allows to tolerate also the heat capacity of a metal so that other considerations may be adopted to select the absorber material.
Microcalorimeters for calorimetric measurements of the tritium, \Re, or \Ho, decay spectra must contain the unstable isotope in their absorbers.
As it will be discussed in more details in the following, while tritium and \Ho\ can be included by various means in materials with no special relation with hydrogen or holmium, \Re\ is naturally found in physical and chemical forms suitable for making LTDs, i.e. superconducting metal and dielectric compounds.
In addition to the electronic and phononic heat capacities considered above, other contributions caused by nuclear heat capacity or by impurities may become important in certain conditions \cite{enss2005low}. As it will be discussed later, this  can be the case for metallic rhenium and embedded \Ho.

The above leads to the conclusion that there is large a flexibility in the choice of the material for the absorber of microcalorimeters for calorimetric neutrino mass experiments: dielectrics and normal or superconducting metals have all indeed been used.
In spite of this apparent flexibility, in such experiments it turned out that the ideal energy resolution (\ref{eq:TLriso})  is quite hard to achieve
because of the details of the chain of physical processes which transform the energy deposited as ionization into the altered equilibrium thermal distribution of phonons -- i.e. the $\Delta T$ above --  sensed by the thermometers. This chain -- also called the {\it thermalization process} -- is responsible for the introduction of a fluctuation in the deposited energy $E$ which is finally converted in the measured  $\Delta T$: the so-called {\it thermalization noise}.
The chain starts with the hot electron-hole pairs created by the primary ionizing interaction: on a time scale of 10$^{-10}$\,s, this energy is degraded and partitioned between colder electronic and phononic excitations by means of electron-electron and electron-phonon scattering. 
The chain then proceeds with the conversion of the electronic excitations into phonons accompanied by a global cooling of all excitations, and it ends
with the new thermal distribution of phonons which corresponds to a temperature increase $\Delta T$ above the equilibrium operating temperature.
The total time scale of this latter process and its details strongly depend on the material. 
Only when the time elapsed between the primary interaction and the signal formation is long enough to allow the phonon system to relax to the new quasi-equilibrium distribution, the detector works really as a calorimeter. In commonly used thermal sensors, the measured physical quantity is sensitive
to the temperature of sensor electrons: therefore, at the end of the \textit{thermalization} there must be a last heat flow from the absorber phonons to the sensor electrons through a link which ultimately acts as a throttle for the signal rise.  
The extra noise shows up every time the deposited energy is not fully converted into heat \cite{moseley1984thermal}, i.e. into the new quasi-equilibrium thermal distribution, and gets trapped in long -- compared to thermalization and signal formation time scales -- living excitations.
In a simplified picture, if $\eta$ is the fraction of deposited energy $E$ which actually goes into heat, the achievable energy resolution may be written as
\begin{equation}
\label{eq:thermnoise}
\Delta E_{rms}^2= \xi^2 k_BT_c^2C_0  + (1-\eta)^2 EF\omega
\end{equation}
where $F$ is the Fano factor and $\omega$ is the average excitation energy of the long living states. The second term is given by the statistical fluctuation
of the number of long living states.
The parameters $\eta$ and $\omega$ are peculiar of each type of material. The parameter $\eta$ may depend on the operating temperature and on the signal
time scale: often, the {\it thermalization} slows down at low temperatures and the signal time scale must be adapted accordingly. 
Under all these respects, metals are the ideal material because they show fast and complete {\it thermalization} at every temperature  -- i.e. $\eta=1$ is achieved on time scales of the order of nanoseconds or less -- thanks to the strong interactions between electrons and phonons.
Microcalorimeters with metallic absorbers in electrical contact with the sensor are often called \textit{hot-electron} microcalorimeters \cite{nahum1993hot,Irwin1996hot}. In \textit{hot-electron} microcalorimeters the {\it thermalization} ultimately warms up the absorber electronic system and the hot absorber electrons can directly warm up the sensor electrons without throttling, therefore showing a very fast response time.

On the contrary, dielectrics often suffer of a large {\it thermalization noise} translating in a degraded energy resolution which increases with the deposited energy. In dielectrics, impurities and defects can act as traps which lie energetically inside the forbidden band-gap. Following the primary
ionization created by the incident particle, electrons and holes can get trapped before their recombination to phonons. Experimentally it is found that
$\omega$ can be as large as few tens of electronvolts, so that the second term (\ref{eq:thermnoise}) may easily dominate the energy resolution also
for values $\eta$ approaching unity.
Semiconductors may be better than dielectrics, owing to their smaller band-gap. 
But only metals, semi-metals (such as bismuth) and zero-gap semiconductors (such as HgTe) have been successfully employed in microcalorimeters showing energy resolutions close to the {\it thermodynamical limit} (\ref{eq:TLriso}) \cite{ullom2015review}.

In principle, superconductors should provide a further improvement thanks to their band-gap of few millielectronvolts: unfortunately, the {\it thermalization} in superconductors is a complex process in which $\eta$ can be very small. 
In superconductors the electronic excitations produced in the {\it thermalization} process described above are broken Cooper pairs, also called
quasi-particles \cite{van1991probability,zehnder1995response,kozorezov2000quasiparticle}.
Microscopic calculations from the Bardeen-Cooper-Schrieffer theory predict that, indeed, a large part of the energy released inside the
absorber can be trapped in quasi-particles states which can live for many seconds at temperatures below 0.1\,K. 
The energy release inside a superconductor leads to a long living state far from equilibrium in
which many Cooper pairs are continuously broken by phonons produced when quasi-particles recombine. 
A model describing this situation was proposed by S.B.\,Kaplan \cite{kaplan1976quasiparticle},
who found that the time the quasi-particles need to recombine ($\tau_{qp}$) would be somewhere between 1 and 10 seconds. Analogous results were
obtained by the analysis of A.G.\,Kozorezov \cite{kozorezov2000quasiparticle}.
The global result of these models is that in superconductors $\eta$ is expected to be very small on a time scale useful for a LTD.
Despite of these theoretical considerations, it is an experimental fact that some superconducting materials perform well as absorbers in cryogenic detectors. 
Indeed, deviations from the predicted temperature dependence of quasi-particle lifetime $\tau_{qp}$ have been reported: for example,
tin has been used for making LTDs with an energy resolution approaching the {\it thermodynamical limit}, thanks to a fast and complete energy {\it thermalization}. This is apparently one characteristic shared also by other {\it soft} superconductors such as lead an indium.
So far, no generally accepted explanation has been given for these apparent discrepancies between experimental results and theory, and the topic
of quasi-particles recombination in LTDs remains an active field of research.

There are two other important sources of energy resolution degradation which are often observed in LTDs  \cite{moseley1984thermal}.
The first is the escape from the absorber of high energy phonons during the first stages of the {\it thermalization} process, which adds another fluctuation component to the finally thermalized energy.
The second is the accidental direct detection of high energy phonons by the thermal sensors, which determines an excess systematic broadening of the energy resolution because its probability varies with the interaction position.  

\subsubsection{Temperature sensors, read-out, and signal processing}
\label{sec:sensors}
The LTDs used for neutrino mass calorimetric measurements fall in the category of the low temperature microcalorimeters and are designed to 
provide energy resolutions better than about 10\,eV, possibly approaching the \textit{thermodynamical limit}.
As shown in \S\,\ref{sec:calosens}, the detector \textit{speed} -- i.e. the detector signal bandwidth, or its rise time $\tau_R$ -- is another 
parameter guiding the design. 
Furthermore, neutrino mass experiments with LTDs need to use large arrays of detectors. This calls for ease of both fabrication and signal read-out.
Along with the selection of the absorber material containing the source, the above are the main guidelines for the design of an LTD based neutrino mass experiment. The choice of the sensor technology is one of the first steps in the design.
To date, only three technologies have been exploited. These are the semiconductor thermistors, the transition edge sensors, and the magnetic metallic sensors, and they will be briefly discussed here (more details can be found in \cite{enss2005cryogenic}). The possibility of employing other technologies, such as the one of superconducting microwave microresonators, is also investigated but its perspectives are not clear yet \cite{faverzani2012developments}. 
The application of LTDs to the spectroscopy of \Re\ and \Ho\ decays fully overlaps the range of use of microcalorimeters developed for soft X-ray  
spectroscopy; therefore, in the following the discussion will be restricted to thermal sensors for X-ray detection.
\begin{figure}[t]
 \centering
 \includegraphics[clip=true,width=0.90\textwidth ]{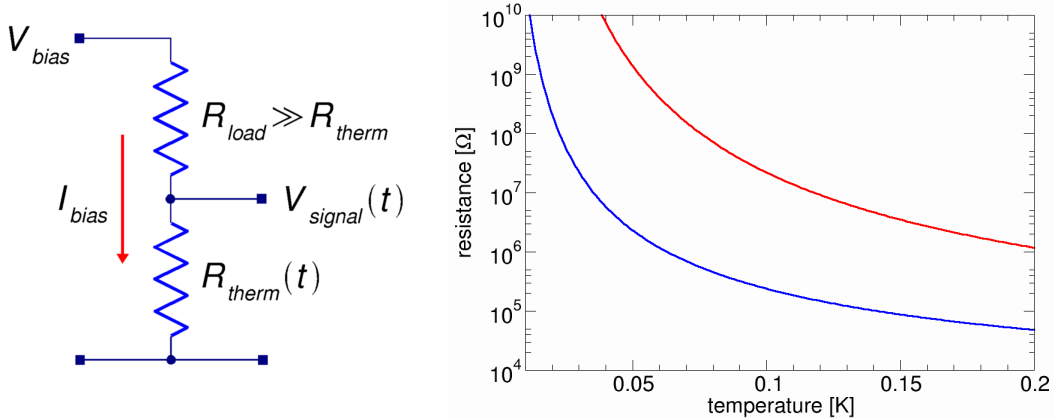}
 \caption{Constant current biasing of thermistors (left). Temperature dependence of the thermistor resistivity for a $T_0$ of 3\,K (blue) and 10\,K (red) (right).}
 \label{fig:thermistor-r}
 \label{fig:thermistor-readout}
\end{figure}
\paragraph{Semiconductor thermistors}
These sensors are resistive elements with a heavy dependence of the resistance on the temperature. 
Usually, they consist  of small crystals of germanium or silicon with a dopant concentration slightly below the metal-to-insulator transition \cite{moseley1984thermal}. 
The sensor low temperature resistivity is governed by variable range hopping (VRH) conduction and it is often well described by the
expression $\rho(T)=\rho_0 \exp(T_0/T)^{1/2}$, where $T_0$ and $\rho_0$ are parameters controlled by the doping level \cite{zhang1993hopping} (Fig.\,\ref{fig:thermistor-r}). 
Semiconductor thermistors are high impedance devices -- 1-100\,M$\Omega$ -- and are usually parameterized by the sensitivity $A$, defined as $-d \log R / d \log T$, which typically ranges from 1 to 10. Semiconductor thermistors can be realized also in amorphous film form, like NbSi. 
Silicon thermistors are fabricated using multiple ion implantation in high purity silicon wafers to introduce the dopants in a thin box-like volume defined by photolithographic techniques. Germanium thermistors are fabricated starting from bulk high purity germanium crystals doped by means of neutron irradiation (nuclear transmutation doping, NTD) \cite{haller1984ntd,haller1994advanced}. Single NTD germanium sensors are obtained by dicing and further
processing using a combination of craftsmanship and thin film techniques.
In early times, the weak coupling to the heat sink was provided by the electrical leads used for the read-out; nowadays, microelectronic planar technologies  and silicon micro-machining are used to suspend the sensors on thin silicon nitride membranes or thin silicon beams. 
Thermistors are read-out in a constant current biasing configuration which allows to convert the thermal signal $\Delta T$ in a voltage signal $\Delta V$ (Fig.\,\ref{fig:thermistor-readout}). Because of their high impedance, thermistors are best matched to JFETs.
Semiconductor thermistor present few drawbacks. First of all their high impedance requires the JFET front end to be placed as close as possible -- centimeters -- to the devices to minimize microphonic noise, and bandwidth limitations due to signal integration on parasitic electrical capacitance. Since commonly used silicon JFET must operate at temperatures not lower than about 110\,K, this becomes quickly a technical challenge when increasing the number of detectors. Secondly, it has been experimentally observed that conductivity of semiconductor thermistors deviates from linearity at low temperatures \cite{zhang1998non,aubourg1993measurement}. The deviation is understood in terms of a finite thermal coupling between electrons and phonons, whose
side effect is to intrinsically limit the signal rise times to hundreds of microseconds for temperatures below 0.1\,K.    
Semiconductors are now an established and robust technology, and arrays of microcalorimeters based on these devices have been widely used for X-ray spectroscopy \cite{enss2005cryogenic} achieving energy resolutions lower than 5\,eV with tin or HgTe absorbers.

\begin{figure}[t]
 \centering
 \includegraphics[clip=true,width=0.90\textwidth ]{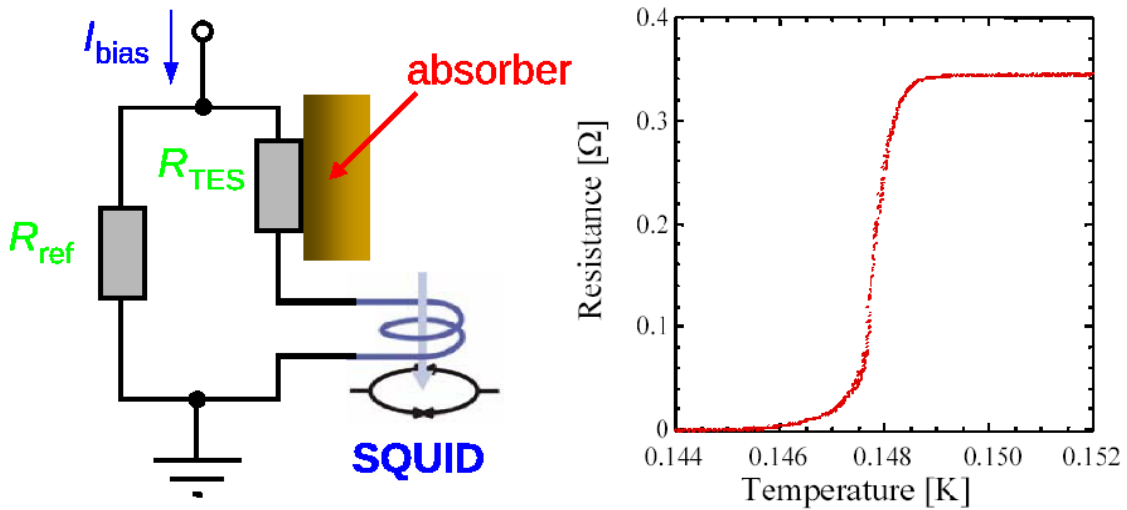}
 \caption{Constant voltage biasing of a TES (left). Temperature dependence of the TES resistivity at $T_c$ (right)}
 \label{fig:tes-readout}
 \label{fig:tes-r}
\end{figure}
\paragraph{Superconducting transition edge sensors (TESs)}
TES are also resistive devices made out of thin films of superconducting materials whose resistivity changes sharply from 0 to a finite value in a very narrow temperature interval around the critical temperature $T_c$ (Fig.\,\ref{fig:tes-r}). 
The superconducting material can be an elemental superconductor (such as tungsten or iridium), although it is more often a bilayer made of a normal metal and a superconductor. 
With bilayers, the $T_c$ of the superconductor is reduced by the proximity effect and can be controlled by adjusting the relative thicknesses of the two layers.
Common material combinations used to fabricate TES bilayer with a $T_c$ between 0.05 and 0.1\,K are Mo/Au, Mo/Cu, Ti/Au or Ir/Au.  
TES fabrication exploits standard thin film deposition techniques, photolithographic patterning, and micro-machining. Sensors can be designed to have,  at the operating point, a sensitivity $A$ as high as 1000 and a resistance usually less than 1\,$\Omega$. The most common ways to isolate TES microcalorimeters
from the heat sink are the use of thin silicon nitride membranes or thin silicon beams.
TES are read-out at a constant voltage and their low impedance is ideal to use SQUIDs to amplify the current signal induced by a particle interaction (\ref{fig:tes-readout}). The constant voltage biasing provides the condition to achieve the extreme electro-thermal feedback (ETF) regime \cite{irwin1995etf} which leads to substantial improvements in resolution, linearity, response speed, and dynamic range. This regime also eases the operation of large pixel count arrays because ETF produces a self-biasing effect that causes the temperature of the film to remain in stationary equilibrium within its transition region. 
With respect to semiconductor thermistors, TESs offer many advantages: 1) large arrays can be fully fabricated with standard micro-fabrication processes, 2) the larger electron-phonon coupling allows signal rising as fast as few microseconds, and 3) the low impedance reduces the sensitivity to environmental mechanical noise.  
The main drawbacks of TESs are the limited dynamic range, the adverse sensitivity to magnetic fields of TES and SQUID, and the not fully understood physics of superconducting transitions and excess noise sources \cite{ltd15proceedings}.
TES microcalorimeter arrays are being actively developed as X-ray spectrometers for many applications, which include material analysis and X-ray astrophysics \cite{ullom2015review}. TES sensors are particularly well suited to be coupled to metallic (gold) or semi-metallic (bismuth) absorbers, providing fast response and energy resolutions lower than few electronvolts (Fig.\,\ref{fig:tes-de}).
\begin{figure}[t]
 \centering
 \includegraphics[clip=true,width=0.90\textwidth ]{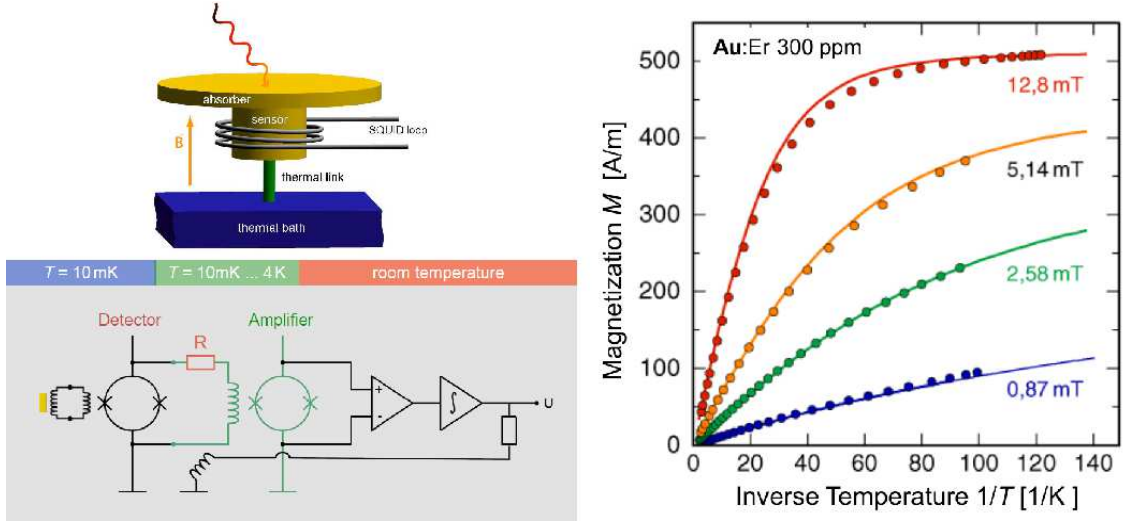}
 \caption{Read-out of a paramagnetic sensor (left). Temperature dependence of the sensor magnetization (right)}
 \label{fig:mmc-readout}
\end{figure}
\paragraph{Magnetic metallic sensors}
These sensors are quite different from the previous two, and their successful development is more recent \cite{Fleischmann2009AIPC}.
They are paramagnetic sensors exposed to a small magnetic field. The temperature rise $\Delta T$ causes a change in the sensor magnetization,
which is sensed by a SQUID magnetometer. The non dissipative read-out scheme avoids the noise sources typical of dissipative systems, such as
the Johnson noise of semiconductor thermistors and of TESs.
State of the art sensors use Er$^+$ paramagnetic ions localized in a Au metallic host (Au:Er sensors). The use of a metallic host 
ensures a very fast sensor response time, since the spin-electron relaxation time for Au:Er is around 0.1\,\mus\ at about 0.05\,K.
Microcalorimeters with magnetic metallic sensors (Magnetic Metallic Calorimeters, MMCs) are usually fully made out of gold to obtain both a fast and  
efficient energy thermalization to the absorber electronic system and a quick equilibration with the sensor electrons. Despite  its high sensitivity, the paramagnetic sensor has an intrinsically large heat capacity: therefore the gold absorbers may be relatively large without adversely affecting the MMC performance.
These microcalorimeters, in general, do not need special measures for thermally isolating the devices from the heat sink because the signal is predominantly
developed in the electronic system and the electron-phonon coupling is rather weak and slower at low temperatures.
An interesting feature of MMCs is the availability of a complete and successful modeling, which allow a precise design tailored to each specific application.
The micro-fabrication of MMCs is somewhat more cumbersome than for TES microcalorimeters but, for the large part, can be carried out with a standard micro-fabrication process \cite{hsieh2008fabrication,burck2008microstructured}.
Presently, the most used design for arrays of MMCs has planar sensors on meander shaped pickup coils and achieves record energy resolutions of few electronvolts for soft X-rays (Fig.\,\ref{fig:mmc-de}) accompanied by large dynamic range and good linearity. 
\begin{figure}[t]
 \centering
 \includegraphics[clip=true,width=0.46\textwidth ]{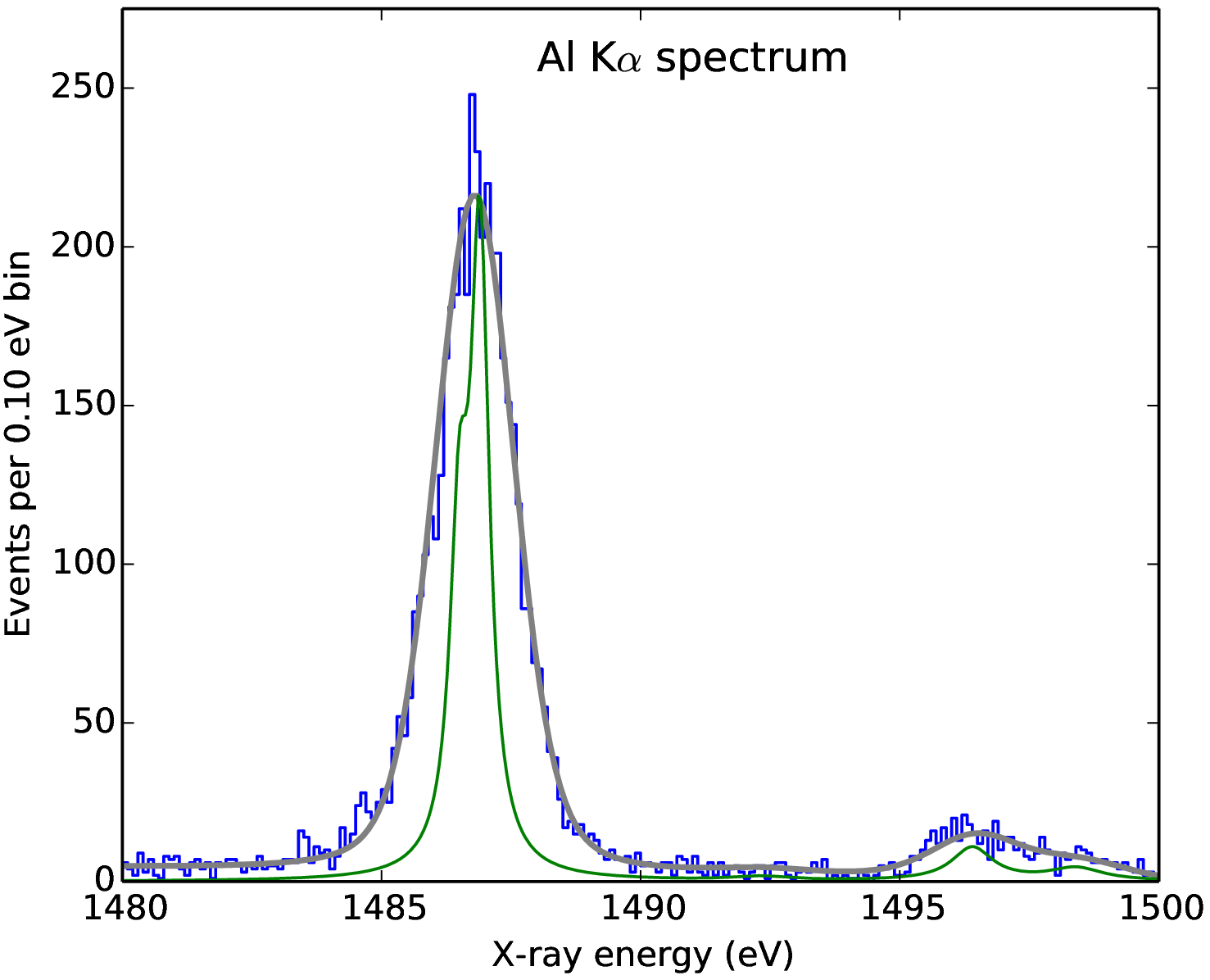}\hfill
 \includegraphics[clip=true,width=0.49\textwidth ]{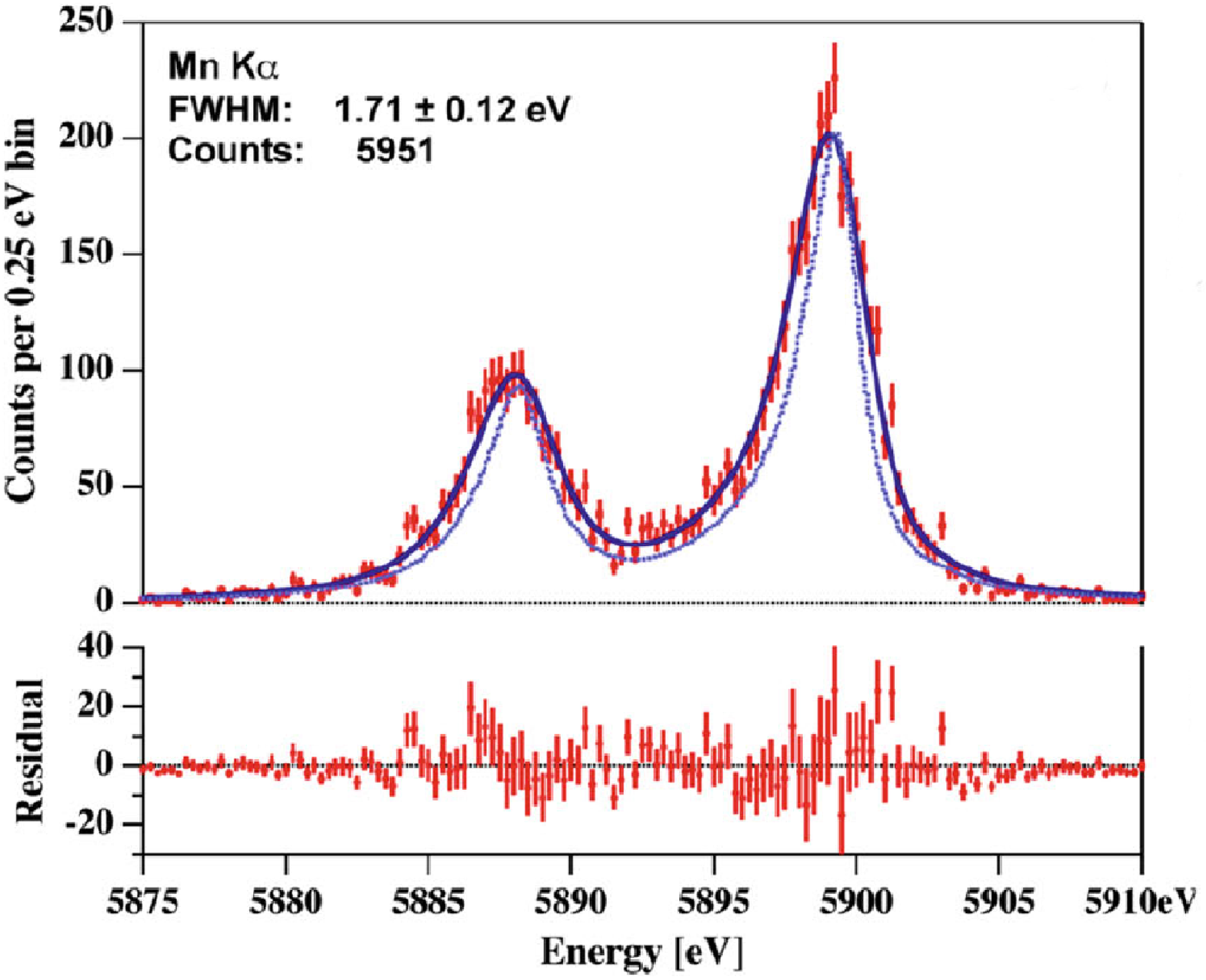}
 \caption{Energy resolutions achieved with a TES ($\Delta E_\mathrm{FWHM}=1.5$\,eV, courtesy of NIST) (left) and with an MMC \cite{porst2014characterization} (right).}
 \label{fig:tes-de}
 \label{fig:mmc-de}
\end{figure}

\paragraph{Signal read-out}
Neutrino mass experiments are necessarily carried out using LTD arrays with a large pixel count, and this calls for the implementation of an efficient multiplexing system for reading out many sensors with the smallest possible number of amplifiers. This, in turn, reduces the number of read-out leads from
room temperature to the array and the power dissipation at low temperature. 
Therefore, in order to be of some use for future experiments, a sensor technology must be compatible with some sort of multiplexed read-out, 
not causing restrictions on the available signal bandwidth and degradation of the resolving power. This makes the semiconductor thermistors 
not appealing for sensitive neutrino mass experiments. The opposite is true for the other two technologies owing to the use of a SQUID read-out.

TES arrays with SQUID read-out can be multiplexed according to three schemes  \cite{ullom2015review}: Time Division (TDM) \cite{reintsema2003prototype}, Frequency Division (FDM)  \cite{dobbs2012frequency} and Code Division (CDM) \cite{stiehl2012code}. The three schemes differ by the set of orthogonal modulation functions used to encode the signals. TDM and FDM (in the MHz band) are the most mature ones, and they have been already applied to the read-out of many multi-pixel scientific instruments. The more recently developed CDM combines the best features of TDM and FDM, and is useful for applications demanding fast response and high resolution.

\begin{figure}[ht]
 \centering
 \includegraphics[clip=true,width=0.49\textwidth ]{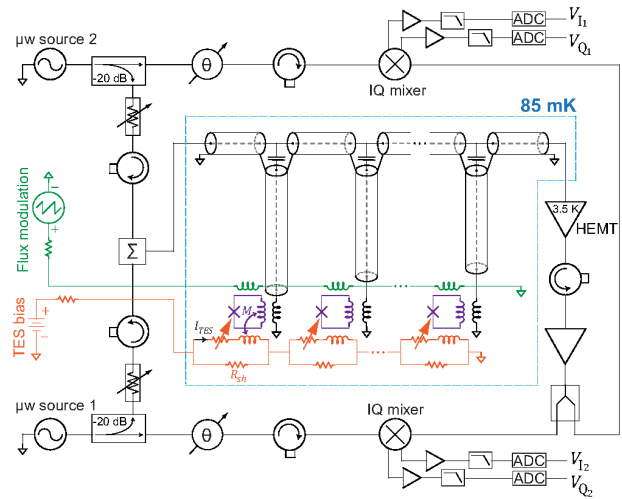}\hfill
 \caption{Circuit schematic for a two channel microwave multiplexed read out of a TES array (From  \cite{noroozian2013high}).}
\label{fig:umux}
\end{figure}

Recent advancements on microwave multiplexing ($\mu$MUX) suggest that this is the most suitable system for neutrino mass experiments, since it provides a larger bandwidth for the same multiplexing factor (number of multiplexed detector signals).
It is based on the use of rf-SQUIDs as input devices, with flux ramp modulation \cite{noroozian2013high} (Fig.\,\ref{fig:umux}). The modulated rf-SQUID signals are read-out by coupling the rf-SQUID to superconducting LC resonators in the GHz range and using the homodyne detection technique. By tuning the LC resonators at different frequencies it is straightforward to multiplex many RF carriers.
The feasibility of this approach has been demonstrated in \cite{noroozian2013high} only with two channels, but it is making quick progresses
as shown with the multiplexed arrays of TES bolometers for millimeter astronomy of MUSTANG2 \cite{dicker2014mustang2}. 

The $\mu$MUX is suitable for a fully digital approach based on the Software Defined Radio (SDR) technique \cite{mchugh2012readout,bourrion2013high}. The comb of frequency carriers is generated by digital synthesis in the MHz range and up-converted to the GHz range by $IQ$-mixing. The GHz range comb is sent to the cold $\mu$MUX chips coupled to the TES array through one semi-rigid cryogenic coax cable, amplified by a cryogenic low noise High Electron Mobility Transistor (HEMT) amplifier, and sent back to room temperature through another coax cable.
The output signal is down-converted by $IQ$-mixing, sampled with a fast analog-to-digital converter, and digital mixing techniques are used to recover the signals of each TES in the array ({\it channelization}). 

Because of their excellent energy resolution combined with a very fast response time, the multiplexed read-out of MMCs is more demanding than for TESs.  To date, although some results have been obtained also with TDM, $\mu$MUX is the most promising approach for multiplexing MMCs \cite{kempf2014umltiplexed}, even if its development for these devices is still in progress.
\paragraph{Signal processing}
One of the conditions to obtain a thermodynamically limited energy resolution is to process the microcalorimeter signals with the Optimal Filter (OF) \cite{gatti1986processing,moseley1984thermal}. For this purpose the signal waveforms must be fully digitized and saved to disk for further
off-line processing. This approach allows also to apply various specialized signal filters to the same waveform, with the aim of improving time resolution -- thereby reducing $f_{pp}$ --, rejecting spurious events, and gating background induced events with a coincidence analysis. 
The storage of the raw data needed for off-line signal processing and for building the energy spectrum to be analyzed sets a practical limit to the lower energy limit of the final energy spectrum.
While there is no issue for LTD experiments, such as dark matter or \bbn\ searches, to save digitized waveforms for later off-line analysis,
this becomes quickly unpractical for sub-eV neutrino mass experiments. The pulses collected over the whole spectrum would amount to about $10^{14}$, and digitizing pulses
with 512 samples with 16\,bit depth would translate in a storage of about 100\,PB (un-compressed), way more than LHC data.
The only viable strategy is then to save only the relevant event parameters calculated by the pulse processing software. These parameters
must include at least the energy, the arrival time, and a couple of other useful shape parameters. 
Assuming one can limit the useful parameters to 5 and that each is saved as a 4 bytes number, then the required storage reduces to about 1\,PB,
which is still quite a large but manageable number.
It looks therefore likely that only fractions of the spectrum of the order of 10\% will be available for a neutrino mass analysis.  
For \Re\ and \Ho, this means that the analysis will be forcibly limited to an energy interval which extends, respectively, about 1200\,eV and 750\,eV -- just right of the main peak of the spectrum (\S\,\ref{sec:holmium}) -- below the spectrum end-point.

\subsection{Additional direct neutrino measurements with LTDs}
\label{sec:other}
Thanks to the acquisition of the full energy spectrum, LTD calorimeters offer the opportunity to perform other interesting investigations on the data collected for the neutrino mass measurement: these include the searches for massive sterile neutrinos and for the cosmic relic neutrinos 
(Cosmic Neutrino Background, C$\nu$B). These by-products of the neutrino mass measurements are largely out of the scope of the present work
and are  discussed here briefly only for the sake of completeness.

The finite neutrino mass manifesting in neutrino oscillations is already an important breach in the Standard Model of fundamental interactions, but the neutrino sector could hold more surprises. In fact, recent re-analysis of existing 
data from reactor oscillation experiments together with some anomalies observed in short baseline accelerator oscillation experiments (LSND, MiniBOONE) and in solar experiment calibration with neutrino sources (GALLEX), point to the existence of at least a fourth generation of neutrinos \cite{Men11}.
These hypothetical neutrinos would be {\it sterile} in the sense that they would feel only gravitational interactions, along with those induced by mixing with the other ordinary neutrinos.
Combined analysis of the available data from various sources leads to an additional mass splitting of  $\Delta m_{sterile}^2 \approx 1$\,eV, with
a mixing parameter of about $\sin^2 (2\theta) \approx 0.1$.
Sterile Right Handed neutrinos are indeed introduced naturally when one tries to extend the Standard Model to include the mass of active neutrinos ($\nu$MSM) \cite{abazajian2012light}. Moreover sterile neutrinos in the keV mass range are perfect candidate as Warm Dark Matter (WDM) particles \cite{destri2013fermionic}.

The calorimetric spectra of \Re\ or \Ho\ are suitable to investigate  the emission of heavy sterile neutrinos with a mixing angle $\theta$. Assuming the electron neutrino $\nu_e$ is a mixture of two mass eigenstates $\nu_H$ and $\nu_L$, with masses $m_H \gg m_L$, then $\nu_e =  \nu_L \cos \theta +   \nu_H \sin \theta$ and the measured energy spectrum is $N(E,m_L,m_H,\theta) = N(E,m_L) \cos^2 \theta + N(E,m_H) \sin^2 \theta$. The emission of heavy neutrinos would manifest as a kink in the spectrum at an energy of $Q - m_H$ for heavy neutrinos $\nu_H$ with masses between about 0 and $Q-E_{th}\lesssim2.5$\,keV, where $E_{th}$ is the experimental low energy threshold. It is worth noting that the strategy is of course the very analogous of the one adopted by J.J.\,Simpson \cite{simpson1981limits,simpson1985evidence} and which started off the already mentioned saga of the 17\,keV neutrino. Moreover, such search may be affected by systematic uncertainties due to the background and to the ripple observed in the \Re\ spectrum and caused by the BEFS (\S\,\ref{sec:stat-sys}).
An alternative and possibly more robust approach to the search of sterile neutrino emission in \Ho\ has been proposed in \cite{filianin2014kev}. 

Cosmology predicts that there are about 55 neutrinos/cm$^3$ in the universe as  left-overs of the Big Bang. Their average temperature today is about 1.95\,K and therefore their observation is extremely difficult. It has been proposed that the C$\nu$B could be detected via the induced beta decay on beta decaying isotopes: for example $\nu_e +^3$H$ \rightarrow^3$He$ + $e$^-$. 
This reaction could be detected as a peak at an energy of $Q+m_{\nu_e}$ in beta decay spectra. The expected rate can be calculated starting from the beta decay lifetime,
and for 100\,g of tritium it would be of about 10 counts/year \cite{lazauskas2008charged}. Unfortunately 100\,g is 10$^6$ times the amount of tritium contained in KATRIN, and the situation is not more favorable for other isotopes like \Re\ or \Ho.
The reactions $\nu_e +^{187}$Re$ \rightarrow^{187}$Os$ + $e$^-$ and $\nu_e + $e$^- +^{163}$Ho$ \rightarrow^{163}$Dy$^*$ are expected to give
yearly about $10^{-10}$ \cite{cocco2007probing,hodak2011towards} and $10^{-5}$ events  per gram of target isotope \cite{lusignoli2011relic,vignati2012163ho}, respectively. 
Recently, a dedicated experiment called PTOLEMY has been proposed \cite{betts2013development}: it combines a large area surface-deposition tritium target, the KATRIN magnetic/electrostatic filtering, LTDs, RF tracking and time-of-flight systems.  

The possibility to detect heavy sterile neutrino dark matter (WDM) via the above induced beta decays in \Re\ and \Ho\ has also been investigated but, again, the expected rates are hopelessly low \cite{li2011possible,li2011captures}.

\section{Past Experiments}
\subsection{Rhenium experiments with LTDs}
\Re\ was mentioned in \cite{derujula1981new} as interesting alternative to tritium because its transition energy of about 2.5\,keV is one of the lowest known. Thanks to this characteristic, the useful fraction of events close to the end-point is $\sim$350 times higher for \Re\ than for tritium. 
In addition, the half lifetime of about $4\times 10^{10}$\,years together with 
the large natural isotopic abundance (62.8~\%) of $^{187}$Re allows to get useful beta sources without any isotopic separation process. The beta decay rate in natural rhenium is of the order of $\sim$1\,Bq/mg, almost ideally suited to calorimetric detection with LTDs. 

As soon as the idea of developing LTDs for X-ray spectroscopy with energy resolutions caught, metallic rhenium became one of the 
most appealing materials also for making the X-ray absorbers. 
First of all metallic rhenium is a superconductor with a critical temperature $T_c$ of about 1.69\,K therefore, ideally, it is a good candidate for photon detection free of thermalization noise. 
Then the combination of high $Z$, high density ($\rho=21.02$\,g/cm$^3$), and high  Debye temperature ($\Theta_D\approx 417$\,K) makes metallic rhenium a unique material for designing X-ray detectors with low heat capacity $C$ -- i.e. high sensitivity -- and high photon stopping power.
Unfortunately, it became soon clear that metallic rhenium absorbers do not behave as expected and metallic rhenium was abandoned in favor of other more friendly materials as absorber for X-ray microcalorimeters.
The result of early efforts on rhenium absorbers for X-ray microcalorimeters are reported in \cite{stahle1992phd,stahle1992adapting}. 
The long time constants (up to 100\,ms) and a significant deficit in the signal amplitude are the distinguishing features of microcalorimeters
with metallic rhenium absorbers. In the same years the Genova group was finding similar results, as discussed below.

Although the explanation of the observed behavior most probably resides in the superconductivity of rhenium, also the heat capacity
may contribute to poor and inconsistent performance.
In fact, according to \cite{gregers1971sign} the specific heat of rhenium in the normal state is given by
\begin{equation}
c(T) = 40.6\,T^{-2} + 0.034\,T^{-3} +  2290\,T + 27 \,T^3 \hbox{$\mu$J/(mol K)}
\end{equation}
where the last two terms are the contributions from normal conduction electrons and phonons, respectively.
The first two terms are due to the nuclear heat capacity, which arises from the interaction between
the  large nuclear quadrupole moment of the two natural isotopes of rhenium  -- both have nuclear spin 5/2 -- and
the electrical field gradient at the nucleus in the non-cubic rhenium lattice.
When rhenium is in the superconducting state the nuclear heat capacity term should vanish
since the slow spin-lattice relaxation thermally isolates the nuclear spin system.
In the superconducting state a non-reproducible small fraction of the
normal-state nuclear heat capacity may still be observed if trapped magnetic flux causes regions in the specimen to remain normal.

In spite of these difficulties, research on LTDs with metallic rhenium went on for the purpose of making detectors for calorimetric neutrino mass experiments, although other dielectric materials were also tested (see \S\,\ref{sec:mibeta}).

\subsection{The \Re\ beta decay spectrum}
\label{sec:respe}
\begin{figure}[ht]
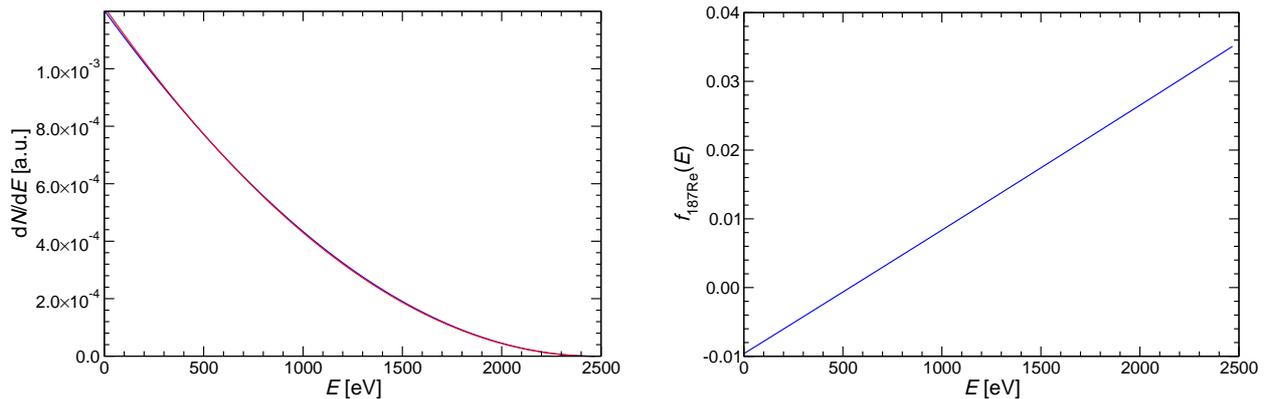

 \centering
 \includegraphics[clip=true,width=0.49\textwidth ]{spe-re187}\hfill
 \includegraphics[clip=true,width=0.47\textwidth ]{deviation-from-quad}
 \caption{Theoretical shape of the \Re\ beta decay spectrum (left). Deviation of the \Re\ spectrum from a simple quadratic form (right).}
 \label{fig:re187}
 \label{fig:corrbeta}
\end{figure}

The \Re\ beta decay
\begin{equation}
^{187}\mathrm{Re} (5/2^+) \rightarrow ^{187}\mathrm{Os} (1/2^-) + e^- + \overline\nu_e \\
\end{equation}
is a unique first forbidden transition. Unlike non-unique transitions, the nuclear matrix element is computable, even if the calculation is not straightforward as in the case of tritium. In the literature, it is possible to find detailed calculations of both the matrix element and  the Fermi function for this process \cite{dvornicky2011absolute,de2013role}. The electron and the neutrino are emitted in the $p_{3/2}$ and the $s_{1/2}$ states, respectively, or vice versa. Higher partial waves are strongly suppressed because of the low transition energy.
The distribution of the kinetic energies $E$ of the emitted electrons, calculated neglecting the neutrino mixing is (according to \cite{dvornicky2011absolute})
\begin{eqnarray}
\label{eq:reniospe}
N(E,m_{\nu_e})&=& N^{p_{3/2}}(E,m_{\nu_e})+N^{s_{1/2}}(E,m_{\nu_e})\nonumber\\
&=& C p E (E_0 -E)^2 [ F_1(Z,E)p^2+F_0(Z,E)p_\nu^2] \sqrt{1 - \frac{m_{\nu_e}^2}{(E_0-E)^2}} 
\end{eqnarray}
where $p$ is the electron momentum, $p_\nu^2=((E_0-E)^2-m_{\nu_e}^2)$ is the neutrino momentum, and $F_0(Z,E)$ and $F_1(Z,E)$ are the relativistic Coulomb factors, which take into account the distortion of the electron wave function due to the electromagnetic
interaction of the emitted electron in $s_{1/2}$ and $p_{3/2}$ states with the atomic nucleus.
In general, the Coulomb factor takes the form
\begin{eqnarray}
F_{k-1}(Z,E)  =  \left( \frac{\Gamma (2k+1)}{\Gamma(k) \Gamma(1+2\gamma_k)}  \right)^2 (2pR )^{2(\gamma_k -k)}  |\Gamma (\gamma_k +iz)|^2 e^{\pi z} 
\end{eqnarray}
with
\begin{eqnarray}
\gamma_k &=& \sqrt{k^2-(\alpha Z)^2} \nonumber \\
z &=&  \alpha Z \frac{E}{p}.
\end{eqnarray}
where $\Gamma$ is the gamma function, $\alpha$ is the fine structure constant, and $R$ is the nuclear radius.
It can be found numerically that the $p_{3/2}$ component of the spectrum is dominant, i.e.  \cite{dvornicky2011absolute}
\begin{equation}
\frac{I_{s_{1/2}}}{I_{p_{3/2}}}=\frac{\int_0^{E_0}N^{s_{1/2}}(E,m_{\nu_e})d \! E}{\int_0^{E_0}N^{p_{3/2}}(E,m_{\nu_e})d \!E} \approx 10^{-4}
\end{equation}  
This has been confirmed experimentally in \cite{arnaboldi2006measurement} (see \S\,\ref{sec:mibeta}).
It can also be shown that  (\ref{eq:reniospe}) can be approximated by the expression
\begin{equation}
N(E,m_{\nu_e})= C^\prime (1 + f_{^{187}Re}(E))(E_0-E)^2 \sqrt{1- \frac{m_{\nu_e}}{(E_0-E)^2}}
\end{equation}  
where the correction factor $f_{^{187}Re}(E)$ is shown in Fig.\,\ref{fig:corrbeta}.

\subsection{Statistical sensitivity and systematics}
\label{sec:stat-sys}
An accurate assessment of calorimetric neutrino mass experimental sensitivity requires the use of Monte Carlo frequentist approach \cite{nucciotti2010expectations}. 
The parameters describing the experimental configuration are the total number of \Re\ decays $N_{ev}$,
the FWHM of the Gaussian energy resolution \de \fwhm, the fraction of unresolved pile-up events $f_{pp}$, and the radioactive background $B(E)$.
The total number of events is given by \mbox{$N_{ev} = N_{det}A_\beta t_M$}, where $ N_{det}$, $A_\beta$ and  $t_M$ are the total number of detectors, the beta decay rate in each detector, and the measuring time, respectively. As discussed in \S\,\ref{sec:calosens}, $f_{pp}=\tau_R A_\beta$, where $\tau_R$ is the time resolution of the detectors. 
The $B(E)$ function is usually taken as a constant, $B(E)=bT$, where $b$ is the average background count rate for unit energy
and for a single detector, and $T=N_{det}\times t_M$ is the experimental exposure.
A set of experimental spectra are simulated and fitted with $m^2_\nu$, $E_0$, $N_{ev}$, $f_{pp}$ and $b$ as
free parameters. The 90\% C.L. $m_{\nu_e}$ statistical sensitivity $\Sigma_{90}(m_{\nu_e})$ of the simulated experimental 
configuration can be obtained from the distribution of the \mnesq\ found by fitting the spectra.
The statistical sensitivity is then given by $\Sigma_{90}(m_{\nu_e}) = \sqrt{1.7 \sigma_{m_{\nu_e}^2}}$, where $\sigma_{m_{\nu_e}^2}$ is the standard deviation of the \mnesq\ distribution.

\begin{table*}[ht!]
\caption{\label{tab:sens} Experimental exposure required for the target statistical sensitivity in the second column with $b=0$.}
\begin{center}
\begin{tabular}{ccccccc}
\hline 
isotope & sensitivity & $A_\beta$ &	$\tau$ &	\de\ &	$N_{ev}$ &	exposure $T$ \\[0pt] %
& [eV] & [Hz] & 	[$\mu$s] &	[eV] &	[counts] &	[detector$\times$year] \\
\hline 
\Re\ & 0.2 & 10&	3&	3&	$1.3\times10^{14}$&	$4.1\times10^{5}$\\
\Re\ & 0.1 & 10&	1&	1&	$10.3\times10^{14}$&	$3.3\times10^{6}$\\
\hline 
\end{tabular} 
\end{center}
\end{table*}
The symbols in Fig.\,\ref{fig:sensre} are the results of Monte Carlo simulations for various experimental parameters and for $b=0$ compared to the analytic estimate.
 
As an example, Tab.\,\ref{tab:sens} reports two experimental configurations which could allow to achieve statistical sensitivities of about 0.2 and 0.1\,eV, respectively. The two sensitivities could be attained measuring for 10 years, respectively,  $4\times 10^4$ and $3.2\times 10^5$ detectors, while the total mass of natural metallic rhenium in the two cases would be about 400\,g and 3.2\,kg, respectively.

A flat background remains almost negligible as long as it is much lower than the pile-up contribution at the end-point, i.e. 
$b \ll \approx A_\beta f_{pp} / (2 E_0)$. For the two experiments in Tab.\,\ref{tab:sens} this translates in a constant background,
lower than about $1\times10^{-2}$ and $4\times10^{-3}$\,counts/day/eV, respectively, which should be achievable without operating the arrays in the extreme low background conditions of an underground laboratory.

Given the strong dependence of the sensitivity on the total statistics, for a fixed experimental exposure $T$ -- that is, for a fixed measuring time and a fixed experiment size -- and for fixed detector performance -- \de\fwhm\ and $\tau_R$ --, it always pays out to increase the single detector activity $A_\beta$ as high as technically feasible, even at the expenses of an increasing pile-up level. 
Of course, since the rhenium specific activity is practically fixed,  the ultimate limit to  $A_\beta$ is set by the
tolerable heat capacity of the absorber.

With the same Monte Carlo approach it is also possible to investigate the source of
systematic uncertainties peculiar to the calorimetric technique.
As shown in \cite{nucciotti2010expectations}, it appears that the most crucial and worrisome
sources of systematics are the uncertainties related to the Beta Environmental Fine Structure (BEFS), the theoretical spectral shape of the \Re\ beta decay, the energy response function, and the radioactive background. These sources are briefly discussed in the following. 

\begin{figure}[t]
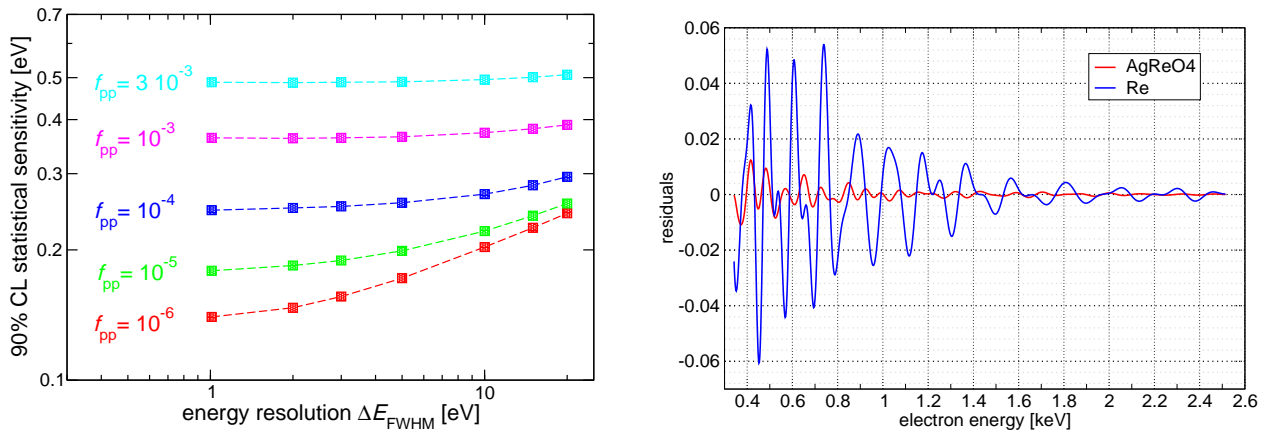

 \centering
 \includegraphics[clip=true,width=0.47\textwidth ]{f_mc-new}
\hfill
 \includegraphics[clip=true,width=0.48\textwidth ]{perrenato-prl}
 \caption{Monte Carlo computed statistical sensitivity of a \Re\ decay calorimetric measurement for different experimental configurations and for $N_{ev}=10^{14}$ (left). BEFS ripple prediction for metallic rhenium and for \agre\ for an infinite instrumental resolution (right). }
 \label{fig:mcsens}
 \label{fig:befs-re-agre}
\end{figure}

The BEFS is a modulation of the beta emission probability due to the atomic and molecular surrounding of the decaying nuclei \cite{koonin1991environmental}, which is 
explained by the electron wave structure in terms of reflection and interference.
The BEFS oscillations depend on the inter-atomic distance, while their amplitude is tied to the electron-atom scattering cross-section: although the phenomenon is completely understood, its description is quite complex and the parameters involved are not all known \textit{a priori}. 
So far it could be detected only in the low energy region -- $E\lesssim1.5$\,keV -- of the \Re\ spectra, where both the beta rate and the BEFS are larger, but  
as far as the effect on the neutrino mass determination is concerned,  its effect extends way up to the end-point (Fig.\,\ref{fig:befs-re-agre}). 
For a safe extrapolation up to the end-point and to minimize the systematic uncertainties, the 
BEFS must be characterized using much higher statistics beta spectra and independent EXAFS analysis of the material containing rhenium.

The theoretical description of the \Re\ decay spectrum given in \S\,\ref{sec:respe} is slightly contradicted by experimental observation, since
available high statistics spectra are in fact better interpolated as $N(E)=(E_0-E)^2$, i.e. with $f_{^{187}Re}(E)\approx 1$. This deviation from theory 
has not found a plausible explanation yet\footnote{Fedor \v{S}imkovic, private communication}, and it will become troublesome when larger
statistics experiments will call for more accurate description of the spectrum.

The detector response function is probed by means of X-ray sources which are not exactly monochromatic and which do not replicate
the same type of interactions in the absorber as the beta decay. In fact, the X-ray interactions happen at a shallow depth, whereas the beta decays are uniformly distributed in the volume; moreover, in the case of X-rays the energy is deposited by a primary photo-electron followed by a cascade of secondary X-rays and Auger electrons, whereas in beta decay all the energy is deposited along one single track.
It is therefore extremely important -- yet challenging -- to fully understand the measured response function in order to disentangle the contributions to its shape caused by the external X-rays.

In calorimetric experiments, since the beta source cannot be switched off, the environmental and cosmic background in the energy
range of the beta spectrum cannot be directly assessed. Therefore a constant background is usually included in the fit model as the safest hypothesis. This hypothesis may happen to be not accurate enough for future high statistics measurements.

\subsection{MANU}
\label{sec:manu}
The research program in Genova which lead to the MANU experiment started in 1985 \cite{vitale1985renio} with a focus on the
use of metallic rhenium absorbers. At that time there was absolutely no knowledge about the behavior of this material as
absorber for LTDs. Therefore, the first years were devoted to study the heat capacity and the thermalization efficiency $\eta$ of
metallic rhenium.  

The outcomes of the preliminary phase are summarized in \cite{cosulich1993further}. The thermalization efficiency $\eta$ was
studied for many superconductors in form of small
single crystals (cubic millimeters) of Al, Pb, In, Ti, Nb, Va, Zn and Re, and a quasi universal dependence on the
ratio $T/\Theta_D$ was found, with $\eta$ dropping sharply for $T$ lower than about  $2\times 10^{-4}\Theta_D$. 
In particular rhenium thermalization  was investigated for single- and poly-crystals
between 50\,mK and 200\,mK. The rise-time was limited to about 200\,$\mu$s, preventing to assess the thermalization efficiency at shorter
time scales. Also, for rhenium it was found that full thermalization is attained for an operating temperature above about 83\,mK.
The effect of magnetic fields was also investigated, and an unexpected and unexplained reduction of $\eta$ for magnetic fields increasing
up to 20\,Gauss was found.

\begin{figure}[ht]
 \centering
 \includegraphics[clip=true,width=0.49\textwidth ]{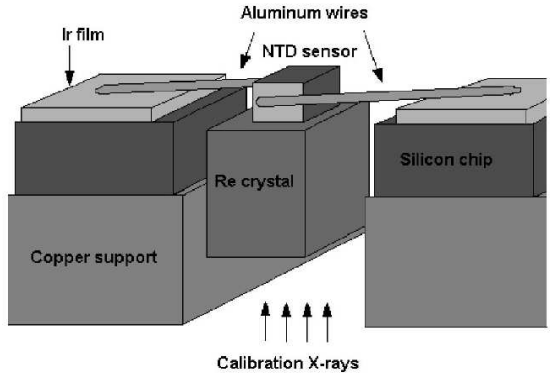}
 \caption{The MANU experiment detector (from \cite{galeazzi2000end}).}
 \label{fig:manu1}
\end{figure}

\begin{figure}[ht]
 \centering
 \includegraphics[clip=true,width=0.50\textwidth ]{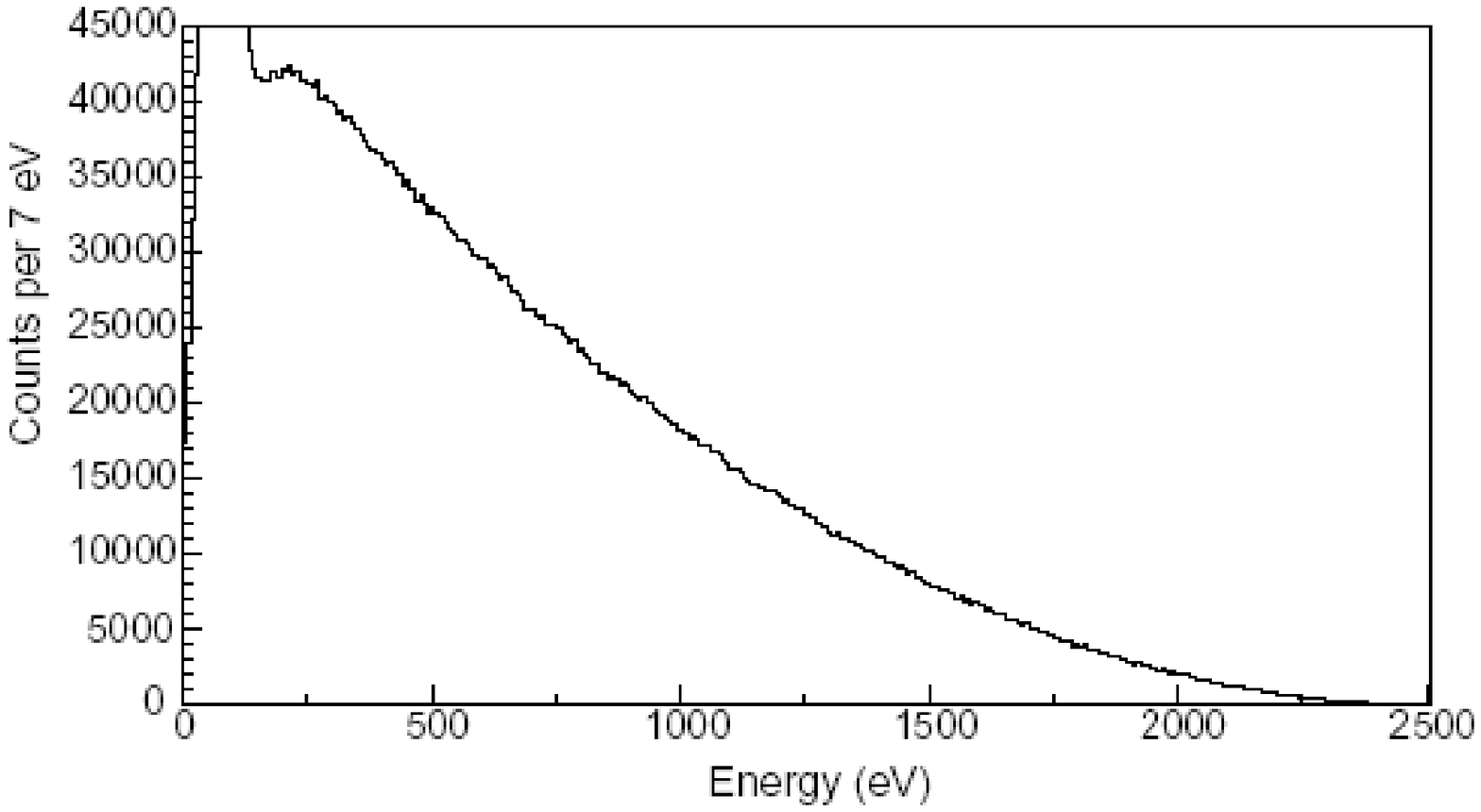}\hfill
 \includegraphics[clip=true,width=0.48\textwidth ]{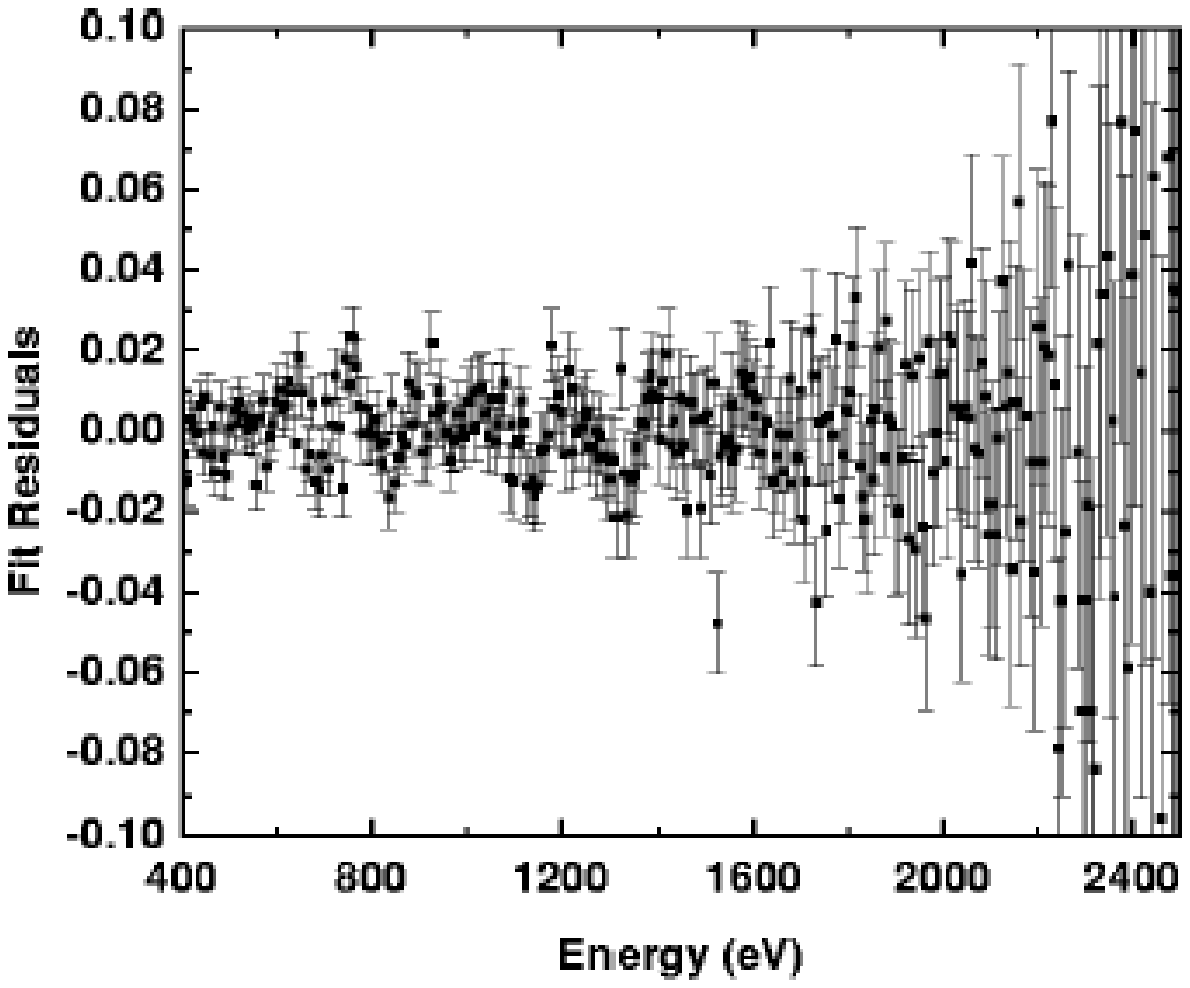}
 \caption{The MANU experiment final spectrum (left) and its fitting residuals  showing the BEFS ripples (right) (from \cite{galeazzi2000end}). }
 \label{fig:manu2}
\end{figure}

\begin{figure}[ht]
 \centering
 \includegraphics[clip=true,width=0.48\textwidth ]{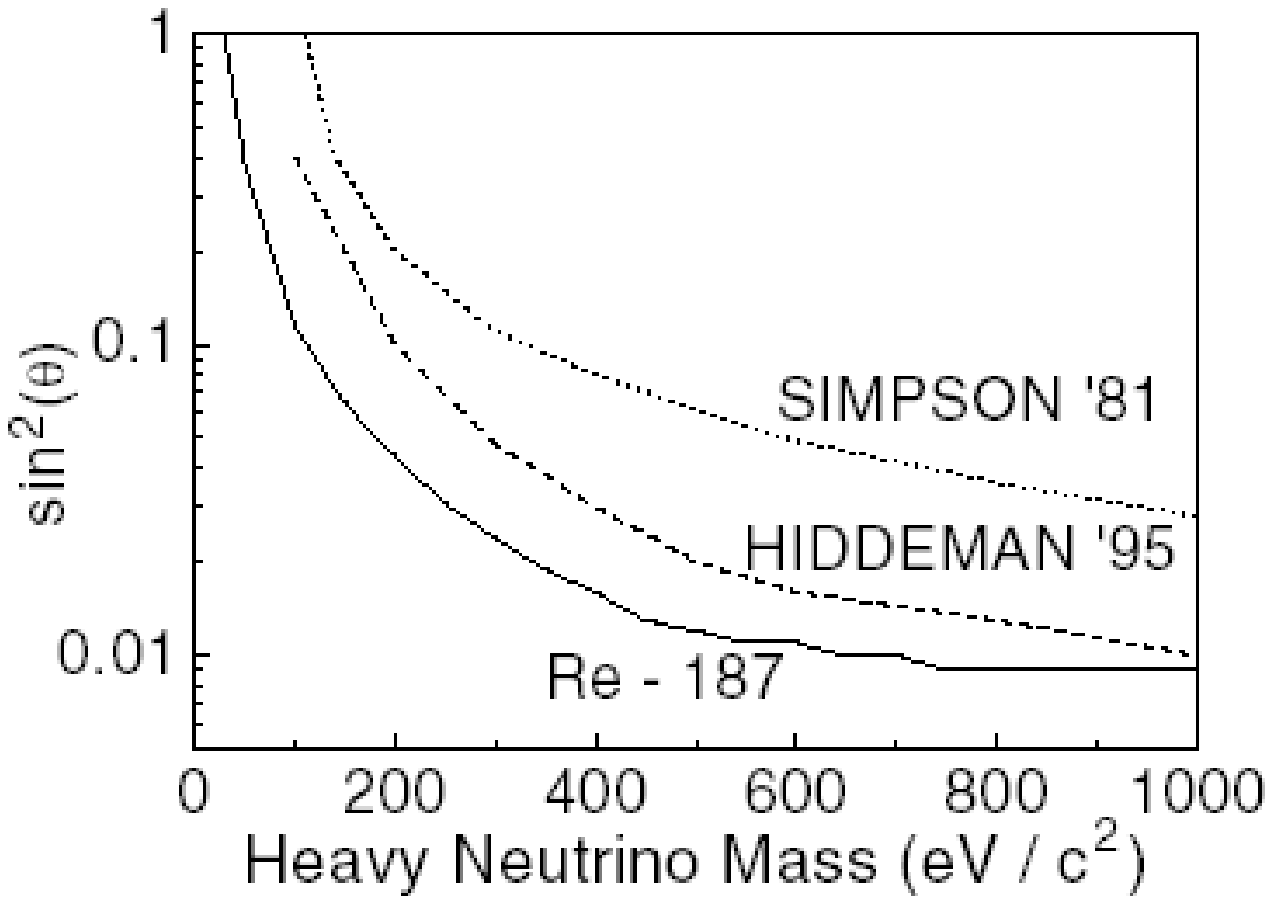}
 \hfill
 \includegraphics[clip=true,width=0.51\textwidth ]{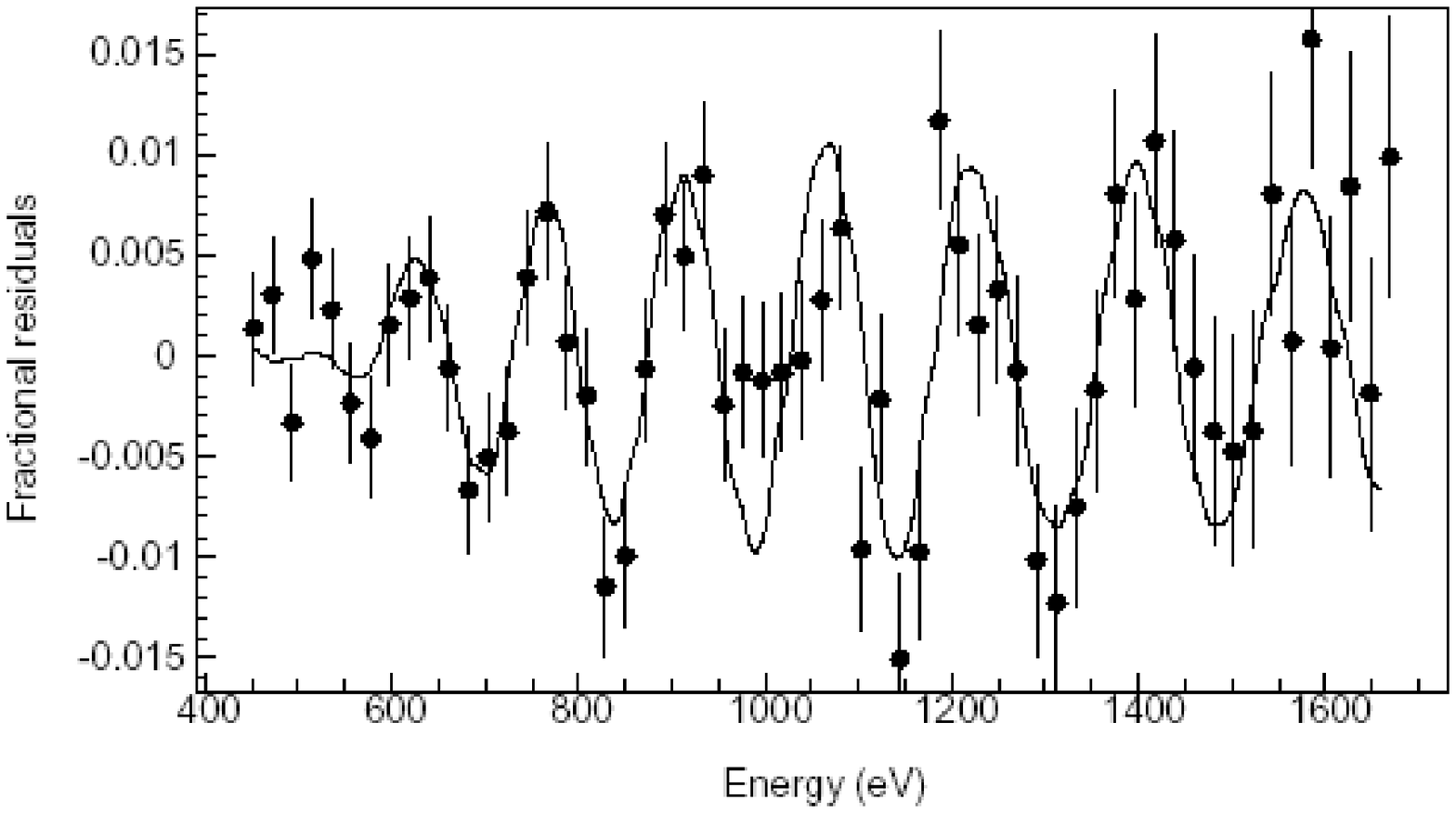}
 \caption{Heavy sterile mass exclusion plot of MANU \cite{galeazzi2001limits} (left). BEFS in the MANU spectrum \cite{gatti1999detection} (right).}
 \label{fig:manu3}
\end{figure}

The first observation of \Re\ spectrum was reported in \cite{cosulich1992detection}.
After this, a period was spent to optimize the microcalorimeter performance, also exploiting the gained understanding of
metallic rhenium absorbers: energy resolutions as good as about 30\,eV were demonstrated with small -- about $50$\,$\mu$g -- rhenium absorbers.
In 2001 the results of the high statistics measurement of MANU were published \cite{galeazzi2000end}.
The MANU experiment's microcalorimeter was a NTD germanium thermistor coupled with epoxy resin to a
1.572\,mg rhenium single crystal. Two ultrasonically bonded aluminum wires provided both the path for the electrical signal and the thermal contact to the heat sink at 60\,mK (Fig.\,\ref{fig:manu1}).
The detector had a thin shield against environmental radiation made out of ancient Roman lead.
A weak $^{55}$Fe source ($10^{-3}$\,counts/s) allowed to monitor the gain stability during the measurement, while the energy calibration
was established through a removable fluorescence source emitting the K lines of Cl, Ca, and K.
Signals were read-out by a cold stage with unitary gain using a JFET at about 150\,K,  digitized at 12\,bit in 1024 long records, and processed with an optimal filter. Further processing was used to detect pile-up events \cite{fontanelli1999data}.
The high statistics measurement lasted for about 3\,months and the detector performance are listed in Tab.\,\ref{tab:renioexp} \cite{galeazzi2000end}.
The $^{55}$Fe K$_\alpha$ line had a perfectly Gaussian shape with tails lower than 0.1\%. The calibration with the fluorescence source showed
that the energy resolution is practically constant with energy, and the deviation from linearity of the energy response was of about 0.16\% at
the \Re\ spectrum end-point.

The fit of the spectrum  (Fig.\,\ref{fig:manu2}) gave a squared
neutrino mass of $m_{\nu_e}^2=-462^{+579}_{-679}$\,eV$^2$, which translated in an upper limit $m_{\nu_e} \le26$\,eV at 95\% C.L., or 19\,eV at 90\% C.L. \cite{gatti2001microcalorimeter}.

The results reported in \cite{galeazzi2000end} were the most precise measurements of the transition energy $E_0$  and of the \Re\ half-life
at the time of publishing; the \Re\ half-life in particular is of great interest for geochronology for determining the age of minerals and meteorites.

This high statistics measurement allowed also to observe for the first time the BEFS \cite{gatti1999detection} and to set a limit for the emission of sterile neutrinos with masses below 1\,keV \cite{galeazzi2001limits} (Fig.\,\ref{fig:manu3}).

\subsection{MIBETA}
\label{sec:mibeta}
The Milano program for a neutrino mass measurement with \Re\ started in 1992 with an 
R\&D to fabricate silicon implanted thermistors in collaboration with FBK \cite{alessandrello1999fabrication}. The final objective was to make large arrays 
of high resolution microcalorimeters using micro-machining \cite{faes2004fabrication}. 
NTD germanium based microcalorimeters were also tested.
In the light of the encouraging results obtained at Genova, at first the program concentrated on metallic rhenium absorbers.
Many single- and poly-crystalline samples were tested with disappointing results: small signals, long time constants, and inconsistently varying pulse shapes. A possible correlation with the sample purity and with residual magnetic fields was individuated, but this was not enough to
The research program therefore moved onto the systematic testing 
of dielectric rhenium compounds  as microcalorimeter absorbers.
From the beginning, the most suitable compounds looked like those based on the ReO$_{4}^{-}$ (perrhenate) anion. A non exhaustive list of tested compounds includes: Re$_{2}$(CO)$_{10}$, K$_{2}$ReCl$_{6}$, (NH$_{4}$)ReO$_{4}$, KReO$_{4}$, AgReO$_{4}$. The tests with Re$_{2}$(CO)$_{10}$ failed since this compound sublimates in vacuum at room temperature. The second-to-fourth materials, despite the good theoretical expectations and the large signal-to-noise ratio, showed a quite poor energy resolution -- exceeding 100\,eV at 6\,keV -- 
which could be explained as due to a large thermalization noise. 

Silver perrhenate (AgReO$_{4}$), on the other hand, immediately exhibited good properties with limited thermalization noise. 
The calibration peaks were sufficiently symmetric, and energies resolutions as good as 18\,eV FWHM at 6\,keV were achieved. AgReO$_{4}$ crystals are transparent, crumbly, and slightly hygroscopic, with a specific \Re\ activity of about $5.4 \times 10^{-4}$Hz/\mug\ \cite{alessandrello1999bolometric}. 

\begin{figure}[htb]
 \centering
 \includegraphics[clip=true,width=0.5\textwidth ]{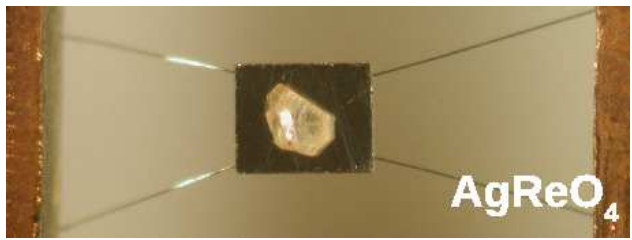}
 \caption{One of the MIBETA microcalorimeters with \agre.}
 \label{fig:agreo4}
\end{figure}

The MIBETA experiment ran between 2002 and 2003 an array of 10 \agre\
microcalorimeters for a high statistics measurement, which was preceded by a campaign of measurements dedicated to tuning the set-up and to reducing the background \cite{arnaboldi2003bolometric}.

\begin{figure}[htb]
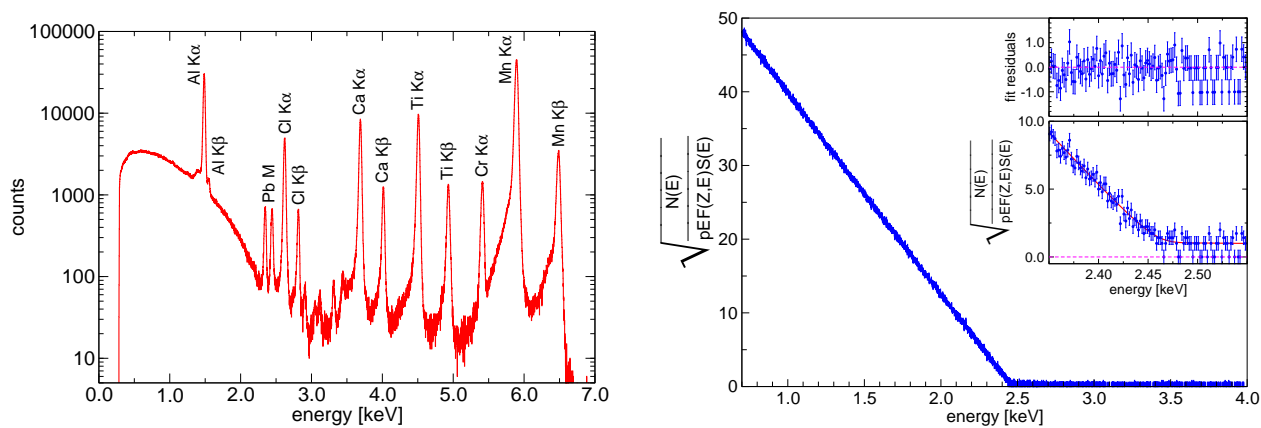

 \centering
 \includegraphics[clip=true,width=0.48\textwidth ]{cal-tot-B1B10}\hfill
 \includegraphics[clip=true,width=0.48\textwidth ]{kurie-tot-fit}
 \caption{Final MIBETA calibration spectrum (left). Kurie plot of the final MIBETA data (right).}
 \label{fig:mibeta}
\end{figure}

The array was made of \agre\ crystals with masses ranging from 250 to 300\,$\mu$g to
limit event pile-up, for a total mass of 2.683\,mg. The crystals were attached to silicon implanted thermistors with epoxy resin, and four ultrasonically bonded aluminum wires were used both as signal leads and  heat links to the heat bath, stabilized at 25\,mK.
The 10 microcalorimeters were enclosed in two copper holders without lead shieldings, to avoid the background caused by lead fluorescence at 88\,keV, which in turn provokes escape peaks in \agre\ very close to the beta end-point.
The stability and performance of all detectors were monitored with 
a movable multi-line fluorescence source at 2\,K, which was activated for 25\,min every 2\,h to emit the
K lines of Al, Cl, Ca, Ti, and Mn. When not used for the calibration, the primary \Fe\ source was pulled
inside a massive shield of ancient Roman lead \cite{alessandrello1998measurements}, 
in order to minimize the contribution to the radioactive background caused by the IB of \Fe. 
The data acquisition program controlled the movements of the source and tagged the events collected
during the calibrations.
The first stage of the electronic chain used 10 JFETs cooled to about 120\,K and placed few centimeters 
below the detectors. A 16-bit data acquisition system digitalized and saved to disk the signals for an Optimal Filter based off-line 
analysis.

The high statistics measurement of MIBETA lasted for about 7 months. In the final analysis, the data from 
two detectors, with poorer energy resolution, were not included. The total active mass was therefore 2.174 mg,
for a \Re\ activity of 1.17\,Bq. The final beta spectrum obtained from the sum of the 8 working 
detectors corresponds to about 8745 hours$\times$mg \cite{sisti2004new}.
The performance of the detectors were quite stable during the run and are reported in Tab.\,\ref{tab:renioexp}.  

All X-ray peaks in the calibration spectrum showed tails on the low energy side, and the thermalization noise of \agre\ caused
their width to increase with the energy (Fig.\,\ref{fig:mibeta}). 
The fit of spectrum (Fig.\,\ref{fig:mibeta}) gave a squared 
neutrino mass of $m_{\nu_e}^2=-112\pm 207_{stat} \pm 90_{sys}$, which translates in an upper limit $m_{\nu_e} \le15$\,eV at 90\% C.L.
The systematic error was dominated by the uncertainties on the energy resolution function, on the background, and on the theoretical
shape of the spectrum.

\begin{figure}[htb]
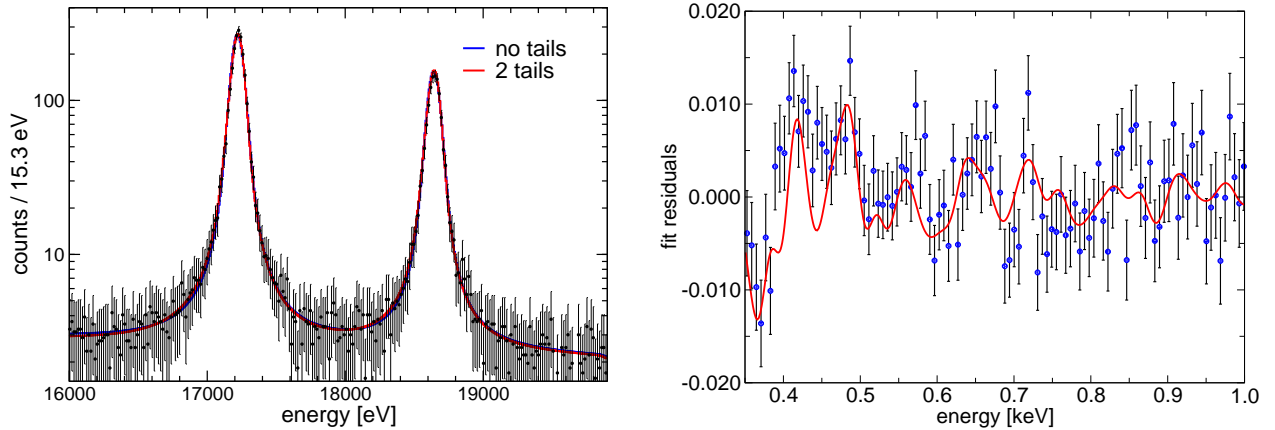

 \centering
 \includegraphics[clip=true,width=0.48\textwidth ]{fit-esc12_2}
 \hfill
 \includegraphics[clip=true,width=0.48\textwidth ]{befs-350}
 \caption{Escape peaks due to the exposure of \agre\ microcalorimeters to a $^{44}$Ti source  \cite{ferri2012investigation} (left) BEFS in the MIBETA final spectrum \cite{arnaboldi2006measurement} (right).}
 \label{fig:ti44esc}
 \label{fig:befsmibeta}
\end{figure}

Additional lower statistics measurements with the same set-up were carried out to study and reduce the background, and to investigate the energy response function. In particular, using the escape peaks caused at about 17\,keV by the irradiation with a $^{44}$Ti gamma source  (see Fig.\,\ref{fig:ti44esc}) as comparison, it was possible to partly understand the complex shape of the X-ray calibration peaks and to establish that at least the longest of the observed tails were due to surface effects \cite{ferri2012investigation}.

Although BEFS (see \S \ref{sec:stat-sys} and Fig.\,\ref{fig:befs-re-agre}) is almost one order of magnitude fainter in AgReO$_{4}$ 
than in metallic rhenium, it was observed also in the high statistics spectra of MIBETA \cite{arnaboldi2006measurement} (Fig.\,\ref{fig:befsmibeta}). In particular, the BEFS ripple interpolation allowed to determine that $p_{3/2}$ to $s_{1/2}$ branching ratio in the \Re\ beta emission is  $0.84\pm0.30$, which is compatible with the expected prevalent $p_{3/2}$ emission (see \S\,\ref{sec:respe}).

\begin{table}
\caption{Comparison of the MANU and MIBETA experiments.}
 \begin{tabular}{rcc}\hline
\label{tab:renioexp}
~
 &
MANU &
MIBETA\\\hline
source / absorber &
metallic Re &
AgReO4\\
sensor &
NTD &
Si implanted thermistor\\
number of detectors &
1 &
8\\
total mass [$\mu$g] &
1572 (Re)
 &
2174 (\agre)
\\
measuring time $t_M$ [h] &
$\approx 2800$
 &
$\approx 8700$
\\
total activity $A$ [Bq] &
1.1 (above 350\,eV)
 &
1.17 (above 700\,eV)
\\
energy resolution \de \fwhm [eV] &
96 (at 5.9\,keV)
 &
28.5 (average at \Re\ end-point)
\\
rise time [$\mu$s] &
 $\approx 1000$
 &
 492 (average 10-90\%)
\\\hline
statistics $N_{ev}$  &
$6\times 10^{6}$ (above 420\,eV)
 &
$6.2\times 10^{6}$ (above 700\,eV)
\\pile-up fraction $f_{pp}$ &
-- 
 &
$2.3\times 10^{-4}$
\\
$E_0$ [eV] &
 $2470\pm 1_{stat} \pm 4_{sys}$
 &
 $2465.3\pm 0.5_{stat} \pm 1.6_{sys}$

\\
$\tau_{1/2}$ [$10^{10}$\,y] &
$4.12\pm 0.02_{stat} \pm 0.11_{sys}$
 &
$4.32\pm 0.02_{stat} \pm 0.01_{sys}$
\\
$m_{\nu_e}^2$ [eV$^2$] &
$-462^{+579}_{-679}$
 &
$-112\pm 207_{stat} \pm 90_{sys}$
\\
\mne\ 90\% [eV] &
19
 &
15
\\
background $b$ [c/eV/day] &
$3\times 10^{-4}$
 &
$1.7\times 10^{-4}$
\\
\hline
\end{tabular}
\end{table}

\subsection{MARE}
\label{sec:mare}
The MANU and MIBETA results, together with the constant advance in the LTD technology, 
made it reasonable to propose a larger scale project: the Microcalorimeter  Arrays for a Neutrino Mass Experiment (MARE).
The ambition of MARE was to establish a sub-eV neutrino mass sensitivity through a gradual deployment approach.
The project was started in 2005 by a large international collaboration \cite{mareproposal, Nucciotti2012155} and it was organized
in two phases.

\begin{figure}[ht]
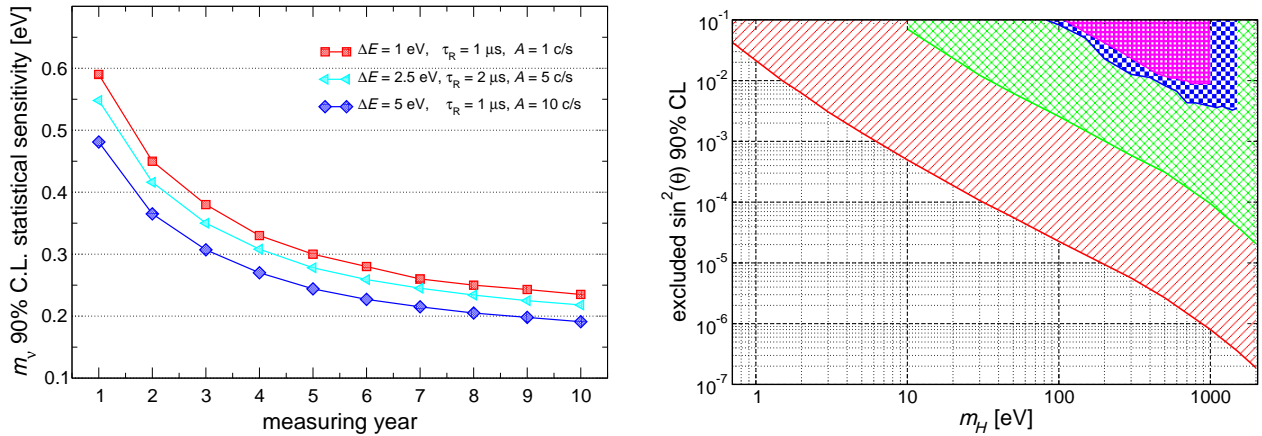

 \centering
  \includegraphics[clip=true,width=0.48\textwidth ]{fase2}
  \hfill
 \includegraphics[clip=true,width=0.48\textwidth ]{sens-heavynu_mare1-2}
 \caption{Statistical sensitivity of MARE final experiment. Curves are calculated for the deployment of 10000 pixels per year for the first 5 years (left). Statistical sensitivity to the emission of heavy neutrinos with mass $m_H$. From lower to upper curve: Monte Carlo simulation with $N_{ev}=10^{14}$, $f_{pp}=10^{-6}$ and $\Delta E = 1.5$\,eV; Monte Carlo simulation with $N_{ev}=8\times10^{9}$, $f_{pp}=10^{-5}$ and $\Delta E = 15$\,eV; MIBETA experiment (unpublished); MANU experiment \cite{galeazzi2001limits} (right).}
  \label{fig:fase2}
  \label{fig:maresterile}
\end{figure}

The final objective of a sub-eV statistical sensitivity on the electron neutrino mass was the goal of the second phase. To accomplish this, the program was to gradually deploy several large arrays -- about $10^4$ elements each -- of detectors, with energy and time resolutions of the order of 1\,eV and 1\,\mus, respectively. Each pixel was planned to have a
source activity of about few counts per second in order  to collect a total
statistics of about $10^{14}$ beta decays in  up to ten years of measurement time  (see Fig.\,\ref{fig:fase2}) \cite{mareproposal}. Fig.\,\ref{fig:maresterile} shows also the MARE sensitivity to the emission of heavy sterile neutrinos with masses below 2\,keV (\S\,\ref{sec:other}). 

Phase 1 -- also called MARE-1 -- had the task to ascertain the most suitable technical approach for the final experimental phase, also with the help of smaller scale experiments.
An R\&D program was started with the aim to improve the understanding of the superconducting rhenium absorbers and of their
optimal coupling to sensors, and to develop the appropriate array technology and multiplexed read-out scheme \cite{mareproposal}.
At the same time, two intermediate size experiments carried out with the available technologies aimed to reach a 
neutrino mass sensitivity of the order of 1\,eV, and to improve the understanding of all the systematics peculiar of the calorimetric approach with \Re.
Furthermore, MARE-1 started to explore the alternative use of \Ho\ for a calorimetric measurement of the neutrino mass.
Given the unavoidable competition with the KATRIN experiment, the time schedule for MARE was quite tight. 

The physics of metallic rhenium as absorber for the MARE detectors was the focus of the Genova and the Heidelberg groups.
The best technologies available for the MARE-2 arrays were: 1) the Transition Edge
Sensors (TES) with Frequency Division Multiplexing, investigated by the Genova group and Physikalisch-Technische Bundesanstalt (PTB, Berlin, Germany); 2) the Metallic Magnetic Calorimeters (MMC) with Microwave Squid Multiplexing, developed by the Heidelberg group; and 3) Microwave Kinetic Inductance Detectors (MKID) with Microwave Multiplexing, explored by the Milano group. 
The Genova group, in collaboration with Miami and Lisbon, planned an experiment consisting of an array of 300 TES detectors, with a total mass of about 1\,mg of rhenium single crystals \cite{Pergolesi2006349}.
With energy and time resolutions of about 10\,eV and 10\,\mus, respectively, the sensitivity attainable in 3
years of measuring time was estimated to be around 1.8\,eV at 90\% C.L., for a statistics of about $3\times 10^{10}$ decays.
The Milano group, together with the NASA/GSFC and Wisconsin groups, deployed an array of silicon
implanted thermistors coupled to \agre\ absorbers.
The experiment used 8 of the 36 pixel arrays that NASA/GSFC had developed for the XRS2 instrument \cite{nucciotti2006comparison}.
With 288 pixels attached to about 500\,\mug\ \agre\ crystals, and with energy and time resolutions of about 25\,eV and 250\,\mus,
respectively, a sensitivity around 3.3\,eV at 90\% C.L. was expected in 3
years of measuring time, with a statistics of about $7 \times 10^{9}$ decays.

Unfortunately, the MARE-1 outcomes were quite disappointing, and MARE-2 ended up canceled before taking off. 
The lack of success of the MARE initiative was mostly the consequence of the final acknowledgment of the impossibility
to fabricate rhenium microcalorimeters matching the specifications set by the aimed sub-eV sensitivity.
The systematic investigations carried out at Heidelberg with rhenium absorbers coupled to MMC, despite some noteworthy
progress, arrived to conclusions similar to that of past works: rhenium absorbers behave inconsistently, showing
a large deficit in the energy thermalization accompanied by long time constants \cite{ranitzsch2012development}.
Therefore the challenging idea of improving and scaling up the pioneer experiments
using metallic rhenium absorbers turned out to be a dead end road.

Indeed, also the other experimental efforts of MARE-1 encountered several difficulties \cite{ferri2014preliminary}.
For example, the setting up of arrays of \agre\ crystals turned out to be more troublesome than expected. 
The freshly polished surfaces of \agre\ crystals shaped to cuboids resulted to be incompatible with the sensor coupling methods used  successfully in MIBETA with as-grown small crystals. Despite the use of a micro-machined array of silicon implanted thermistors, the performance of the pixels were irreproducible and, while gradually populating and testing the XRS2 array with \agre\ crystals, the performance of the instrumented pixels started to degrade.
This made the array finally unusable. Given the sorts of the MARE project, also this branch of the MARE-1 program was thus dropped in 2013.
\begin{figure}[ht]
 \centering
 \includegraphics[clip=true,width=0.9\textwidth ]{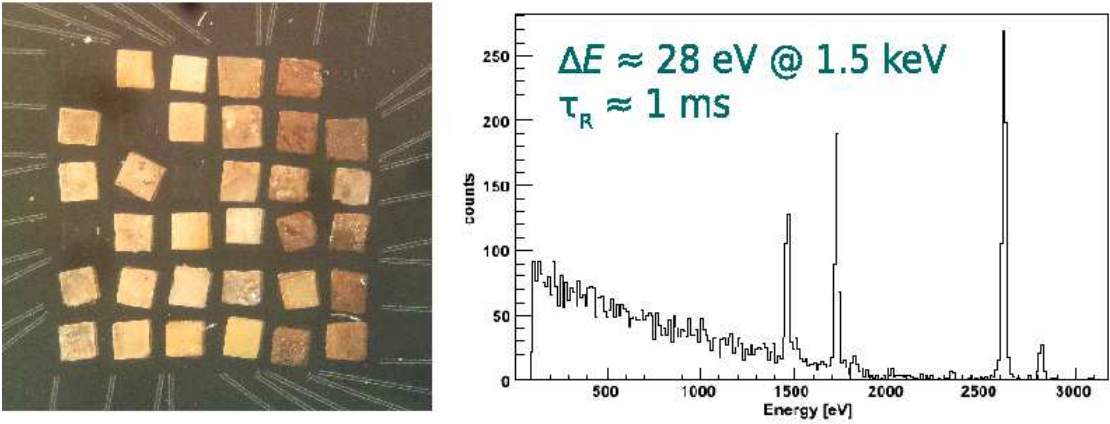}
 \caption{31 \agre\ crystal glued on the first XRS2 array of MARE-1 (left).
 The 16 usable pixel give \de $\approx 47$\,eV at 2.6\,keV, $\tau_R \approx 1$\,ms. Spectrum measured with the best pixel (right).}
 \label{fig:mare-1}
\end{figure}

\subsection{Future of rhenium experiments}
From the MARE experience, it is clear that a large scale neutrino mass experiment based on \Re\ beta decay is not foreseeable in the
near future. It would require a major step forward in the understanding of the superconductivity of rhenium but,  after more than 20
years of efforts, this is not anymore in the priorities of the LTD scientific community.
Besides the intrinsic problems of metallic rhenium, there are other considerations which make rhenium microcalorimeters not quite an appealing choice for high statistics measurements.
Because of the large half life of \Re, the specific activity of metallic rhenium is too low to design pixels with both high performance and
high intensity beta sources. The \Re\ activity required by a high statistics experiment must be therefore distributed over a large number of pixels -- of the order of $10^5$ -- while the difficulties inherent with the production of high quality metallic rhenium absorbers contrast with the  full micro-fabrication of the arrays. MARE-1 also demonstrated that \agre\ is not a viable alternative to metallic rhenium.
For these reasons the new hope for a calorimetric neutrino mass experiment with LTDs is \Ho. 

\section{Current Experiments}
\subsection{Calorimetric absorption spectrum of \Ho\ EC}
\label{sec:holmium}
A.\,De Rujula introduced the idea of a calorimetric measurement of \Ho\ EC decay already in 1981 \cite{derujula1981new}, but
it was only one year later that this idea was fully exploited in the paper written with M.\,Lusignoli \cite{de1982calorimetric}. 
The EC decay 
\begin{equation}
^{163}\mathrm{Ho} + e^- \rightarrow ^{163}\mathrm{Dy} + \nu_e
\end{equation}
has the lowest known $Q$ value, around 2.5\,keV, and its half-life of about 4750\,years is much shorter than the $^{187}$Re one. 
In \cite{de1982calorimetric} the authors compute the calorimetric spectrum and give also an estimate of the statistical sensitivity to the  neutrino mass at the spectrum end-point, including the presence of the pile-up background. 
Unfortunately, at that time the experimental measurements of the $Q$ value were scattered between 
2.3\,keV to 2.8\,keV causing a large uncertainties on the achievable statistical sensitivity. 

\begin{figure}[ht]
\centering
\includegraphics[clip=true,width=0.505\textwidth ]{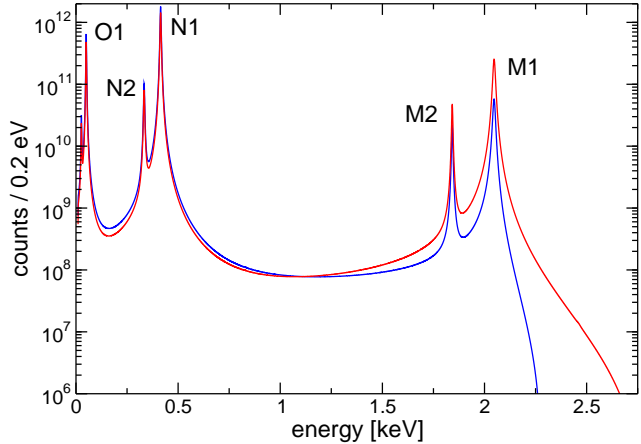}
 \caption{Calculated \Ho\ EC calorimetric spectrum for $Q=2.3$\,keV (blue) and $Q=2.8$\,keV (red), and for $\Delta E=2$\,eV and $N_ev=10^{14}$. }
 \label{fig:ho163spe}
\end{figure}
A calorimetric EC experiment records all the de-excitation energy and therefore it measures the escaping neutrino energy $E_\nu$  -- see (\ref{eq:q_ec}). The de-excitation energy $E_c$ is the energy released by all the atomic radiation emitted in the process of filling the vacancy left by the EC decay, mostly electrons with energies up to about 2\,keV 
(the fluorescence yield is less than $10^{-3}$) \cite{derujula1983seeec}. The calorimetric spectrum has lines at the ionization energies $E_i$ of the captured electrons. These lines have a natural width $\Gamma_i$ of a few eV, therefore the actual spectrum is a continuum with marked peaks with Breit-Wigner shapes (Figure\,\ref{fig:ho163spe}). 
The spectral end-point is shaped by the same neutrino phase space factor $(Q-E)\sqrt{(Q-E)^2-m_{\nu_e}^2}$ that appears in a beta decay spectrum, with the total de-excitation energy $E_c$ replacing the electron kinetic energy $E$. 
For a non-zero $m_{\nu_e}$, the de-excitation (calorimetric) energy $E_c$ distribution is expected to be
\begin {eqnarray}
\label{eq:E_c-distr}
N_{EC}(E_c,m_{\nu_e}) = {G_{\beta}^2 \over {4 \pi^2}}(Q-E_c) \sqrt{(Q-E_c)^2-m_{\nu}^2} \sum_i n_i  C_i \beta_i^2 B_i {\Gamma_i \over 2\,\pi}{1 \over (E_c-E_i)^2+\Gamma_i^2/4} 
\end{eqnarray}
where $G_{\beta} = G_F \cos \theta_C$ (with the Fermi constant $G_F$ and the Cabibbo angle $\theta_C$), $E_i$ is the binding energy of the $i$-th atomic shell, $\Gamma_i$ is the natural width, $n_i$  is the fraction of occupancy, $C_i $ is the nuclear shape factor, 
$\beta_i$ is the Coulomb amplitude of the electron radial wave function (essentially, the modulus of the wave function at the origin) and $B_i$ 
is an atomic  correction for electron exchange and overlap. 
The sum in (\ref{eq:E_c-distr}) runs over the Dy shells which are accessible to the EC with the available $Q$ (M1, M2, N1, N2, O1, O2, and P1). 
The expression (\ref{eq:E_c-distr}) is derived in \cite{de1982calorimetric}, where numerical checks to test the validity of the approximations made are also presented. 

Until about 2010 only three calorimetric absorption measurements were reported in the literature:
\begin{enumerate}
\item{the ISOLDE collaboration used a Si(Li) detector with an implanted source \cite{ravn1984mn,Laegsgaard154030};}
\item{Hartman and Naumann used a high temperature proportional counter with organometallic gas \cite{hartmann1992high};}
\item{Gatti et al. used a cryogenic calorimeter with a sandwiched source \cite{gatti1997calorimetric}.}
\end{enumerate}
However, none of these experiments had the sensitivity required for an 
end-point measurement, therefore they all gave results in terms of  capture rate ratios. 
The most evident limitations of these experiments were statistics and energy resolution.
One further serious trouble for the Si(Li) and the cryogenic detectors was the incomplete energy detection caused by implant damages and  weak 
thermal coupling of the source, respectively.

Recently a new generation of calorimetric holmium experiments has been stimulated by the MARE project.
In fact, despite the shortcomings of previous 
calorimetric experiments, and theoretical and experimental uncertainties, a \Ho\ calorimetric absorption 
experiment seems the only way to achieve sub-eV sensitivity for the neutrino mass.
Moreover, low temperature X-ray microcalorimeters have reached the necessary maturity to 
be used in a large scale experiment with good energy and time resolution, hence they are the detectors of choice for a 
sub-eV holmium experiment.

Thanks to the short \Ho\ lifetime, the limited number of nuclei needed for a neutrino mass experiment --  $10^{11}$\,nuclei for 1 decay/s --
can be introduced in the energy absorber of a low temperature microcalorimeter.
Therefore, holmium experiments can leverage the microcalorimeters development for high energy resolution soft X-ray spectroscopy, whereas rhenium experiments would need a dedicated development of detectors with metallic rhenium absorbers. 
Small footprint kilo-pixel arrays can be fully fabricated with well established micro-fabrication techniques.

Indeed, in microcalorimeters with metallic absorbers such as gold, the relatively high concentration of holmium (J = 7/2) could cause an excess heat capacity due to hyperfine level splitting in the metallic host \cite{enss2005low} and thereby degrade the microcalorimeter performance. Low temperature measurements have been already carried out in the framework of the MARE project to assess the gold absorber heat capacity (at temperatures $<150$\,mK), both with holmium and erbium implanted ions \cite{prasai2013thermal}. Those tests did not show any excess heat capacity, but more sensitive investigations need to be carried out.

The Genova group pioneered the application of LTDs to the measurement of the \Ho\ calorimetric spectrum \cite{gatti1997calorimetric}
and continued this research until it converged in the MARE project \cite{Gastaldo2004224}. For long, the focus has been on the production of the \Ho\ isotope, on the chemistry of metallic holmium, and on the techniques to embed the isotope in the detector absorbers.

The new experiments, now ready to start the  production  of high resolution detectors for the high statistics
calorimetric measurement of the \Ho\ EC decay spectrum, will be the subject of next sections.

\subsection{The $Q$ value of \Ho\ decay}
\label{sec:holmesQ}
\begin{figure}[ht]
\centering
\includegraphics[clip=true,width=0.505\textwidth ]{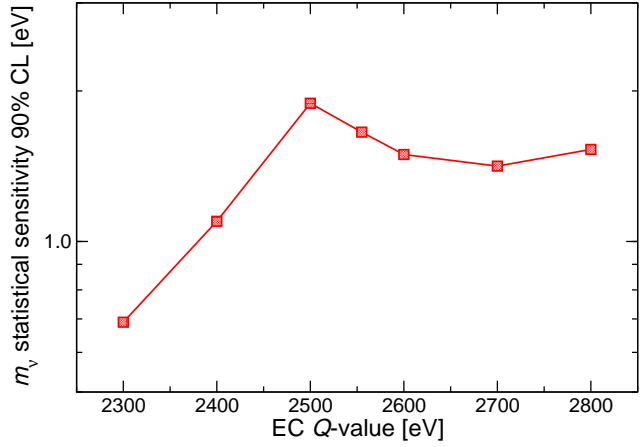}
 \caption{Monte Carlo computed statistical sensitivity as function of $Q$ for \Nev$= 3\times10^{13}$ and with \fpp$=3\times10^{-4}$, \de\fwhm$= 1$\,eV, and no background.}
 \label{fig:holmes}
\end{figure}

Until very recently, the question of the exact $Q$ value of the \Ho\ EC decay was not settled.
Although the results showed a general tendency to accumulate around 2.8\,keV, especially restricting to the calorimetric measurements \cite{Laegsgaard154030,gatti1997calorimetric,ranitzsch2012development}, the reliability of the capture ratios as tool for determining $Q$ remained questionable.
Indeed, the $Q$ has never been measured directly from the end-point of the \Ho\ EC spectrum, but only from the capture ratios $\lambda_i/\lambda_j$, whose accuracy is limited (\S\,\ref{sec:betadecay}). The  currently recommended value of $Q$ is 2.555$\pm$0.016 keV \cite{reich2010nuclear}, but it is deduced from a limited set of data.
The statistical sensitivity of a \Ho\ experiment depends strongly on how close the end-point and the M2 capture peak are.
To a good degree of approximation, the Lorentzian tail of the M1 peak centered at $E_{M1}=B(\mathrm{M1})$ dominates the end-point, and for \mne\ equal to zero one has
\begin{equation}
N_{EC}(E_c,m_{\nu_e}=0) \propto \frac{(Q-E_c)^2}{(E_{M1}-E_c)^2} = \frac{(Q^\prime-E^\prime)^2}{E^{\prime\,2}}
\end{equation}
where $E^\prime = E_c-E_{M1}$ and $Q^\prime = Q - E_{M1}$. It can be shown that, in these conditions, the neutrino mass sensitivity is $\Sigma_{EC}(m_{\nu_e}) \propto Q^\prime$.
The uncertainty on $Q$, therefore, turns into the difficulty to design a \Ho\ experiment and to predict its sensitivity reach (Fig.\,\ref{fig:holmes}).
Indeed, the shift of attention from \Re\ to \Ho\ has been eased by the reasonable hope that a very low $Q$ could greatly enhance the achievable sensitivity of an \Ho\ experiment. 

Very recently, the $Q$ value was determined from a measurement of \Ho-\Dy\ mass difference using the Penning trap mass spectrometer SHIPTRAP  \cite{eliseev2015direct}. The measured value $\Delta m = 2833\pm 30_{stat}\pm15_{sys}$ confirms the most recurrent $Q$ measured in recent calorimetric experiments, although chemical shifts may still be expected for \Ho\ embedded in the LTD absorbers. 
The knowledge of the $Q$ value is indeed a crucial ingredient for the optimal design of an experiment, while its limited precision and accuracy
prevent from using it as fixed parameter when the experimental data are interpolated to assess the neutrino mass \cite{otten2006q}. 
Nevertheless, a comparison of the $Q$ from the interpolation with a value obtained with an independent measurement -- such as the \Ho-\Dy\ mass difference -- is a powerful tool to pinpoint systematic effects.

In any case, the direct assessment of $Q$ from the end-point of the calorimetric spectrum, remains the first important goal of 
upcoming high statistics measurements.

\begin{figure}[ht]
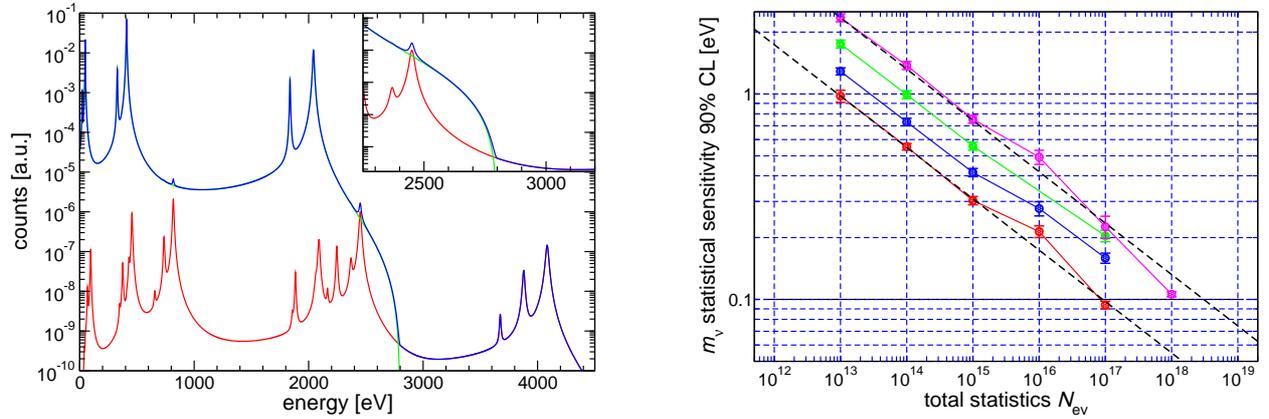

 \centering
 	  \includegraphics[clip=true,width=0.47\textwidth ]{Ho-singlehole-pup}\hfill
 	  \includegraphics[clip=true,width=0.45\textwidth ]{q28}
 \caption{Calculated experimental \Ho\ EC calorimetric spectrum for $Q=2.8$\,keV, $\Delta E=2$\,eV, $f_{pp}=10^{-4}$, and $N_ev=10^{14}$ (blue). The pile-up spectrum is the red curve (left). Monte Carlo estimated statistical sensitivity for $\Delta E=1$\,eV, $\tau_R=1$\,$\mu$s, and for $f_{pp}=10^{-3}$, $10^{-4}$, $10^{-5}$, and $10^{-6}$ (from top to bottom). To two dashed lines correspond to (top)
$A_{EC}=1000$\,Bq and $N_{dey}\times t_M=10^7$\,detector$\times$year and (bottom) $A_{EC}=1$\,Bq and $N_{dey}\times t_M= 3\times10^9$\,detector$\times$year
(right).
}
\label{fig:statsensHo}
\end{figure}

\begin{figure}[htb]
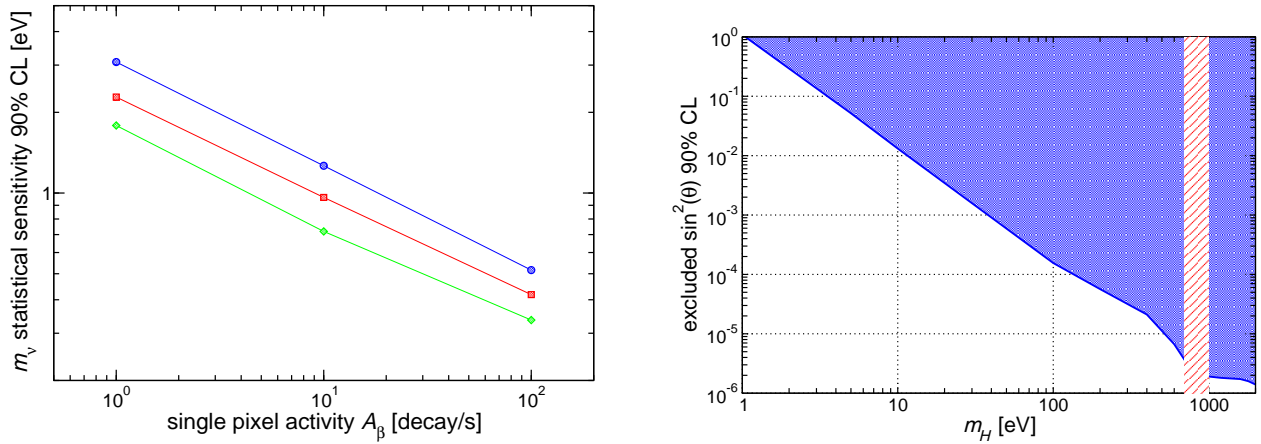

 \centering
 	  \includegraphics[clip=true,width=0.47\textwidth ]{sens_const-exp}\hfill
 	  \includegraphics[clip=true,width=0.47\textwidth ]{runHH006}
 \caption{Calculated experimental \Ho\ EC calorimetric spectrum for $Q=2.8$\,keV, $\Delta E=1$\,eV, for a constant exposure of $10^5$\,detector$\times$year, and for (top to bottom) $\tau_R=10$\,$\mu$s, $1$\,$\mu$s, and $0.1$\,$\mu$s (left). Sensitivity to heavy sterile neutrinos detected from kinks in a \Ho\ calorimetric spectrum with $Q=2.8$\,keV, $N_{ev}=3\times10^{13}$, $\Delta E=1$\,eV, and $f_{pp}=3\times10^{-4}$.
 (right).
}
\label{fig:hoheavy}
\end{figure}
\subsection{Statistical sensitivity}
While the complexity of both the EC and the pile-up spectra make an analytical estimate of the statistical sensitivity an impossible task, a
Monte Carlo approach analogous to the one described in \S\,\ref{sec:stat-sys} can give useful results \cite{nucciotti2014statistical}.
Most of the considerations made for \Re\ are also valid in the case \Ho. The general conclusion about the importance of the total statistics
is well exemplified by Fig.\,\ref{fig:statsensHo} for the now established $Q$ value of 2800\,eV: indeed, the high $Q$ value raises the stakes of the experimental challenge and the prospects for a  sub-eV sensitivity are  scaled down. 
Table\,\ref{tab:sensHo} shows the exposures required for two possible experiments aiming at a \mne\ sensitivity of 0.2 and 0.1\,eV, respectively (see also Fig.\,\ref{fig:hoheavy}).
Although it may be possible to design microcalorimeters with a high \Ho\ activity, sub-eV sensitivity will likely require arrays with total
number of channels of the order of $10^6$. Indeed, there are several limitations to the possible activity \Aec, such as, for example, the effect of the \Ho\ nuclei on the detector performance or  detector cross-talk and dead time considerations. 
As shown in \S\,\ref{sec:stat-sys}, a high activity \Aec\ causes an increase of $f_{pp}$ and thereby a reduced sensitivity to radioactive
background. This, along with the relative thinner aspect-ratio of microcalorimeters with \Ho, 
makes it likely that it is not strictly necessary to operate arrays in an underground laboratory \cite{nucciotti2014statistical}.

No high statistics measurement faced so far with the task of a careful estimation of systematic uncertainties. Nevertheless it is fair to say that there are some substantial differences between the systematic uncertainties expected for \Re\ and \Ho\ experiments, which are worth a mention.
To avoid spectral distortions due to the escape of radiation, the absorbers must provide a $4\pi$ encapsulation with a minimum thickness of few microns. For gold absorbers, Monte Carlo simulations indicate a thickness of 2\,$\mu$m for a 99.99998\% (99.927\%) absorption of 2\,keV electrons (photons). 

Furthermore, the M1 and M2 peaks in the calorimetric spectrum provide a useful tool to evaluate the detector response function overcoming the problems related to the use of an external X-ray source (\S\,\ref{sec:stat-sys}). The same peak can also be exploited for energy calibration, for tracking and correcting gain drifts, and for easing the summation of the spectra measured with the many pixels of the arrays.

\begin{table}[]
 \caption{\label{tab:sensHo} Experimental exposure required for various target statistical sensitivities, with no background and two
 different sets of detector parameters.}
 \begin{center}
\begin{tabular}{ccccccc}
\hline 
$Q$ & target sensitivity & $A_{EC}$ & $\Delta E$ & $\tau_R$ & $N_{ev}$ & exposure $T$ \\[0pt]
[eV] & [eV]  &	[counts/s] & [eV] & [ $\mu s$ ] & counts & [detector$\times$year] \\
\hline 
 2800 & 0.2 & 100 & 1 & 0.1 &    $9.8\times10^{15}$ & $3.1\times10^6$\\
 2800 & 0.1 & 100 & 0.3 & 0.1 &   $1.9\times10^{17}$ & $5.9\times10^7$\\
\hline 
\end{tabular} 
\end{center}
\end{table}

\subsection{A more precise description of the \Ho\ EC spectrum}
\label{sec:ho163spe}
While the question of the actual $Q$ value of the \Ho\ EC transition is now settled, many authors are still debating about the 
precise shape of the calorimetric spectrum.
Indeed, (\ref{eq:E_c-distr}) is only an approximation. Already in the original work \cite{de1982calorimetric}, it was demonstrated the applicability of two approximations: the neglection of possible interference between the capture from different levels,
and the inclusion of transitions with  off-shell intermediate states such as K and L.

Riisager in 1988 \cite{riisager1988low} discussed the distortions of the Lorentzian peak shape 
expected when considering 
that, in the atomic radiation cascade, the atomic phase space available at each step is altered by the natural width of previous transitions.

More recently, beginning with Robertson papers \cite{robertson2014can,robertson2015examination}, some authors started to recognize that the sum in (\ref{eq:E_c-distr}) must be 
extended to more transitions which initially were deemed as negligible \cite{de2013two}. 
This is caused by the incomplete overlap between the Ho and Dy atomic wave functions. 
Recalling that calorimeters measure the neutrino energy $E_\nu$, while writing (\ref{eq:E_c-distr}) it was assumed  
\begin{equation}
E_\nu=Q-B(\mathrm{H})
\end{equation}
where $B(\mathrm{H})$ is the binding energy of shell H in a Dy atom, i.e. the energy to fill the hole H in the Dy$^+$ atom.
But this is not correct. The hole H is in a neutral Dy atom with an extra (the eleventh) 4f electron, because the parent Ho atom 
has an electronic configuration which differs from the one of Dy in the number of $4f$ electrons (11 vs. 10) (see also \cite{springer1985enhanced}). 
Following \cite{Bambynek} this can be expressed as
\begin{equation}
E_\nu=Q-B(\mathrm{H})-E_R 
\end{equation}
where $E_R$ is a correction which accounts for the imperfect atomic wave functions overlap. So the capture peaks in the calorimetric
spectrum are expected to be shifted by a small amount which Roberston \cite{robertson2015examination} calculated as $E_R\approx B(4f)_\mathrm{Ho}$, where $B(4f)_\mathrm{Ho}$ is the binding energy of the $4f$ electron in the Ho atom.
The atomic wave function mismatch goes along with the possibility of shake-up and shake-off processes, adding more 
final states to the \Ho\ EC transition and, therefore, more terms in the sum in (\ref{eq:E_c-distr}).
These processes are the ones responsible for the presence in the final states of two (or more) vacancies created in the Dy atom, along with the extra $4f$ electron.
The second vacancy is left by an atomic electron which has been shaken by the sudden change in the wave functions to an higher bound unoccupied state (shake-up) or to the continuum (shake-off).
In the case of shake-up processes the neutrino energy is given by \cite{robertson2015examination}
\begin{equation}
E_\nu = Q-B(\mathrm{H1})-B(\mathrm{H2}) - E_R
\end{equation}
and the contribution to (\ref{eq:E_c-distr}) it is just another Lorentzian peak term with $E_i = B(\mathrm{H1})+B(\mathrm{H2}) +E_R$.
The case of the shake-off process is more complex because it is a three-body process
\begin{equation}
\label{eq:shakeoff}
^{163}\mathrm{Ho}  \rightarrow ^{163}\mathrm{Dy^{H_1H_2}} + e^- + \nu_e
\end{equation}
and
\begin{equation}
E_\nu + E_e  = Q-B(\mathrm{H1})-B(\mathrm{H2}) - E_R
\end{equation}
The corresponding contribution to (\ref{eq:E_c-distr}) is not a narrow line since $E_e$ adds up to the observable atomic dis-excitation $B(\mathrm{H1})-B(\mathrm{H2}) - E_R$. The actual shape of the energy spectrum of the shaken-off electrons can be calculated as shown in \cite{de2015calorimetric}. 
In general, the probability for the multi-hole processes is small and it can be estimated to be of the order of $10^{-5}$ \cite{de2013two}.
The precise calculation of the 2- or 3-hole processes probability is treated in many recent papers \cite{robertson2015examination,faessler2015determination,faessler2015improved}, with the purpose 
to improve past results from \cite{carlson1973calculation}, although, so far, all calculations apparently consider only shake-up processes.

\begin{figure}[ht]
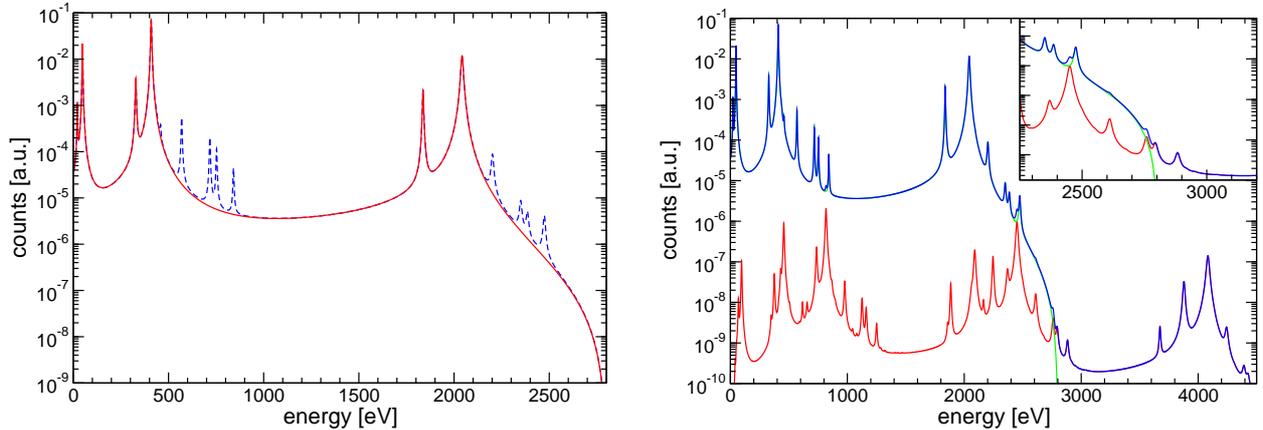

 \centering
  	  \includegraphics[clip=true,width=0.48\textwidth ]{Ho-cfr-doublehole}\hfill
	  \includegraphics[clip=true,width=0.48\textwidth ]{Ho-doublehole-pup}
 \caption{Predicted single and double holes calorimetric spectrum of the  EC decay of $^{163}$Ho with $Q=2.8$\,keV (left). 
The experimental spectrum analogous to the one in Fig.\,\ref{fig:statsensHo} with double hole transitions included (right).}.
 \label{fig:fig1}
 \label{fig:Ho-doublehole-pup}
\end{figure}

In Figure\,\ref{fig:fig1} the dashed line is the \Ho\ EC calorimetric spectrum calculated including 2-holes excitations and using the parameters
calculated in \cite{faessler2015determination}. 

The awareness of the above corrections to (\ref{eq:E_c-distr}) triggered some skepticism about the actual feasibility of a neutrino mass measurement from the end-point
of the calorimetric \Ho\ EC spectrum. The main argument is that, since the neutrino mass is searched as the difference between 
the observed experimental spectrum and the theoretical one for $m_{\nu_e}=0$, the {\it a priori} knowledge of the latter one
is an absolute condition.

However, it can be argued that all the above corrections are immaterial for the neutrino mass measurement since, at the end-point, they
have no other effect than altering the overall rate, i.e. the spectrum normalization, while the shape remains determined by the phase space factor. 
The change in the rate may be even in the direction of a favorable increase. 
This is indeed the beauty of the calorimetric experimental approach.

For what concerns the additional 2- or 3-holes transitions, a more subtle threat to the neutrino mass measurement with \Ho\ is brought 
by the underlying pile-up spectrum: as it can be seen in Fig.\,\ref{fig:Ho-doublehole-pup}, these higher order transitions cause additional peaks to appear in the end-point vicinity.

\subsection{$^{163}$Ho production}
\Ho\ is not a naturally occurring isotope: it was discovered at Princeton in a sample of \Er\ that was neutron irradiated in a nuclear reactor \cite{naumann1960preparation}.
To carry out neutrino mass experiments with \Ho, the isotope must be produced in fairly large amounts. Upcoming medium size experiments will have to contain about $10^{16}$  nuclei of \Ho\ -- i.e. about 3\,$\mu$g -- for a total activity of the order of $10^5$\,Bq. The isotope production and separation are critical steps in every plan for an ambitious holmium neutrino mass experiment. 
There are many nuclear reactions which can be exploited to produce \Ho. A comprehensive critical evaluation of all possible \Ho\ production routes is presented in \cite{engle2013evaluation} although, presently, not all the cross-sections of the considered processes are experimentally known.

In general, the production process starts with a nuclear reaction, which can be either direct -- such as $^{nat}\mathrm{Dy}(p,xn)^{163}\mathrm{Ho}$ -- or indirect -- such as $^{162}\mathrm{Er}(n,\gamma)^{163}\mathrm{Er}\rightarrow^{163}\mathrm{Ho}$.
These reactions unavoidably co-produce other long-living radioactive species -- also owing to the presence of un-necessary isotopes in the target material -- which need to be removed to prevent interferences to the neutrino mass measurement. Chemical separation of holmium can remove most of them, with the notable exception of the beta decaying isomer $^{166m}$Ho ($\tau_{1/2}=1200$\,years, $E_0=1854$\,keV). 
Geant4 Monte Carlo simulations performed for gold absorbers show that each Bq of $^{166m}$Ho can contribute about 1\,count/eV/day to the background level in the end-point region of the \Ho\ spectrum \cite{alpert2015holmes}. Therefore, $^{166m}$Ho must be removed  by means of a further isotope mass separation step. 
The key parameters of the entire process are the \Ho\ isotope production rate, the $^{166m}$Ho/\Ho\ ratio, and the efficiencies of  chemical and mass separations. They determine the amount of starting material that is required to have the target number of \Ho\ nuclei  to be embedded in the detector absorbers. Of course, also the final embedding process causes further isotope losses which must be considered, although in some approaches the embedding is part of the production process -- e.g. when the embedding is achieved by means of the same accelerator used for mass separation. 
When all efficiencies entering in the process are considered, the \Ho\ needed for the next high statistic measurements is likely to increment to tens or hundreds of MBq.
 
Early experiments used the same process with which the isotope was discovered, i.e. neutron irradiation of \Er. Another route used for past experiments is based on proton spallation with Ta targets.
The experiments presented in the following use either neutron irradiation of
enriched \Er\ targets or proton irradiation of natural Dy targets.

Neutron irradiation of an enriched \Er\ sample is a very efficient route. The starting material is usually enriched \ero\ which is available as by-product of the production of isotopes for medical applications. The large thermal neutron cross-section $\sigma(\mathrm{n},\gamma)\approx 20$\,barns together with the availability of high thermal neutron flux nuclear reactors -- as the one of the Institut Laue-Langevin (ILL, Grenoble, France)  with a thermal neutron flux of about $1.3\times10^{15}$\,n/s/cm$^2$ \cite{koster2012319} -- give an estimated \Ho\ production rate of about 50\,kBq(\Ho)/week/mg(\ero), for \ero\ enriched at 30\% in \Er. 
This rate may be reduced by the yet unknown cross-section of the burn-up process $^{163}\mathrm{Ho}(n,\gamma)^{164}\mathrm{Ho}$. 
Neutron irradiation causes also the production of $^{166m}$Ho owing to the presence of impurities such as $^{164}$Er and $^{165}$Ho in the enriched \ero\ target. If the $^{164}$Er production route prevails, for a 10\% isotopic abundance in the \ero\ target a co-production of  about $A(^{166m}\mathrm{Ho})/A(^{163}\mathrm{Ho})\approx 0.001$ can be expected. 
One drawback of this route is the cost for the enriched \Er\ procurement.

The \Ho\ production via proton irradiation of natural Dy target depends on the proton energy and has a production rate which is not competitive with high-flux reactors, especially for large amounts. 
In \cite{engle2013evaluation} the production rate as a function of the total cumulative charge is estimated to be about few Bq($^{163}$Ho)/$\mu$Ah/g($^{nat}$Dy) for 24\,MeV protons.
$^{166m}$Ho is produced by the neutrons from the reaction $^{164}\mathrm{Dy}(p,n)$ in $(n,\gamma)$ captures on $^{164}\mathrm{Dy}$ or on $^{165}\mathrm{Ho}$ contaminations. Monte Carlo simulations give a co-production lower than $A(^{166m}\mathrm{Ho})/A(^{163}\mathrm{Ho})\approx 10^{-6}$.
In spite of its low efficiency, the use of a natural target and the limited $^{166m}$Ho co-production make this route appealing for small scale experiments.

\begin{figure}[tb]
 \centering
 	  \includegraphics[clip=true,width=0.9\textwidth ]{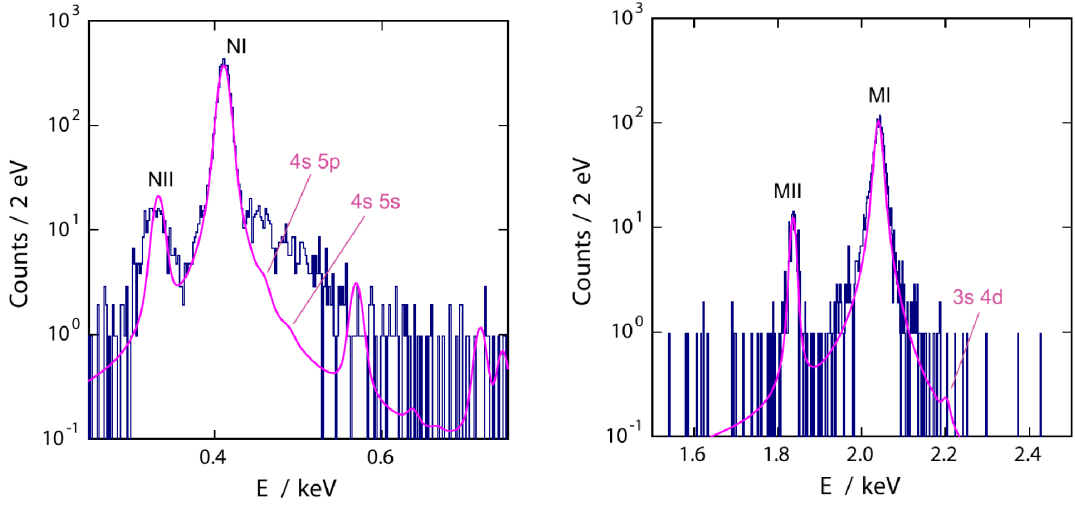}
 \caption{Latest measured spectrum from ECHo, see text (from \cite{gastaldo2015recent}).}
 \label{fig:echospe}
\end{figure}
\subsection{ECHo}
ECHo is a project carried out by the Heidelberg group in collaboration with many other European and Indian groups \cite{echowww}. The mid term goal of this project -- ECHo-1k -- is a medium scale experiment with an array of 1000 MMCs, each implanted with 1\,Bq of \Ho\ \cite{gastaldo2014electron}. 
With energy and time resolutions of at least 5\,eV and 1\,$\mu$s, respectively, a statistical sensitivity of about 20\,eV at 90\% C.L. is expected after one year of measurement (Tab.\,\ref{tab:holmioexp}). The microcalorimeters are derived from the gold detectors with Au:Er sensors designed and fabricated by the Heidelberg group for soft X-rays spectroscopy. 

So far, the results of two prototypes with \Ho\ in the absorbers have been presented.
For the first prototype the \Ho\ isotope was implanted at the isotope separation on-line facility ISOLDE (CERN). Here the \Ho,
produced by spallation with protons on Ta, was accelerated, mass separated, and implanted in the absorbers of four detectors. A total  \Ho\ activity of $10^{-2}$\,Bq was enclosed between two gold films with dimensions $190\times190\times5$\,\mum$^3$.
The results of the characterization of these detectors are reported in \cite{ranitzsch2012development,gastaldo2013characterization}, and include
an energy resolution of about 8\,eV and a remarkable rise time of about 130\,ns.
In the high statistics spectrum, the peaks due to a contamination of co-produced $^{144}$Pm are visible, although decaying with time.
In addition there are structures on the high energy side of the N1 peak which are tentatively interpreted as due to higher order EC transitions.
From this measurement, the intensities of the N1 and M1 lines give $Q=2.800\pm0.080_{stat}$\,keV \cite{ranitzsch2012development}.

For the second prototype the \Ho\ isotope is produced at the ILL high flux nuclear reactor by neutron
irradiating an enriched \ero\ target. The \ero\ sample is chemically purified at Mainz both before and after irradiation.
The \Ho\ in the target is then mass separated and implanted off-line at ISOLDE in the absorbers of two maXs-20 chips.
The maXs-20 chips are arrays of 16 MMCs designed and optimized for soft X-ray spectroscopy \cite{pies2012maxs}. 
About 0.2\,Bq are encapsulated between two gold layers with  dimensions $250\times240\times5$\,\mum$^3$.
Preliminary measurements  (see Fig.\,\ref{fig:echospe}) show an energy resolution of about 12\,eV and a strong reduction of the background, and confirm the structures
on the right side of the N1 \cite{gastaldo2015recent}. The persistence of these structures, in spite of the improvements in the background and in the instrumental line shape, supports their interpretation as due to processes related to the \Ho\ EC decay.
Another preliminary analysis discussed in \cite{de2015calorimetric} interprets these as the broad structures expected for shake-off transitions.

Present ECHo activities are aimed at running ECHo-1k in the next years (2016--2018) and include the development of the microwave multiplexed read-out of the MMCs \cite{kempf2014umltiplexed}, the optimization of MMCs design, and the production of 10\,MBq of high purity \Ho.

\begin{figure}[tb]
 \centering
 	  \includegraphics[clip=true,width=0.9\textwidth ]{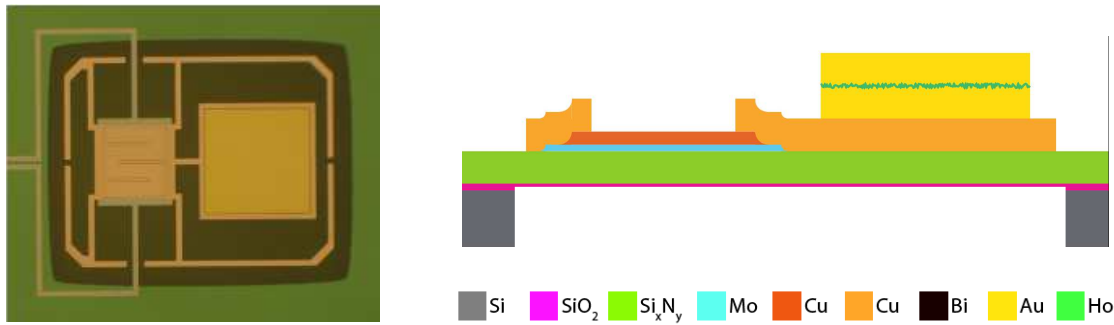}
 \caption{HOLMES TES pixel}
 \label{fig:holmespix}
\end{figure}
\subsection{HOLMES}
HOLMES is an experiment carried out by the Genoa and Milano groups in collaboration with NIST, ILL, PSI, and Lisbon \cite{holmeswww,alpert2015holmes}.
The baseline program is to deploy an array of about 1000 TES based microcalorimeters each with about 300\,Bq of \Ho\ fully embedded in the absorber, with the goal of energy and time resolutions as close as possible to 1\,eV and 1\,\mus, respectively (Tab.\,\ref{tab:holmioexp}).
In this configuration, HOLMES can collect about $3\times10^{13}$ decays in 3 years of measuring time and the expected \mne\  statistical sensitivity is about 1.5\,eV at 90\% C.L. 
The choice of this configuration is driven by the aim to collect the highest possible statistics with a reasonable exposure. Despite the 
high pile-up level and the technical challenge that derives from it, this provides a net improvement on the achievable \mne\ sensitivity and a lower impact of the radioactive background.   

The amount of \Ho\ needed for the experiment is estimated to be about 100\,MBq and it is being produced at ILL by neutron irradiation  of an enriched \ero\ target, subjected to chemical pre-purification and post-separation at PSI (Villigen, Switzerland). 
A custom ion implanter is being set-up in Genova to embed the isotope in the detector absorbers. It consists of a Penning sputter ion source, a magnetic/electrostatic mass analyzer, an acceleration section, and an electrostatic scanning stage. The full system is being designed to achieve an optimal mass separation of \Ho\ vs. $^{166m}$Ho. The implanter will be integrated with a vacuum chamber for the simultaneous evaporation of gold, first  to control the \Ho\ concentration, and then to deposit a final Au layer to prevent the \Ho\ from oxidizing. 
The cathode of the ion source will be made of high purity metallic holmium to avoid end-point deformations due to the different $Q$ shifts in diverse chemical species. The metallic holmium  will be obtained by thermal reduction at about 2000\,K, using the reaction Ho$_2$O$_3$+2Y({\it met})$\rightarrow$2Ho({\it met})+ Y$_2$O$_3$ \cite{ferri2014preliminary}.

HOLMES uses TES microcalorimeter arrays with $\mu$MUX read-out, both fabricated at NIST (Boulder, USA). The DAQ exploits the
Reconfigurable Open Architecture Computing Hardware (ROACH2) board equipped with a FPGA Xilinx Virtex6 \cite{mchugh2012readout}, which has been developed in the framework of CASPER (Collaboration for Astronomy Signal Processing and Electronic Research).

Presently, the collaboration is working on the optimization of the isotope production processes. Two \ero\ samples have been irradiated at ILL and processed at PSI. ICP-MS is used to assess the amount of \Ho\ produced and the efficiency of the chemical separation. From preliminary assessments, the total available \Ho\ activity is about 50-100\,MBq.

The optimization of the pixel design is also in progress \cite{jim2015ltd16} and fig.\,\ref{fig:holmespix} shows the design that best matches HOLMES specifications. The absorber is made of gold and,  to avoid interference to the superconducting transition, it is placed side-by-side with the Mo/Cu sensor on a silicon nitride membrane. The design also includes features to control the microcalorimeter speed. Energy and time resolutions are
within a factor 2-3 of the target ones, owing also to new algorithms for pile-identification \cite{ferri2015ltd16,alpert2015ltd16}.   

HOLMES is expected to start data taking in 2018, but a smaller scale experiment with a limited number of pixels will run in 2017, with the aim to collect a statistics of about $10^{10}$ decays in a few months.

\begin{figure}[ht]
 \centering
 	  \includegraphics[clip=true,width=0.45\textwidth ]{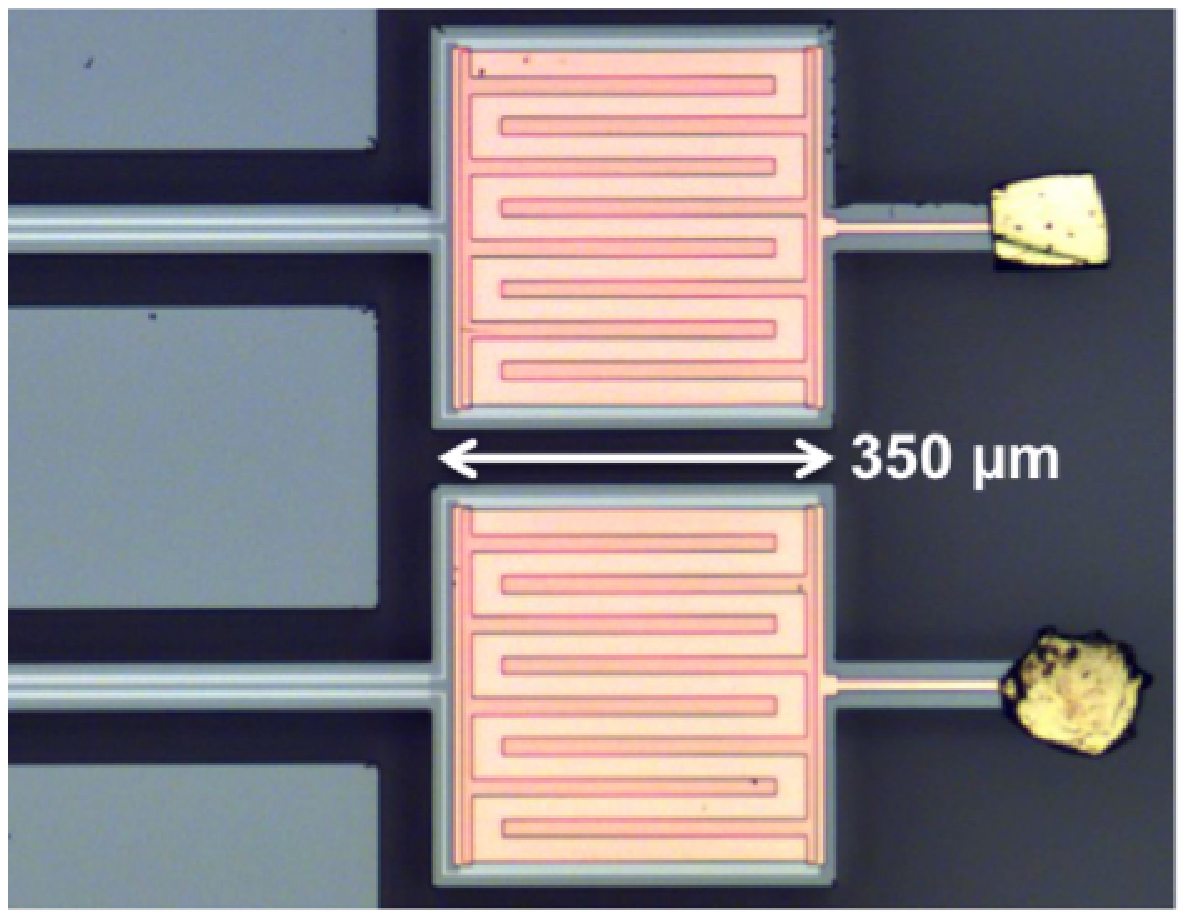}\hfill
 	  \includegraphics[clip=true,width=0.5\textwidth ]{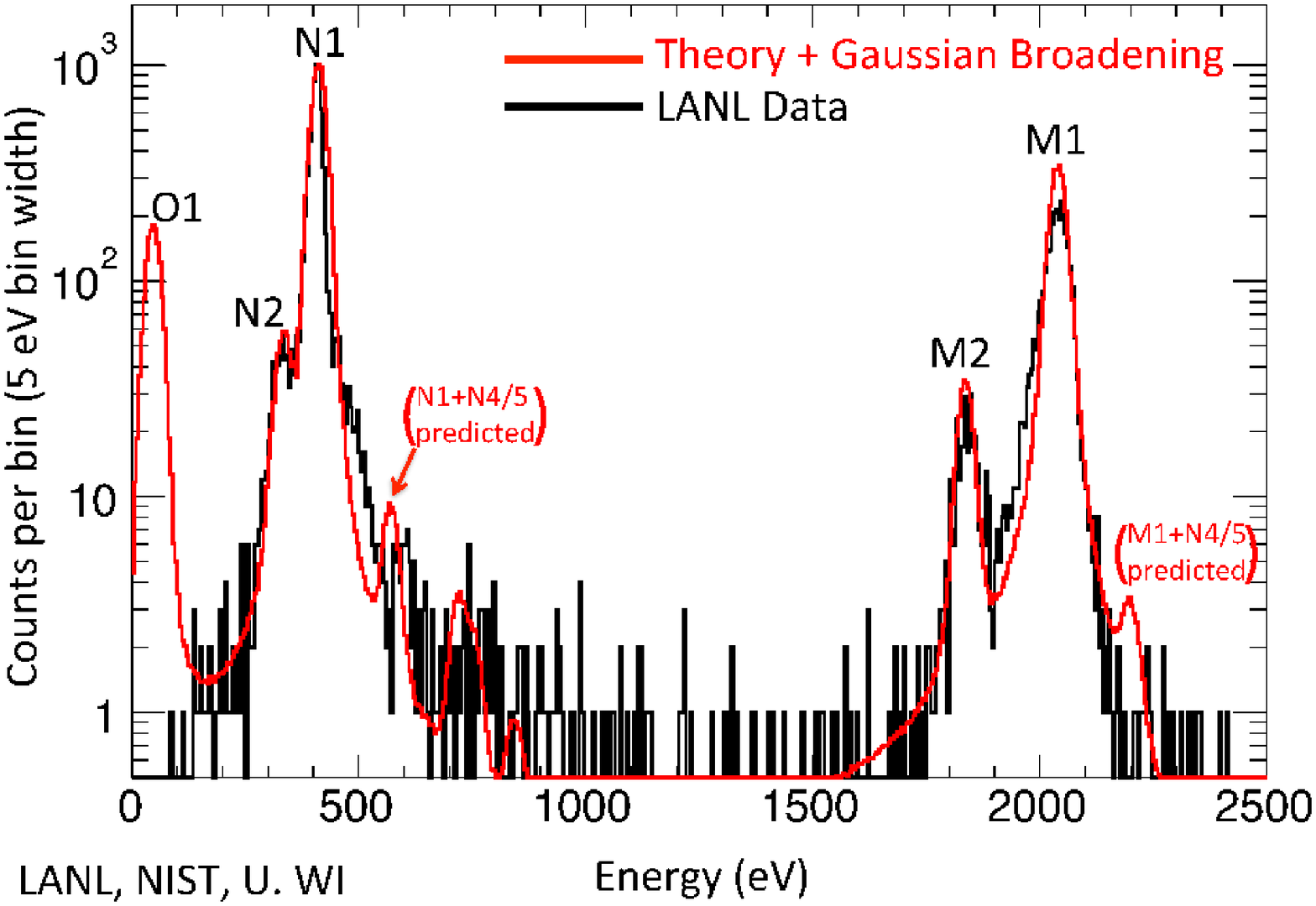}
 \caption{NUMECs TES detectors (left) and the latest spectrum measured from NuMECS (right), see text (from  \cite{croce2015development}).}
 \label{fig:numecsspe}
\end{figure}
\subsection{NuMECS}
NuMECS is a collaboration of several US institutions (LANL, NIST, NSCL, CMU) with the goal to critically assess the
potential of  holmium calorimetric neutrino mass measurements \cite{numecswww}. The NuMECS program includes the validation of the 
isotope production, purification, and sensor incorporation techniques, the scalability  to high resolution LTD arrays, and the 
understanding of underlying nuclear and atomic physics.  

Recent work has successfully tested the \Ho\ production via proton irradiation of a natural dysprosium target.
About 3\,MBq of \Ho\ have been produced by irradiating about 13\,g of high purity natural dysprosium with $3.4\times10^4$\,$\mu$Ah of 25\,MeV protons at the Los Alamos National Laboratory Isotope Production Facility (IPF). At the same time, a cation-exchange high performance liquid chromatography (HPLC) procedure for the chemical separation of holmium has been developed and a separation efficiency of about 70\% has been measured.

For the present testing phase, NuMECS uses TES microcalorimeters fabricated by NIST and specially designed to be mechanically robust.
The TES sensor is at the end of a silicon beam, close to a pad used for testing the attachment of a wide range of absorbers (Fig.\,\ref{fig:numecsspe}). 

To incorporate the isotope in the microcalorimeter absorber, NuMECS exploits the drying of solutions containing the \Ho\ isotope
onto thin gold foils. After testing many procedures, the best results were recently obtained by deposition of an aqueous solution on nanoporous gold on a regular gold foil, followed by annealing in dilute H$_2$ atmosphere at $800^\circ$C. The microcalorimeter absorber is made by folding and pressing a small piece of the gold foil.

Figure\,\ref{fig:numecsspe} shows a spectrum collected in 40\,hours \cite{croce2015development}. The \Ho\ activity is about 0.1\,Bq. 
Peaks have a low energy tail and show an excess broadening, explained as caused by thermalization noise. 
A fit of the M1 peak gives a \de \fwhm\ of about 43\,eV, inclusive of the peak natural width. 
All peaks are found within 1\% of the tabulated positions. Remarkably, the spectrum does not show any of the satellite peaks predicted in \cite{faessler2015electron}, although the statistics is still limited. There is instead an unexplained shoulder on the high energy side of the N1 peak, which
resemble the structure observed by ECHo and interpreted as shake-off transition in \cite{de2015calorimetric}.

NuMECS future plans include the deployment of four 1024 pixel arrays, aiming to a \mne\ statistical sensitivity of about 1\,eV.

\begin{table}[tb]
\centering
\caption{Comparison of the ECHo, HOLMES, and NuMECS projects.}
\newcommand{\mc}[3]{\multicolumn{#1}{#2}{#3}}
{%
\begin{tabular}{rccc}\hline
\label{tab:holmioexp}
× & \mc{1}{c}{ECHo} & \mc{1}{c}{HOLMES} & \mc{1}{c}{NUMECS}\\\hline
\Ho\ production & \mc{1}{c}{$^{162}Er(n,\gamma)$} & \mc{1}{c}{$^{162}Er(n,\gamma)$} & \mc{1}{c}{Dy(p,)}\\
absorber & \mc{1}{c}{gold} & \mc{1}{c}{gold} & \mc{1}{c}{nanoporous gold}\\
sensor & \mc{1}{c}{Au:Er magnetic} & \mc{1}{c}{TES Mo/Cu} & \mc{1}{c}{TES Mo/Cu}\\
\cline{1-4}
\mc{4}{c}{present status}\\
\cline{1-4}
$\Delta E$ at M1 peak [eV] & \mc{1}{c}{12} & \mc{1}{c}{--} & \mc{1}{c}{43 (incl. $\Gamma_\mathrm{M1}$)}\\
$\tau_{rise}$ [\mus] & \mc{1}{c}{--} & \mc{1}{c}{--} & \mc{1}{c}{--}\\
$A_{EC}$ [Bq] & \mc{1}{c}{0.2} & \mc{1}{c}{--} & \mc{1}{c}{0.1}\\
\cline{1-4}
\mc{4}{c}{projected ($E_0=2800$\,eV)}\\
\cline{1-4}
$N_{det}$ & \mc{1}{c}{1000} & \mc{1}{c}{1000} & \mc{1}{c}{4096}\\
$\Delta E$ [eV] & \mc{1}{c}{$<5$} & \mc{1}{c}{1} & \mc{1}{c}{--}\\
$\tau_{rise}$ [\mus] & \mc{1}{c}{$<1$} & \mc{1}{c}{1} & \mc{1}{c}{--}\\
$A_{EC}$ [Bq/detector] & \mc{1}{c}{1} & \mc{1}{c}{300} & \mc{1}{c}{100}\\
$f_{pp}$ & \mc{1}{c}{$10^{-6}$} & \mc{1}{c}{$3\times 10^{-4}$} & \mc{1}{c}{--}\\
$t_M$ [y] & \mc{1}{c}{1} & \mc{1}{c}{3} & \mc{1}{c}{1}\\
$\Sigma_{90}(m_{\nu_e})$ [eV] & \mc{1}{c}{20} & \mc{1}{c}{2} & \mc{1}{c}{1}\\
\hline
\end{tabular}
}%
\end{table}

\section{Summary and outlook}
The use of \Re\ and \Ho\ as an alternative to tritium for the direct measurement of the neutrino mass was 
proposed in the same years when the low temperature detector technology was moving the first steps. 
The idea of making low temperature detectors with rhenium absorbers immediately caught up, both because
it appeared to be of almost immediate realization and because of its appealing impact on X-ray spectroscopy.
Unfortunately, on the long run the  technological difficulties inherent to the use of superconducting rhenium caused
the interest of the low temperature detector community to fade away, and the neutrino mass projects to have the
same fate as the X-ray applications of rhenium detectors.

\Ho\ measurements took more time to take off, as if they were awaiting for the readiness of the technology of microcalorimeter arrays applied to
high resolution spectroscopy of soft X-rays. 
Now, \Ho\ neutrino mass experiments are ready to leverage this mature technology, and the interest of the
low temperature detector community is high, as demonstrated by the number of parallel efforts.
Despite the unluckily high $Q$ value, the good prospects to perform high statistics neutrino mass measurements in the
next couple of years are also attracting the attention of the neutrino physics community as a valid complementary alternative to KATRIN.

\section*{Acknowledgments}
I would like to thank Andrea Giachero, Marco Faverzani, Elena Ferri, Andrei Puiu, and Monica Sisti, who in various ways supported me
with the writing of this paper; Maurizio Lusignoli and Alvaro De Rujula, for the many useful discussions; and Loredana Gastaldo, Michael Rabin and Mark Philip Croce, for providing me with updated informations on their experiments. 
\noindent
\normalem
\bibliography{mybiblio,monbiblio}{}
\bibliographystyle{ieeetr}
\end{document}